\documentclass[1p]{elsarticle}

\usepackage{varioref}%  smart page, figure, table, and equation referencing
\usepackage{wrapfig}%   wrap figures/tables in text (i.e., Di Vinci style)
\usepackage{nomencl}%   nomenclature generation via makeindex
\usepackage{subfigure}
\usepackage[colorlinks]{hyperref}%  hyperlinks [must be loaded after dropping]
\usepackage{amsmath}
\usepackage{amssymb}
\usepackage{cleveref}
\usepackage{bm}
\usepackage{algorithm}
\usepackage{algorithmicx}
\usepackage{algpseudocode}

\usepackage{etoolbox}
\makeatletter
\patchcmd{\ps@pprintTitle}% <cmd>
  {Preprint submitted}% <search>
  {Updated manuscript submitted}% <replace>
  {}{}% <succes><failure>
\makeatother
\journal{arXiv}

\DeclareMathOperator{\sign}{\mathrm{sign}}

\newcommand{\pd}[2]{\frac{\partial{#1}}{\partial{#2}}} %partial derivative

\title{A High-Order Weighted Compact High Resolution Scheme with Boundary Closures for Compressible Turbulent Flows with Shocks}

\author[aa]{Akshay Subramaniam\corref{cor1}\fnref{fn1}}
\ead{akshays@stanford.edu}
\author[aa]{Man Long Wong\corref{cor1}\fnref{fn1}}
\ead{wongml@stanford.edu}
\author[aa,me,ctr]{Sanjiva K. Lele}

\fntext[fn1]{These authors contributed equally to the work.}
\address[aa]{Department of Aeronautics \& Astronautics, Stanford University, Stanford, CA 94305, USA}
\address[me]{Department of Mechanical Engineering, Stanford University, Stanford, CA 94305, USA}
\address[ctr]{Center for Turbulence Research, Stanford University, Stanford, CA 94305, USA}

\begin{document}

\begin{abstract}
We present an improved high-order weighted compact high resolution (WCHR) scheme that extends the idea of weighted compact nonlinear schemes (WCNS's) using nonlinear interpolations in conjunction with compact finite difference schemes for shock-capturing in compressible turbulent flows. The proposed scheme has better resolution property than previous WCNS's. This is achieved by using a compact (or spatially implicit) form instead of the traditional fully explicit form for the nonlinear interpolation. Since compact interpolation schemes tend to have lower dispersion errors compared to explicit interpolation schemes, the proposed scheme has the ability to resolve more fine-scale features while still having the ability to provide sufficiently localized dissipation to capture shocks and discontinuities robustly. Approximate dispersion relation characteristics of this scheme are analyzed to show the superior resolution properties of the scheme compared to other WCNS's of similar orders of accuracy. Conservative and high-order accurate boundary schemes are also proposed for non-periodic problems. Further, a new conservative flux-difference form for compact finite difference schemes is derived and allows for the use of positivity-preserving limiters for improved robustness. Different test cases demonstrate the ability of this scheme to capture discontinuities in a robust and stable manner while also localizing the required numerical dissipation only to regions containing discontinuities and very high wavenumber features and hence preserving smooth flow features better in comparison to WCNS's.
\end{abstract}


\maketitle

% \section*{Nomenclature}
% 
% \begin{tabbing}
%   XXX \= \kill% this line sets tab stop
%   $J$ \> Jacobian Matrix \\
%   $f$ \> Residual value vector \\
%   $x$ \> Variable value vector \\
%   $F$ \> Force, N \\
%   $m$ \> Mass, kg \\
%   $\Delta x$ \> Variable displacement vector \\
%   $\alpha$ \> Acceleration, m/s\textsuperscript{2} \\[5pt]
%   \textit{Subscript}\\
%   $i$ \> Variable number \\
%  \end{tabbing}

\section{Introduction}
Simulations of compressible flows that involve shock waves, contact discontinuities, and turbulence have conflicting requirements. While capturing discontinuities like shock waves, contact surfaces or vortex sheets require numerical dissipation for stabilization, the fine scales of turbulence are severely affected by numerical dissipation. Hence, a method that can adaptively switch between a low dissipation formulation in regions of smooth flow to a formulation that adds sufficient dissipation at discontinuities is of paramount importance. In the past, weighted essentially non-oscillatory (WENO)~\cite{jiang1995efficient, henrick2005mapped, martin2006bandwidth, borges2008improved, hu2010adaptive, hu2011scale} schemes, their variants weighted compact nonlinear schemes (WCNS's)~\cite{deng2000developing, nonomura2007increasing, zhang2008development, liu2015new, wong2017high} and targeted essentially non-oscillatory (TENO)~\cite{fu2016family} scheme have been proposed as methods to provide this adaptation. These schemes capture shocks well and improvements like the WENO6-CU-M2~\cite{hu2011scale} and WCNS6-LD~\cite{wong2017high} schemes localize the numerical dissipation to regions around discontinuities. However, their resolution properties are limited by the underlying explicit reconstruction and interpolation schemes. One way to improve the resolution of the adaptive scheme is to increase the stencil width of the scheme while optimizing the dispersion and dissipation properties under the constraint of same order of accuracy like the TENO scheme with tailored resolution by \citet{fu2017targeted}. Another way for improved resolution is the use of compact or spatially implicit finite difference scheme. \citet{lele1992compact} developed compact finite difference and interpolation schemes that are high order accurate and have spectral-like resolution properties. Although these schemes are well-suited for problems involving turbulence, they cannot be directly used for problems that contain sharp gradient features like shocks unless certain numerical regularization is used. One kind of numerical regularization for compact finite difference schemes is to add numerical dissipation explicitly~\cite{cook2004high, cook2005hyperviscosity, cook2007artificial, bhagatwala2009modified, kawai2010assessment, shankar2010numerical, ghaisas2018unified, subramaniam2018high} in solutions to capture shocks and material interfaces using the localized artificial diffusivity (LAD) first proposed by \citet{cook2005hyperviscosity}. These regularization methods preserve the resolution properties of compact schemes, but are still prone to some mild spurious oscillations near shocks or discontinuities. They also, in some cases, introduce additional time step limitations due to the extra artificial dissipation terms. In addition to adding dissipation terms, solutions typically need to be filtered at every time step for de-aliasing.

As an alternative to adding artificial dissipation explicitly, \citet{deng2000developing} used a compact finite difference scheme with WENO interpolation in the context of WCNS. The process in obtaining flux at midpoints using nonlinear interpolations can be interpreted as a nonlinear filtering process to prevent spurious oscillations near discontinuities. However, the fact that WENO interpolation is explicit limits the effective resolution of the overall scheme even though compact finite difference scheme is used. \citet{ghosh2012compact} developed an upwind-biased compact reconstruction WENO scheme called CRWENO. This method is purely compact, but the scheme is upwind-biased and excessively damps the fine scales of turbulence.

In this paper, we present a newly designed scheme that is based on the WCNS formalism to use a compact finite difference derivative but is also improved with the use of a high-resolution compact nonlinear interpolation scheme. The localized dissipation (LD) nonlinear weights of \citet{wong2017high} are used to provide localized dissipation through adaptive switching between explicit and compact interpolations. Boundary interpolation and derivative schemes are also provided for non-periodic problems. The boundary schemes are conservative, have the same formal order of accuracy as in the interior schemes and are optimized by matching their truncation errors to the interior schemes. The overall improved scheme is shown to have better resolution properties than WCNS's using only explicit interpolations and is also stable and accurate for problems involving inflow-outflow boundaries with significant disturbances when proper boundary treatments are applied.

\section{Numerical methods \label{sec:numerical_method}}
In this section, a scalar conservation law of the following form is considered in a one-dimensional (1D) domain of size $x \in \left[x_a, x_b \right]$ for simplicity:
\begin{equation}
\pd{u}{t} + \pd{F(u)}{x} = 0 \label{eq:scalar_conservation},
\end{equation}
where $u(x,t)$ is a conserved scalar quantity that depends on space $x$ and time $t$ and $F(u)$ is a flux function of $u$. The equation above is discretized on a uniform grid with $N$ cells and the solution $u$ on the cell node at position $x_j = x_a + (j + 1/2) \Delta x$ is denoted by $u_j$, $\forall j \in \{0, \: 1, \: \dots, \:  N-1\}$, where $\Delta x = (x_b - x_a)/N$. The cell midpoints are indexed by half integer values $x_{j+\frac{1}{2}}$, $\forall j \in \{ -1, \: 0, \: 1, \: \dots, \: N-1 \}$. The numerical method described in this section can be easily extended to two-dimensional (2D) and three-dimensional (3D) problems using the method of lines. The extension of the scalar conservation equation to a hyperbolic system of coupled equations such as the Euler equations is discussed in section~\ref{sec:Euler}.

\subsection{Compact and explicit finite difference schemes \label{sec:FD}}
Over the years, various forms of finite difference schemes have been used in WCNS's to obtain the flux derivative in equation~\eqref{eq:scalar_conservation}. \citet{deng2000developing} first used the sixth order compact  midpoint-to-node finite difference (CMD) scheme by \citet{lele1992compact} in following form:
\begin{equation}
\frac{9}{80} \widehat{F}_{j-1}^\prime + \frac{31}{40} \widehat{F}_{j}^\prime + \frac{9}{80} \widehat{F}_{j+1}^\prime = \frac{1}{\Delta x} \left[ \frac{63}{80} \left( \tilde{F}_{j+\frac{1}{2}} - \tilde{F}_{j-\frac{1}{2}} \right) + \frac{17}{240} \left( \tilde{F}_{j+\frac{3}{2}} - \tilde{F}_{j-\frac{3}{2}} \right) \right], \label{eq:CMD}
\end{equation}

\noindent where $\widehat{F}_{j}^\prime$ are numerically approximated first derivatives of flux at cell nodes and $\tilde{F}_{j+\frac{1}{2}}$ are interpolated fluxes at cell midpoints. Since the resolution properties of WCNS's are mainly dominated by the nonlinear interpolations, \citet{nonomura2009effects} suggested using a more efficient explicit sixth order midpoint-to-node finite difference (MD) scheme:
\begin{equation}
\widehat{F}_{j}^\prime = \frac{1}{\Delta x} \left[ \frac{75}{64} \left( \tilde{F}_{j+\frac{1}{2}} - \tilde{F}_{j-\frac{1}{2}} \right) - \frac{25}{384} \left( \tilde{F}_{j+\frac{3}{2}} - \tilde{F}_{j-\frac{3}{2}} \right) + \frac{3}{640} \left( \tilde{F}_{j+\frac{5}{2}} - \tilde{F}_{j-\frac{5}{2}} \right) \right]. \label{eq:MD}
\end{equation}

\citet{nonomura2013robust} later also proposed a robust explicit sixth order midpoint-and-node-to-node finite difference (MND) scheme:
\begin{equation}
\widehat{F}_{j}^\prime = \frac{1}{\Delta x} \left[ \frac{3}{2} \left( \tilde{F}_{j+\frac{1}{2}} - \tilde{F}_{j-\frac{1}{2}} \right) - \frac{3}{10} \left( F_{j+1} - F_{j-1} \right) - \frac{25}{384} \left( \tilde{F}_{j+\frac{3}{2}} - \tilde{F}_{j-\frac{3}{2}} \right) \right], \label{eq:MND}
\end{equation}

\noindent where $F_{j}$ are fluxes at cell nodes\footnote{Equation~\eqref{eq:MND} uses $F_{j-1}$ and $F_{j+1}$ instead of $\tilde{F}_{j-1}$ and $\tilde{F}_{j+1}$ since the fluxes at nodes can be directly evaluated from the conservative variables at nodes and require no interpolation.}.

\subsection{Weighted compact nonlinear schemes (WCNS's)}

In WCNS's, the fluxes at the cell midpoints are obtained with aid of explicit nonlinear interpolations, which can also be interpreted as a nonlinear filtering processes to avoid spurious oscillations near shocks and other discontinuities. For the scalar conservation equation~\eqref{eq:scalar_conservation}, the algorithm to obtain the flux derivative with WCNS's is given below:
\begin{enumerate}
\item Compute a left-biased and a right-biased interpolated solution value $\tilde{u}_L$ and $\tilde{u}_R$ at each cell midpoint using explicit nonlinear interpolations.
\item Compute the flux at the cell midpoints using a flux difference splitting method $\tilde{F}_{j+\frac{1}{2}} = \mathrm{F}_\mathrm{split}(\tilde{u}_L,\tilde{u}_R)$ (typically a Riemann solver).
\item Compute the flux at the cell nodes $F_{j} = F(u_j)$ if the node values of flux are needed in the finite difference scheme e.g. MND scheme in equation~\eqref{eq:MND}.
\item Use the flux(es) $\tilde{F}_{j+\frac{1}{2}}$ (and $F_j$) to compute the flux derivative $F_{j}^\prime$ with a compact or explicit central finite difference scheme.
\end{enumerate}

In this work, only the interpolations of left-biased cell midpoint values are presented. The interpolations of right-biased cell midpoint values are similar due to symmetry and can be obtained by flipping the stencils and corresponding coefficients. It should also be noted that flux vector splitting methods such as Lax--Friedrichs flux splitting can also be used in WCNS's where the flux values are interpolated instead of the solution values, but that is not the procedure followed in this paper.

\begin{figure}[!ht]
 \centering
 \includegraphics[height=0.35\textwidth]{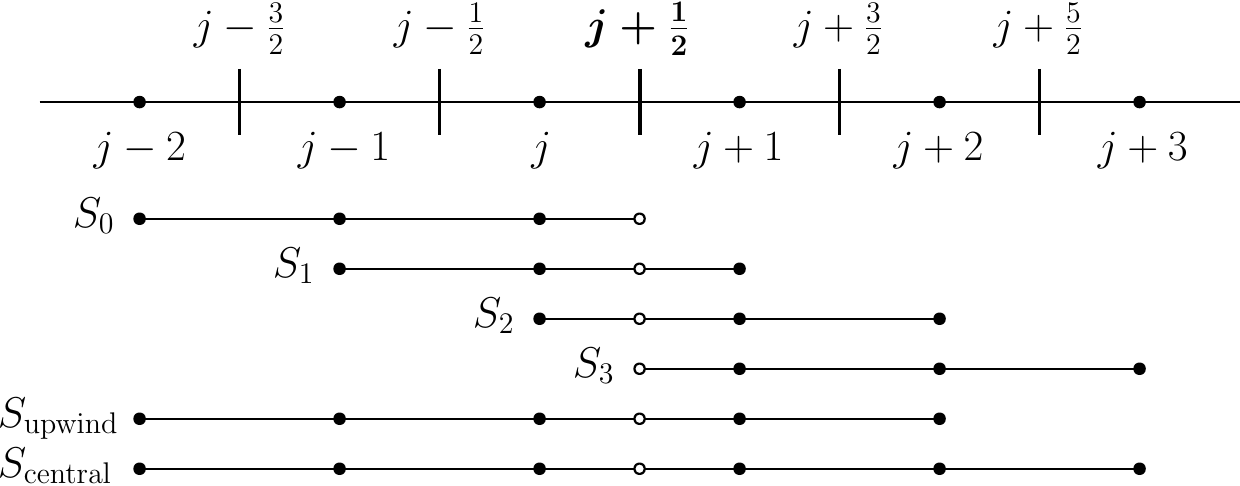}
 \caption{Sub-stencils of WCNS's. The solid circles represent points used in the right hand side of the interpolation stencils, while empty circles represent points used in the left hand side of the interpolation stencils.}
 \label{fig:stencil_WCNS}
\end{figure}

Despite the robustness of interpolations using upwind-biased nonlinear weights in capturing shocks like those by \citet{jiang1995efficient} (JS) and that by \citet{borges2008improved} (Z), they are excessively dissipative in smooth regions. To remedy this, \citet{martin2006bandwidth, hu2010adaptive} proposed a nonlinear interpolation that minimizes dissipation in smooth regions by including the downwind stencil, $S_3$ in figure~\ref{fig:stencil_WCNS}. \citet{wong2017high} further optimized the nonlinear weighting procedure by proposing a localized dissipative (LD) interpolation. The LD interpolation approximates the midpoint values by computing third order linear interpolated values from four different sub-stencils, $S_0$ - $S_3$ (shown in figure~\ref{fig:stencil_WCNS}) and then taking a nonlinear combination of these four values. The interpolated values at the midpoints $\tilde{u}_{j+\frac{1}{2}}$ from the four different explicit interpolations\footnote{Technically, $EI_0$ and $EI_3$ are extrapolations and not interpolations, but we call them interpolations anyway in order to simplify the terminology.} ($EI_k$) are given by:
\begin{align}
	EI_0: \quad \tilde{u}_{j+\frac{1}{2}}^{(0)} =& \frac{1}{8}\left(3u_{j-2} - 10u_{j-1} + 15u_{j} \right), \label{eq:stencil0} \\
    EI_1: \quad \tilde{u}_{j+\frac{1}{2}}^{(1)} =& \frac{1}{8}\left(-u_{j-1} + 6u_{j} + 3u_{j+1} \right), \label{eq:stencil1} \\
    EI_2: \quad \tilde{u}_{j+\frac{1}{2}}^{(2)} =& \frac{1}{8}\left(3u_{j} + 6u_{j+1} - u_{j+2} \right), \label{eq:stencil2} \\
	EI_3: \quad \tilde{u}_{j+\frac{1}{2}}^{(3)} =& \frac{1}{8}\left(15u_{j+1} - 10u_{j+2} + 3u_{j+3} \right). \label{eq:stencil3}
\end{align}

The fifth order linear upwind-biased interpolation $EI_{\mathrm{upwind}}$ and sixth order linear central interpolation $EI_{\mathrm{central}}$ from $S_{\mathrm{upwind}}$ and $S_{\mathrm{central}}$ in figure~\ref{fig:stencil_WCNS} respectively can be obtained from linear combinations of the third order interpolations:
\begin{align}
EI_{\mathrm{upwind}}  =& \sum_{k=0}^{2} d_k^{\mathrm{upwind}}  EI_k,  \label{eq:EI_upwind_1} \\
EI_{\mathrm{central}} =& \sum_{k=0}^{3} d_k^{\mathrm{central}} EI_k,  \label{eq:EI_central_1}
\end{align}
where the linear weights are given by:
\begin{equation}
	d_0^{\mathrm{upwind}} = \frac{1}{16}, \quad
    d_1^{\mathrm{upwind}} = \frac{10}{16}, \quad
    d_2^{\mathrm{upwind}} = \frac{5}{16},
\end{equation}
\begin{equation}
	d_0^{\mathrm{central}} = \frac{1}{32}, \quad
    d_1^{\mathrm{central}} = \frac{15}{32}, \quad
    d_2^{\mathrm{central}} = \frac{15}{32}, \quad
    d_3^{\mathrm{central}} = \frac{1}{32}.
\end{equation}

The expanded form of the linear interpolations from $S_{\mathrm{upwind}}$ and $S_{\mathrm{central}}$ are given by:

\begin{align}
	EI_{\mathrm{upwind}}: \quad \tilde{u}_{j+\frac{1}{2}}^{\mathrm{upwind}}   =& \frac{1}{128} \left(3u_{j-2} - 20u_{j-1} + 90u_j + 60u_{j+1} - 5u_{j+2} \right), \label{eq:EI_upwind_2} \\
	EI_{\mathrm{central}}: \quad \tilde{u}_{j+\frac{1}{2}}^{\mathrm{central}} =& \frac{1}{256} \left(3u_{j-2} - 25u_{j-1} + 150u_j + 150u_{j+1} - 25u_{j+2} + 3u_{j+3} \right). \label{eq:EI_central_2}
\end{align}

The nonlinear LD interpolation is formulated by replacing the linear weights $d_k^{\mathrm{central}}$ in equation~\eqref{eq:EI_central_1} with nonlinear weights $\omega_{k}$ as:
\begin{equation}
	\tilde{u}_{j+\frac{1}{2}} = \sum\limits_{k=0}^{3} \omega_k \tilde{u}_{j+\frac{1}{2}}^{(k)}.
\end{equation}

In smooth regions, the interpolated value given by LD interpolation should converge to the value given by the sixth order linear central interpolation in equation~\eqref{eq:EI_central_2}. The forms of nonlinear weights of the LD interpolation, as well as those of JS and Z interpolations are given in \ref{appendix:nonlinear_weights}.

The CMD scheme (equation~\eqref{eq:CMD}) in conjunction with JS, Z, and LD interpolations are called WCNS5-JS, WCNS5-Z, and WCNS6-LD, respectively. The MND scheme (equation~\eqref{eq:MND}) in conjunction with the three different interpolations are called MND-WCNS5-JS, MND-WCNS5-Z, and MND-WCNS6-LD. The numbers in the names indicate the formal orders of accuracy of the schemes. The difference between the three nonlinear interpolation methods is discussed in \citet{wong2017high}.

% \subsection{Nonlinear explicit-compact interpolation (ECI) and weighted compact high resolution (WCHR) scheme  \label{sec:compact_interpolation}}
\subsection{Weighted compact high resolution (WCHR) scheme  \label{sec:compact_interpolation}}

\subsubsection{Explicit-compact interpolation (ECI)}
WCNS's use explicit interpolations, which typically have larger errors in the real part of the transfer function compared to compact interpolations of the same order of accuracy. In the context of a linear advection equation, this error in the real part of the transfer function manifests itself as a dispersion error. In this sub-section, we propose a new nonlinear explicit-compact interpolation that minimizes the dispersion error by adaptively switching to linear compact interpolations in smooth regions.

\begin{figure}[!ht]
 \centering
 \includegraphics[height=0.35\textwidth]{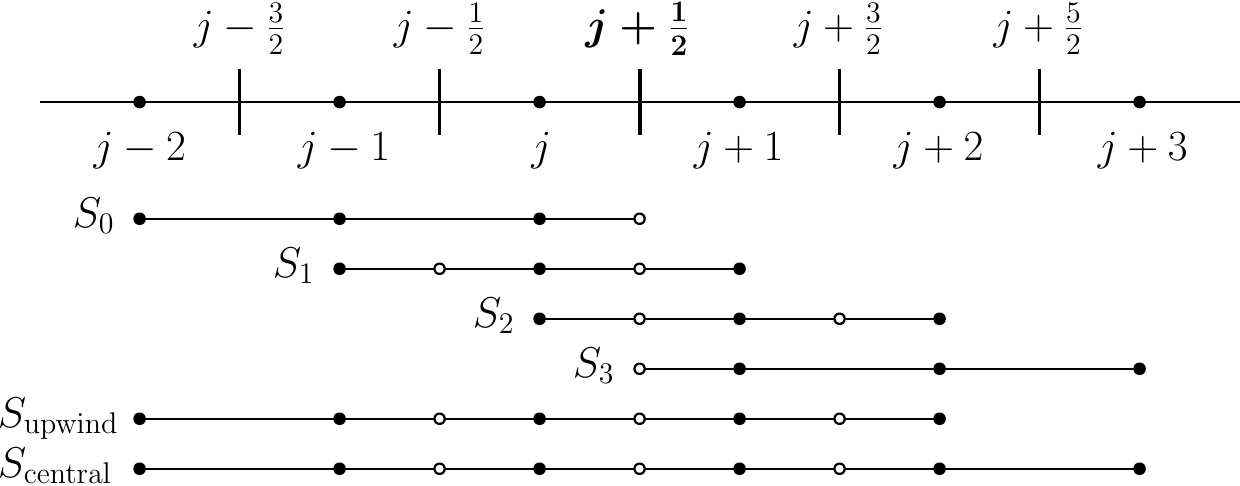}
 \caption{Sub-stencils of the WCHR6 scheme. The solid circles represent points used in the right hand side of the interpolation stencils, while empty circles represent points used in the left hand side of the interpolation stencils.}
 \label{fig:stencil_WCHR}
\end{figure}

Instead of using only explicit interpolations in sub-stencils, the interpolation methods in the central two sub-stencils in figure~\ref{fig:stencil_WCNS} are replaced with compact interpolations. In smooth regions where all the four stencils are used, the interpolation becomes compact and has better resolution properties while near discontinuities where the most left or right biased stencil is used, the interpolation reverts to being explicit for robustness. The interpolation methods ($ECI_k$) in the sub-stencils $S_0$ - $S_3$ of figure~\ref{fig:stencil_WCHR} are given by:

\begin{align}
    &ECI_0:  &\tilde{u}_{j+\frac{1}{2}}^{(0)} &= \frac{3}{8}u_{j-2} - \frac{5}{4}u_{j-1} + \frac{15}{8}u_{j}, \label{eq:WCHR6_S0}\\
    &ECI_1:  &-\left( \xi - 1 \right) \tilde{u}^{(1)}_{j-\frac{1}{2}} + \xi \tilde{u}^{(1)}_{j+\frac{1}{2}} &= - \frac{4 \xi - 3}{8} {u}_{j-1} + \frac{3}{4}{u}_{j} + \frac{4 \xi - 1}{8} {u}_{j+1}, \label{eq:WCHR6_S1}\\
    &ECI_2:  &\xi \tilde{u}^{(2)}_{j+\frac{1}{2}} - \left( \xi - 1 \right) \tilde{u}^{(2)}_{j+\frac{3}{2}} &= \frac{4 \xi - 1}{8} {u}_{j} + \frac{3}{4}{u}_{j+1} - \frac{4 \xi - 3}{8} {u}_{j+2}, \label{eq:WCHR6_S2}\\
    &ECI_3:  &\tilde{u}_{j+\frac{1}{2}}^{(3)} &= \frac{15}{8}u_{j+1} - \frac{5}{4}u_{j+2} + \frac{3}{8}u_{j+3}, \label{eq:WCHR6_S3}
\end{align}
where $\xi$ is a free parameter that can be used to control the dispersion and dissipation characteristics of the scheme. When $\xi = 1$, the explicit-compact interpolations reduce to fully explicit interpolations. In general, $S_1$ and $S_2$ in equations~\eqref{eq:WCHR6_S1} - \eqref{eq:WCHR6_S2} are third order accurate except for $\xi = 5/8$ when they both become fourth order accurate.

The fifth order linear upwind-biased and sixth order linear central interpolations from $S_{\mathrm{upwind}}$ and $S_{\mathrm{central}}$ in figure~\ref{fig:stencil_WCHR} respectively can be obtained from linear combinations of the third order interpolations:
\begin{align}
ECI_{\mathrm{upwind}}  =& \sum_{k=0}^{2} d_k^{\mathrm{upwind}}  ECI_k, \\
ECI_{\mathrm{central}} =& \sum_{k=0}^{3} d_k^{\mathrm{central}} ECI_k,
\end{align}
where the linear weights are given by:
\begin{equation}
d^{\mathrm{upwind}}_0 = \frac{8\xi - 5}{8 \left( \xi + 5 \right)}, \quad d^{\mathrm{upwind}}_1 = \frac{5 \left( 13\xi - 7 \right)}{8\left( \xi + 5 \right)\left( 2\xi - 1 \right)}, \quad d^{\mathrm{upwind}}_2 = \frac{5 \left( 5\xi - 2 \right)}{8 \left( \xi + 5 \right)\left( 2\xi - 1 \right)},
\end{equation}
\begin{equation}
d^{\mathrm{central}}_0 = \frac{8\xi - 5}{16 \left(\xi + 5 \right)}, \quad d^{\mathrm{central}}_1 = \frac{45}{16 \left( \xi + 5 \right)}, \quad d^{\mathrm{central}}_2 = \frac{45}{16 \left( \xi + 5 \right)}, \quad d^{\mathrm{central}}_3 = \frac{8\xi - 5}{16 \left(\xi + 5 \right)}.
\end{equation}

Note that the linear weights for explicit-compact interpolations are in general different from those for explicit interpolations except when $\xi=1$. The expanded form of the linear interpolations from $S_{\mathrm{upwind}}$ and $S_{\mathrm{central}}$ are given by:

\begin{align}
	ECI_{\mathrm{upwind}}: \quad & \alpha^\mathrm{upwind}\tilde{u}^{\mathrm{upwind}}_{j-\frac{1}{2}} + \beta^\mathrm{upwind}\tilde{u}^{\mathrm{upwind}}_{j+\frac{1}{2}} + \gamma^\mathrm{upwind}\tilde{u}^{\mathrm{upwind}}_{j+\frac{3}{2}} = \nonumber \\
    & a^\mathrm{upwind}u_{j-2} + b^\mathrm{upwind}u_{j-1} + c^\mathrm{upwind}u_j + d^\mathrm{upwind}u_{j+1} + e^\mathrm{upwind}u_{j+2}, \label{eq:ECI_upwind}
\end{align}

\begin{align}
	ECI_{\mathrm{central}}: \quad & \alpha^\mathrm{central}\tilde{u}^{\mathrm{central}}_{j-\frac{1}{2}} + \beta^\mathrm{central}\tilde{u}^{\mathrm{central}}_{j+\frac{1}{2}} + \gamma^\mathrm{central}\tilde{u}^{\mathrm{central}}_{j+\frac{3}{2}} = \nonumber \\
    & a^\mathrm{central}u_{j-2} + b^\mathrm{central}u_{j-1} + c^\mathrm{central}u_j + d^\mathrm{central}u_{j+1} + e^\mathrm{central}u_{j+2} + f^\mathrm{central}u_{j+3}.  \label{eq:ECI_central}
\end{align}
The coefficients in equations~\eqref{eq:ECI_upwind} and \eqref{eq:ECI_central} are given in \ref{appendix:ECI_interior_coeffs}.

Figure~\ref{fig:linear_weights} shows the relations between the linear weights of the most upwind stencil $S_0$ in $ECI_{\mathrm{upwind}}$ and $ECI_{\mathrm{central}}$ and $\xi$. For both linear weights to be positive, $\xi$ has to be larger than $5/8$. For increasing $\xi > 5/8$, the linear weights for the most upwind stencil increase linearly.

\begin{figure}[!ht]
\begin{center}
\includegraphics[width=0.5\textwidth]{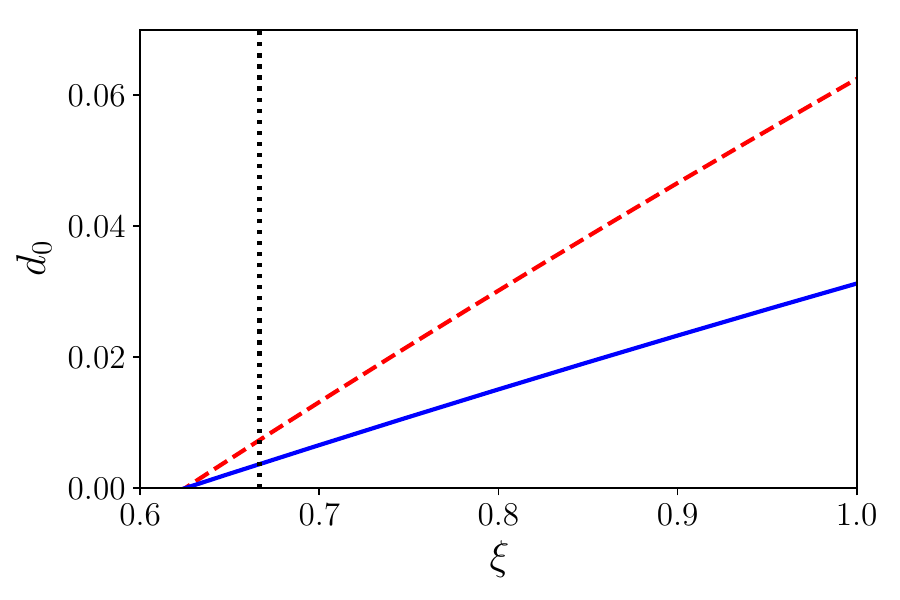}
\caption{Linear weights of sub-stencil $S_0$ of $ECI_{\mathrm{upwind}}$ and $ECI_\mathrm{central}$ against $\xi$. Red dashed line: $ECI_\mathrm{upwind}$; blue solid line: $ECI_\mathrm{central}$. The black dotted vertical line indicates $\xi=2/3$ which is chosen for both ECI's in this work.}
\label{fig:linear_weights}
\end{center}
\end{figure}

\begin{figure}[!ht]
\begin{center}
\subfigure[Dispersion error]{\includegraphics[height=0.32\textwidth]{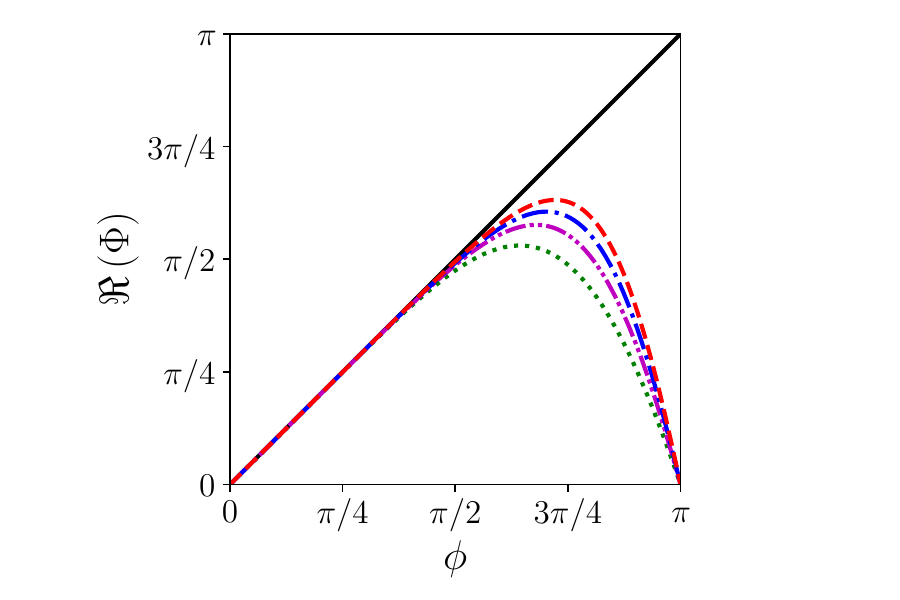} \label{fig:dispersion_upwind}}
\subfigure[Dissipative error]{\includegraphics[height=0.32\textwidth]{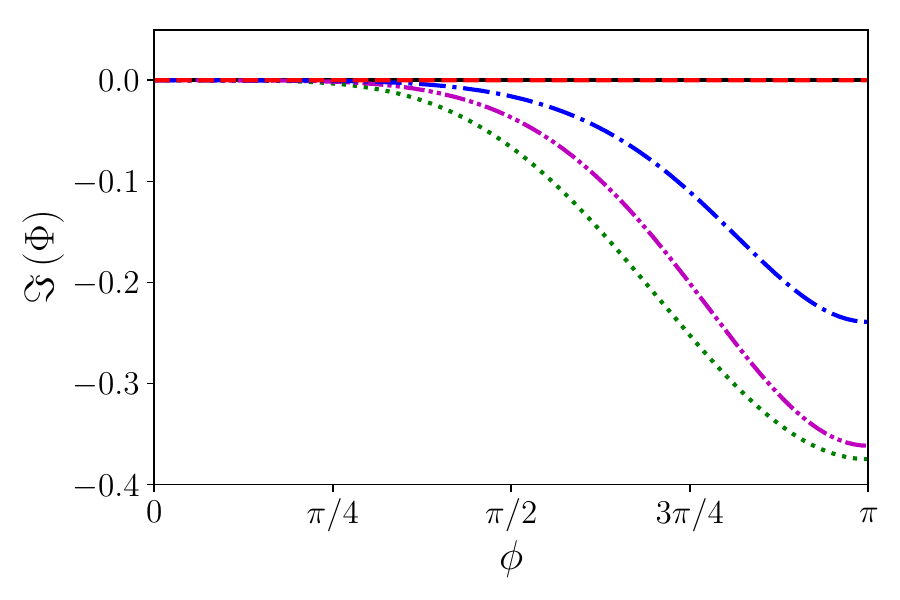} \label{fig:dissipation_upwind}}
\caption{Modified reduced wavenumber, $\Phi$, against reduced wavenumber, $\phi$, of $ECI_{\mathrm{upwind}}$. Black solid line: exact; green dotted line: $\xi = 1$; magenta dashed-dotted-dotted line: $\xi = 3/4$; blue dashed-dotted line: $\xi = 2/3$; red dashed line: $\xi = 5/8$.}
\label{fig:modified_wavenumber_upwind}
\end{center}
\end{figure}

\begin{figure}[!ht]
\begin{center}
\includegraphics[height=0.4\textwidth]{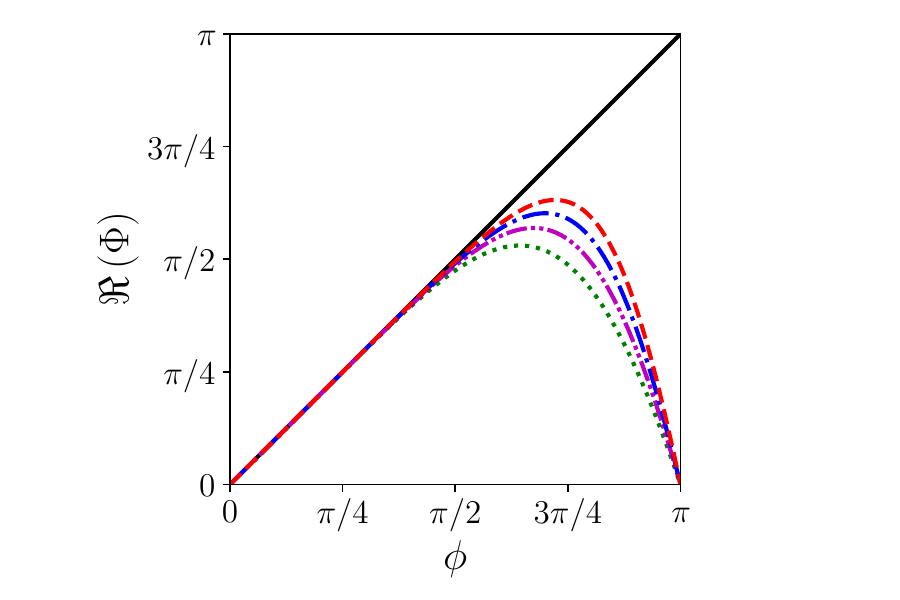}
\caption{Real part of modified reduced wavenumber, $\Phi$, against reduced wavenumber, $\phi$, of $ECI_{\mathrm{central}}$ representing dispersion error. Black solid line: exact; green dotted line: $\xi = 1$; magenta dashed-dotted-dotted line: $\xi = 3/4$; blue dash-dotted line: $\xi = 2/3$; red dashed line: $\xi = 5/8$.}
\label{fig:modified_wavenumber_central}
\end{center}
\end{figure}

Even when used with a perfect derivative scheme, the interpolation transfer function creates dispersion and dissipation errors in a linear advection problem. Figures~\ref{fig:modified_wavenumber_upwind} and \ref{fig:modified_wavenumber_central} show the modified wavenumber of $ECI_\mathrm{upwind}$ and $ECI_\mathrm{central}$ respectively when used with an analytical derivative scheme. When $\xi$ is decreased from $1$ to $5/8$, the resolution increases in both ECI's and the dissipation of $ECI_\mathrm{upwind}$ decreases. It should be noted that the dissipation error of $ECI_\mathrm{central}$ is always zero independent of value of $\xi$ and the dispersion errors of both $ECI_\mathrm{upwind}$ and $ECI_\mathrm{central}$ are the same when $\xi = 5/8$ as both of them become identical. We use a value of $\xi = 2/3$ in this paper. This value of $\xi$ is chosen based on the dispersion relations of the linear schemes as a balance between high resolution and robustness. More rigorous optimization procedures may be used to choose an optimal value of $\xi$ but that is left to future work.

\subsubsection{Weighted compact high resolution (WCHR) scheme}
The finite difference schemes described in section~\ref{sec:FD} may generate spurious oscillations due to Gibbs phenomenon or even be unstable near shocks or discontinuities with either $ECI_\mathrm{upwind}$ or $ECI_\mathrm{central}$. Hence, we use a nonlinear combination of the sub-stencil interpolations with the LD nonlinear weights in equation~\eqref{eq:LD_nonlinear_weights} at any midpoint:
\begin{equation}
	ECI_{\mathrm{nonlinear}} = \sum_{k=0}^{3} \omega_k ECI_k. \label{eq:nonlinear_ECI}
\end{equation}

The CMD scheme in equation~\eqref{eq:CMD} with the nonlinear explicit-compact interpolation ($ECI_{\mathrm{nonlinear}}$) is sixth order accurate in smooth regions and is called weighted compact high resolution scheme, WCHR6, in this paper due to its high resolution property compared to other WCNS's.

The parameters for computing the nonlinear weights in WCNS's and WCHR6 scheme are discussed in \ref{appendix:nonlinear_weights}. The parameter values of each scheme in this work are given in table~\ref{table:parameters}. For a discussion on the choice of parameters in LD nonlinear weights, see~\cite{wong2017high}. The parameters used here for WCHR6 provide stable results while preserving the high resolution property of the underlying compact interpolation scheme. They are also chosen so that numerical dissipation is only locally added to regions containing discontinuities and have minimal effect on regions where the solution is smooth.

\begin{table}[!ht]
  \begin{center}
  \begin{tabular}{ | c | c | c | c | c | c |}
    \hline
    \textbf{Numerical} & \multicolumn{5}{c|}{\textbf{Parameter values}} \\
    \cline{2-6}
    \textbf{schemes}  & $p$ & $q$ & $C$ & $\alpha^\tau_{RL}$ & $\xi$ \\[0.1pc] \hline \hline
    \textbf{WCNS5-JS} & $2$ & $-$ & $-$ & $-$ & $-$ \\[0.1pc] \hline
    \textbf{WCNS5-Z}  & $2$ & $-$ & $-$ & $-$ & $-$ \\[0.1pc] \hline
    \textbf{WCNS6-LD} & $2$ & $4$ & $1.0\mathrm{e}{9}$ & $35.0$ & $-$ \\[0.1pc] \hline
    \textbf{WCHR6}     & $2$ & $4$ & $1.0\mathrm{e}{10}$ & $55.0$ & $2/3$ \\[0.1pc] \hline
  \end{tabular}
  \caption{Parameters for different numerical schemes.}
  \label{table:parameters}
  \end{center}
\end{table}

\iffalse
\begin{table}
  \begin{center}
    \caption{Parameters for computing the nonlinear weights in the WCHR6 scheme.}
    \label{tab:parameters}
  \begin{tabular}{ | c | c | }
    \hline
    \textbf{Parameter} & \textbf{Value} \\[0.1pc] \hline \hline
    $C$ & $2 \times 10^3$  \\[0.1pc] \hline
    $p$ & $2$  \\[0.1pc] \hline
    $q$ & $2$  \\[0.1pc] \hline
    $\alpha^\tau_{RL}$ & $40$  \\[0.1pc] \hline
  \end{tabular}
  \end{center}
\end{table}
\fi

\subsection{Approximate dispersion relation}
\begin{figure}[!ht]
\begin{center}
\subfigure[Dispersion characteristics]{\includegraphics[width=0.48\textwidth]{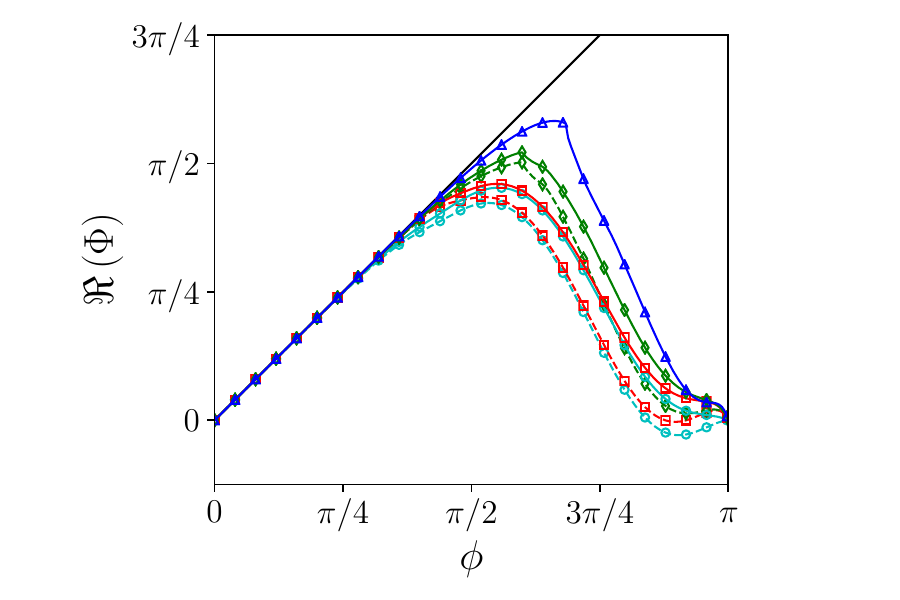}\label{fig:ADR_real}}
\subfigure[Dissipation characteristics]{\includegraphics[width=0.48\textwidth]{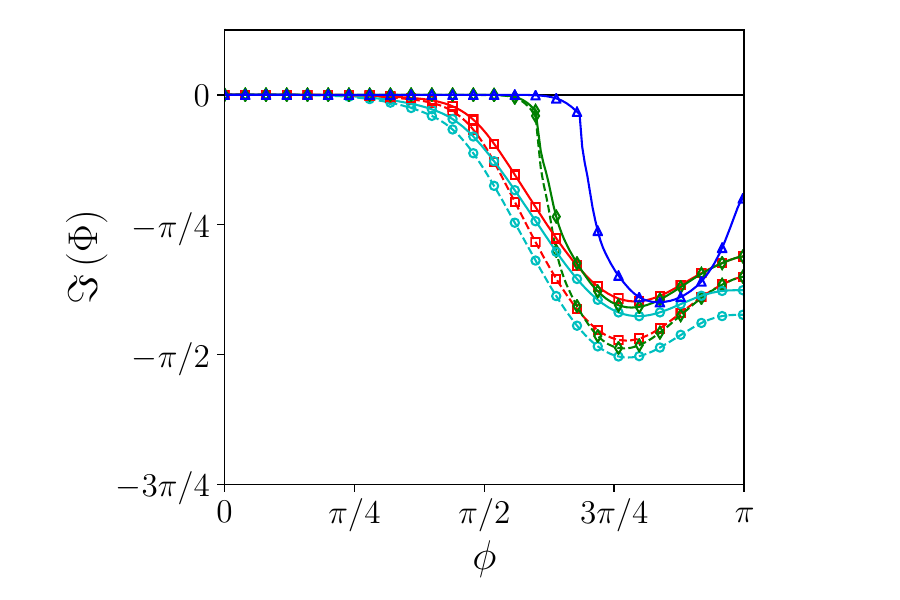}\label{fig:ADR_imag}}
\caption{ADR's of different numerical schemes. Black solid line: spectral; dashed line with cyan circles: MND-WCNS5-JS; solid line with cyan circles: WCNS5-JS; dashed line with red squares: MND-WCNS5-Z; solid line with red squares: WCNS5-Z; dashed line with green diamonds: MND-WCNS6-LD; solid line with green diamonds: WCNS6-LD; solid line with blue triangles: WCHR6.}
\label{fig:ADR}
\end{center}
\end{figure}

\begin{figure}[!ht]
\begin{center}
\includegraphics[width=0.5\textwidth]{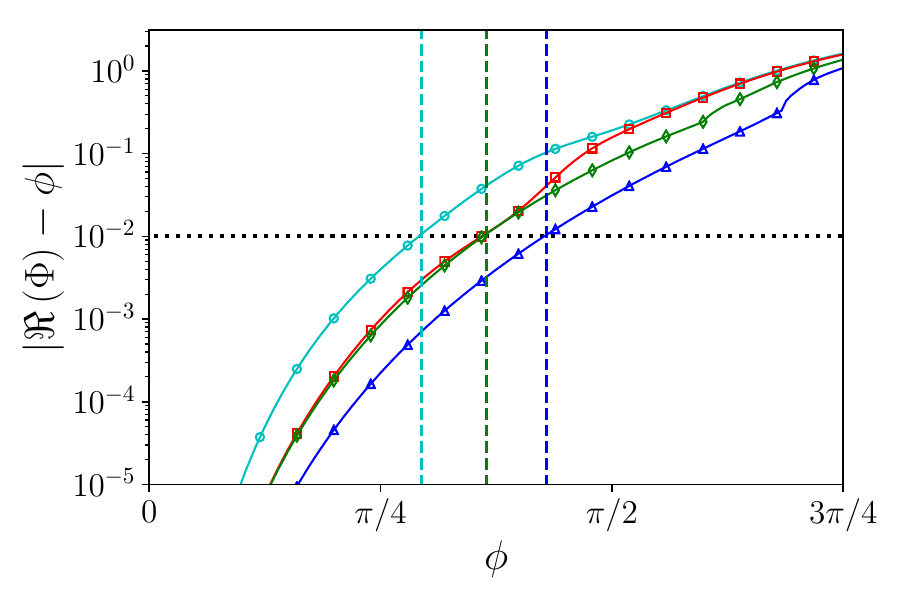}
\caption{Approximate dispersion errors (derivation of real part of modified reduced wavenumber from that of reduced wavenumber) of different numerical schemes. Cyan circles: WCNS5-JS; red squares: WCNS5-Z; green diamonds: WCNS6-LD; blue triangles: WCHR6.}
\label{fig:ADR_dispersion_error}
\end{center}
\end{figure}

For linear schemes, the dissipation and dispersion characteristics can be determined using a dispersion relation analysis discussed by \citet{lele1992compact}. However, this analysis cannot be used for nonlinear schemes. \citet{pirozzoli2006spectral} developed an approximate dispersion relation (ADR) technique to characterize the dispersion and dissipation characteristics of general nonlinear schemes. Results from ADR analysis are shown in figure~\ref{fig:ADR} for the WCHR6 scheme and WCNS's using compact (CMD) and explicit (MND) derivatives. In figure~\ref{fig:ADR_real} where the dispersion characteristics are shown, we can see that the WCHR6 scheme outperforms other schemes in dispersion error. Explicit nonlinear interpolations with CMD in general have higher resolution than those with MND. Figure~\ref{fig:ADR_dispersion_error} shows the dispersion errors for WCHR6 and the WCNS's with explicit interpolations and compact derivative (CMD) on a semi-log plot. Given a threshold $\epsilon_{\mathrm{res}}$ for the maximum tolerable dispersion error, a resolving efficiency of the different schemes can be computed. The resolving efficiency is defined as the fraction of Nyquist wavenumber that the scheme can resolve within the given dispersion error tolerance $\epsilon_{\mathrm{res}}$. In figure~\ref{fig:ADR_dispersion_error}, the horizontal black dashed line represents $\epsilon_{\mathrm{res}} = 0.01$ and the vertical colored dashed lines represent the maximum wavenumber that each scheme can resolve given this threshold. Table~\ref{table:dispersion_error} shows the resolving efficiency for the four different schemes. From the plot, it can be seen that the WCHR6 has much higher resolution ability compared to other schemes of similar orders of accuracy ($\sim 45.8\%$ more than the WCNS5-JS). All schemes considered in figure~\ref{fig:ADR_dispersion_error} use CMD as the flux derivative. This clearly shows the benefit of using compact interpolation to achieve better resolution characteristics. Figure~\ref{fig:ADR_imag} shows the dissipation characteristics of the schemes. In the plot, we see that WCNS5-JS and WCNS5-Z have dissipation over a wide range of wavenumbers while WCNS6-LD has much more localized dissipation only in high wavenumber range. Due to the high resolution characteristic of WCHR6, we choose the parameters in the LD weights such that it has more localized dissipation than WCNS6-LD in the wavenumber space. The high resolution and localized dissipation characteristics of WCHR6 are especially important for problems involving turbulence transition where low resolution and excessive dissipation can curtail the range of scales in the problem.

\iffalse
\begin{figure}[htb]
\begin{center}
\subfigure[]{\includegraphics[width=0.29\textwidth]{Figures_old/ADR-plots_1_0.pdf}\label{fig:adr_a}}
\subfigure[]{\includegraphics[width=0.29\textwidth]{Figures_old/ADR-plots_1_2.pdf}\label{fig:adr_b}}
\subfigure[]{\includegraphics[width=0.4\textwidth]{Figures_old/ADR-plots_1_1.pdf}\label{fig:adr_c}}
\caption{ADR results for four schemes: WCNS5-JS (green), WCNS5-Z (magenta), WCNS6-LD (blue), and WCHR6 (red). (a) Real part of the dispersion relation (b) Imaginary part of the dispersion error characterizing the added dissipation and (c) Solid lines are the dispersion error (deviation of plot (a) from ideal) for the different schemes. The black dashed line is an error threshold level of $10^{-2}$ and the colored dashed lines indicate the reduced wavenumber beyond which the dispersion error exceeds the threshold.}
\label{fig:ADR}
\end{center}
\end{figure}
\fi

\begin{table}[!ht]
  \begin{center}
  \begin{tabular}{ | c | c | c | }
    \hline
    \textbf{Numerical schemes} & \textbf{Resolving efficiency} & \textbf{Improvement over WCNS5-JS} \\[0.1pc] \hline \hline
    WCNS5-JS & $0.294$ & $-$        \\[0.1pc] \hline
    WCNS5-Z  & $0.364$ & $23.7\%$   \\[0.1pc] \hline
    WCNS6-LD & $0.364$ & $23.7\%$   \\[0.1pc] \hline
    WCHR6    & $0.429$ & $45.8\%$   \\[0.1pc] \hline
  \end{tabular}
  \caption{Resolving efficiency of different schemes for $\epsilon_{\mathrm{res}} = 0.01$.}
  \label{table:dispersion_error}
  \end{center}

\end{table}

\subsection{Boundary closures \label{sec:boundary_closures}}

Boundary schemes are essential for interpolation and numerical derivative at the domain boundaries. In this section, we present boundary schemes for both interpolation and conservative derivative that preserve the order of accuracy and have truncation errors matched to those of the interior schemes. The boundary schemes presented here use ghost points at domain boundaries. Specific algorithms to evaluate function values for the ghost points are described in section~\ref{sec:results}.

\subsubsection{Interpolations}

\begin{figure}[!ht]
 \centering
 \includegraphics[height=0.35\textwidth]{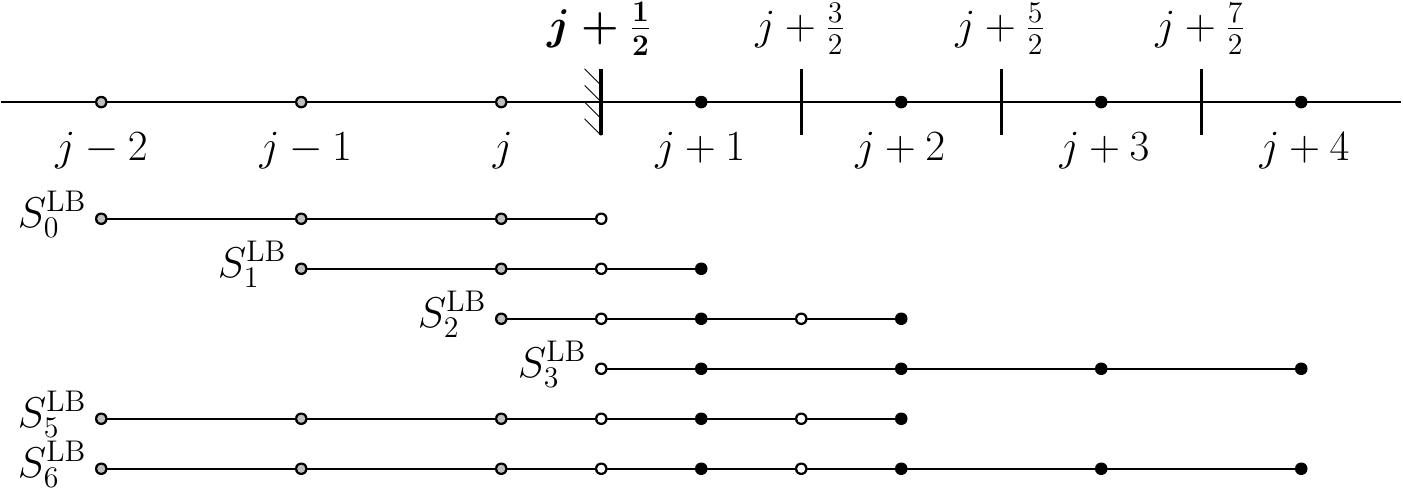}
 \caption{Sub-stencils of the left-biased interpolation scheme at the left boundary (LB). The solid and gray circles represent points used in the right hand side of the compact interpolation stencils, while empty circles represent points used in the left hand side of the interpolation stencils. The solid and gray circles represent the interior points and ghost points respectively.}
 \label{fig:stencil_WCHR_LB}
\end{figure}

Only left-biased interpolations at the left boundary (LB) and right boundary (RB) are discussed in this section. The right-biased interpolations at the left and right boundaries are simply the mirror images of the left-biased interpolations at the right and left boundaries respectively. The sub-stencils of the left-biased interpolation scheme at LB is shown in figure~\ref{fig:stencil_WCHR_LB}.

The four third order interpolations from $S_{0}^{\mathrm{LB}}$-$S_{3}^{\mathrm{LB}}$ in figure~\ref{fig:stencil_WCHR_LB} are given by:
\begin{align}
    &ECI_0^{\mathrm{LB}}:  &\tilde{u}_{j+\frac{1}{2}}^{(0)} &= \frac{3}{8}u_{j-2} - \frac{5}{4}u_{j-1} + \frac{15}{8}u_{j}, \label{eq:WCHR6_LB_S0} \\
    &ECI_1^{\mathrm{LB}}:  &\tilde{u}^{(1)}_{j+\frac{1}{2}} &= -\frac{1}{8}{u}_{j-1} + \frac{3}{4}{u}_{j} + \frac{3}{8}{u}_{j+1},  \label{eq:WCHR6_LB_S1} \\
    &ECI_2^{\mathrm{LB}}:  &a^{\mathrm{LB}} \tilde{u}^{(2)}_{j+\frac{1}{2}} + b^{\mathrm{LB}} \tilde{u}^{(2)}_{j+\frac{3}{2}} &=  c^{\mathrm{LB}} {u}_{j} + d^{\mathrm{LB}} {u}_{j+1} + e^{\mathrm{LB}} {u}_{j+2}, \label{eq:WCHR6_LB_S2} \\
    &ECI_3^{\mathrm{LB}}:  &\tilde{u}_{j+\frac{1}{2}}^{(3)} &= f^{\mathrm{LB}} u_{j+1} + g^{\mathrm{LB}} u_{j+2} + h^{\mathrm{LB}} u_{j+3} + i^{\mathrm{LB}} u_{j+4}. \label{eq:WCHR6_LB_S3}
\end{align}

The fifth order and sixth order linear interpolations from $S_{5}^{\mathrm{LB}}$ and $S_{6}^{\mathrm{LB}}$ in figure~\ref{fig:stencil_WCHR_LB} respectively can be obtained from linear combinations of the third order interpolations:
\begin{align}
ECI_{5}^{\mathrm{LB}}  =& \sum_{k=0}^{2} d_{k}^{(5), \mathrm{LB}}  ECI_k^{\mathrm{LB}}, \label{eq:ECI_LB_upwind} \\
ECI_{6}^{\mathrm{LB}} =& \sum_{k=0}^{3} d_{k}^{(6), \mathrm{LB}}  ECI_k^{\mathrm{LB}}. \label{eq:ECI_LB_central}
\end{align}

 The sub-stencils of the left-biased interpolation scheme at RB is shown in figure~\ref{fig:stencil_WCHR_RB}. The four third order interpolations from $S_{0}^{\mathrm{RB}}$-$S_{3}^{\mathrm{RB}}$ in figure~\ref{fig:stencil_WCHR_RB} are given by:
 
\begin{figure}[!ht]
 \centering
 \includegraphics[height=0.35\textwidth]{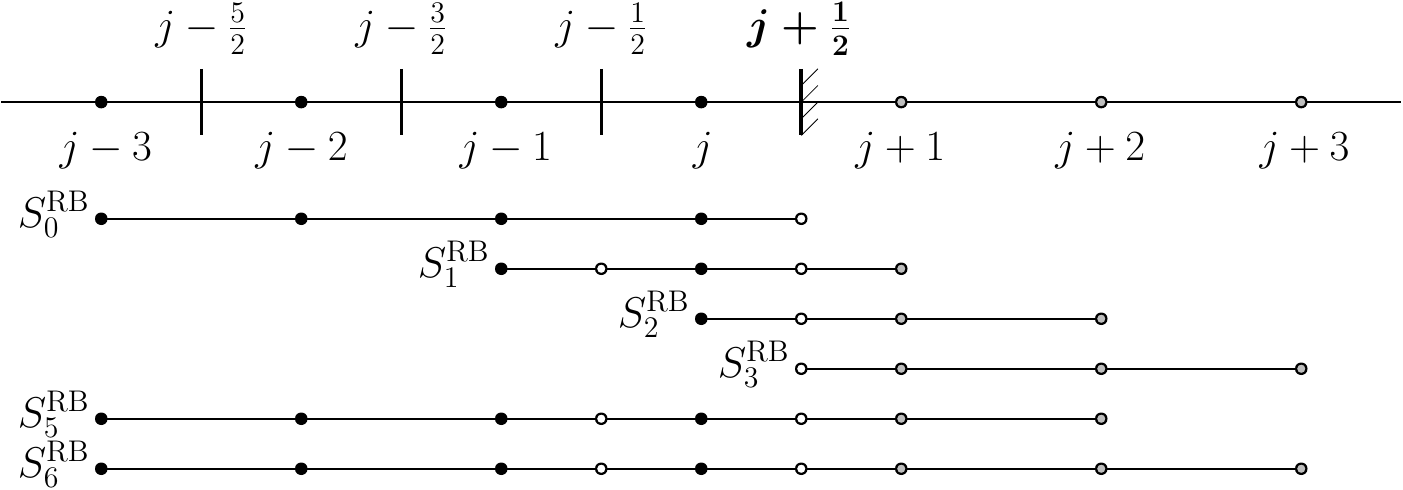}
 \caption{Sub-stencils of the left-biased interpolation scheme at the right boundary (RB). The solid and gray circles represent points used in the right hand side of the compact interpolation stencils, while empty circles represent points used in the left hand side of the interpolation stencils The solid and gray circles represent the interior points and ghost points respectively.}
 \label{fig:stencil_WCHR_RB}
\end{figure}

\begin{align}
    &ECI_0^{\mathrm{RB}}:  &\tilde{u}_{j+\frac{1}{2}}^{(0)} &= a^{\mathrm{RB}} u_{j-3} + b^{\mathrm{RB}} u_{j-2} + c^{\mathrm{RB}} u_{j-1} + d^{\mathrm{RB}} u_{j}, \label{eq:WCHR6_RB_S0}\\
    &ECI_1^{\mathrm{RB}}:  &e^{\mathrm{RB}} \tilde{u}_{j-\frac{1}{2}}^{(1)} + f^{\mathrm{RB}} \tilde{u}_{j+\frac{1}{2}}^{(1)} &= g^{\mathrm{RB}} u_{j-1} + h^{\mathrm{RB}} u_{j} + i^{\mathrm{RB}} u_{j+1}, \label{eq:WCHR6_RB_S1}\\
    &ECI_2^{\mathrm{RB}}:  &\tilde{u}_{j+\frac{1}{2}}^{(2)} &= \frac{3}{8} u_{j} + \frac{3}{4} u_{j+1} - \frac{1}{8} u_{j+2}, \label{eq:WCHR6_RB_S2}\\
    &ECI_3^{\mathrm{RB}}:  &\tilde{u}_{j+\frac{1}{2}}^{(3)} &= \frac{15}{8} u_{j+1} - \frac{5}{4} u_{j+2} + \frac{3}{8} u_{j+3}. \label{eq:WCHR6_RB_S3}
\end{align}

The fifth order and sixth order linear interpolations from $S_{5}^{\mathrm{RB}}$ and $S_{6}^{\mathrm{RB}}$ in figure~\ref{fig:stencil_WCHR_RB} respectively can be obtained from linear combinations of the third order interpolations:
\begin{align}
ECI_{5}^{\mathrm{RB}}  =& \sum_{k=0}^{2} d_{k}^{(5), \mathrm{RB}}  ECI_k^{\mathrm{RB}}, \label{eq:ECI_RB_upwind} \\
ECI_{6}^{\mathrm{RB}} =& \sum_{k=0}^{3} d_{k}^{(6), \mathrm{RB}}  ECI_k^{\mathrm{RB}}. \label{eq:ECI_RB_central}
\end{align}

The coefficients in the sub-stencils and the linear weights of the interpolation schemes at the LB and RB are given in \ref{appendix:ECI_boundary_coeffs}. There are two free parameters for each of the boundary interpolation scheme. The free parameters are set such that the first nonzero truncation errors of $ECI_{5}^{\mathrm{LB}}$/$ECI_{5}^{\mathrm{RB}}$ and $ECI_{6}^{\mathrm{LB}}$/$ECI_{6}^{\mathrm{RB}}$ match those of $ECI_{\mathrm{upwind}}$ and $ECI_{\mathrm{central}}$ of equations~\eqref{eq:ECI_upwind} and \eqref{eq:ECI_central} respectively. To capture discontinuities, the linear weights are replaced with the LD nonlinear weights in \ref{appendix:LD_nonlinear_weights}.

\subsubsection{Derivatives}
A derivative boundary closure for an interior scheme given in equation~\eqref{eq:CMD} is only required at the last boundary point. The boundary derivative schemes at the boundary points are derived by using flux difference formulations of compact finite difference schemes and enforcing discrete conservation. It is proved in \ref{appendix:flux_difference} that any compact or explicit central finite difference scheme can be rewritten in the flux difference form given by:
\begin{equation}
	\left. \widehat{ \frac{\partial F}{\partial x} } \right|_{x=x_j} = \widehat{F}_{j}^\prime =  \frac{1}{\Delta x} \left( \widehat{F}_{j+\frac{1}{2}} - \widehat{F}_{j-\frac{1}{2}} \right),
	\label{eq:flux_reconstruction_form_derivative}
\end{equation}
where $\widehat{F}_{j+\frac{1}{2}}$ are the reconstructed fluxes at midpoints. $\widehat{F}_{j+\frac{1}{2}}$ of the sixth order CMD (equation~\eqref{eq:CMD}) are given by:
\begin{equation}
    \frac{9}{80} \widehat{F}_{j-\frac{1}{2}} + \frac{31}{40} \widehat{F}_{j+\frac{1}{2}} + \frac{9}{80} \widehat{F}_{j+\frac{3}{2}} = \frac{17}{240} {F}_{j-\frac{1}{2}} + \frac{103}{120} {F}_{j+\frac{1}{2}} + \frac{17}{240} {F}_{j+\frac{3}{2}}. \label{eq:flux_reconstruction_form_CMD}
\end{equation}
In deriving the boundary closure for the CMD derivative scheme, we seek for a closure for the flux reconstruction equation such that the truncation error of the boundary derivative scheme is matched to that of the interior derivative scheme up to seventh order. This gives the following boundary scheme at the left boundary with $j = 0$:
\begin{align}
\frac{31}{40} \widehat{F}_{j}^\prime + \frac{9}{80} \widehat{F}_{j+1}^\prime = \frac{1}{\Delta x} \left[ 
\frac{1633}{5376000} F_{j-2} + \frac{9007}{192000} F_{j-1} - \frac{29567}{48000} \tilde{F}_{j-\frac{1}{2}} - \frac{65699}{76800} F_{j} \right. \nonumber \\
\left.  + \frac{44033}{24000} \tilde{F}_{j+\frac{1}{2}} - \frac{26353}{38400} F_{j+1}  + \frac{104579}{336000} \tilde{F}_{j+\frac{3}{2}} - \frac{27233}{768000} F_{j+2}
\right]. \label{eq:CMD_LB}
\end{align}

The derivative scheme for the right boundary at $j=N-1$ can be obtained by mirroring the above derivative scheme:
\begin{align}
\frac{9}{80} \widehat{F}_{j-1}^\prime + \frac{31}{40} \widehat{F}_{j}^\prime = \frac{1}{\Delta x} \left[ \frac{27233}{768000} F_{j-2} - \frac{104579}{336000} \tilde{F}_{j-\frac{3}{2}} + \frac{26353}{38400} F_{j-1} - \frac{44033}{24000} \tilde{F}_{j-\frac{1}{2}} \right. \nonumber \\
\left. + \frac{65699}{76800} F_{j} + \frac{29567}{48000} \tilde{F}_{j+\frac{1}{2}} - \frac{9007}{192000} F_{j+1} - \frac{1633}{5376000} F_{j+2} \right]. \label{eq:CMD_RB}
\end{align}

The relation between finite difference schemes and their flux difference forms, and the details on how to derive the boundary schemes with the flux difference form are further discussed in \ref{appendix:flux_difference}.

\subsection{Extension to Euler equations \label{sec:Euler}}
The inviscid 1D Euler equations are given by:
\begin{equation}
\pd{\bm{Q}}{t} + \pd{\bm{F}(\bm{Q})}{x} = 0,
\end{equation}
where
\begin{equation}
\bm{Q} = \begin{pmatrix}
  \rho  \\
  \rho u \\
  E
 \end{pmatrix} \quad \mathrm{and} \quad \bm{F}(\bm{Q}) = \begin{pmatrix}
  \rho u  \\
  \rho u^2 + p \\
  \left(E + p\right)u
 \end{pmatrix},
\end{equation}
where $\rho$ is the density, $u$ is the velocity, $E$ is the total energy, and $p = (\gamma - 1)\left( E - \rho u^2 / 2 \right)$ is the pressure.

The WCNS's or WCHR6 scheme can be applied to the Euler equations in a similar fashion as the scalar conservation law. Equations~\eqref{eq:CMD}), \eqref{eq:MD}, and \eqref{eq:MND} can be used to get the flux derivatives based on the fluxes at the nodes $\bm{F}_j$ and the fluxes at the midpoints $\tilde{\bm{F}}_{j+\frac{1}{2}} = \bm{\mathrm{F}}_\mathrm{Riemann}(\tilde{\bm{Q}}_L,\tilde{\bm{Q}}_R)$ where $\tilde{\bm{Q}}_L$ and $\tilde{\bm{Q}}_R$ are the left and right interpolated solution vectors at the midpoints and $\bm{\mathrm{F}}_\mathrm{Riemann}$ are the fluxes from a Riemann solver. In this work, the HLLC Riemann solver is used (see \ref{appendix:HLLC_HLL} for details on the Riemann solver) for 1D problems. Although the interpolated solution vectors at the midpoints can be computed by directly interpolating the conserved variables or the primitive variables $\left( \rho, u, p \right)$ using the weighted interpolations, it was found that projecting variables to the local characteristic fields before reconstruction and interpolation can improve the numerical stability at discontinuities. By exploiting the fact that the equations are decoupled in the characteristic space, numerical dissipation is added much more precisely at shocks. The characteristic decomposition and interpolation with the WCHR6 scheme is described in the section below.

\subsubsection{Characteristic decomposition}
For the 1D Euler equation system in primitive form, the three characteristic variables, $\xi^0$, $\xi^1$, and $\xi^2$, at midpoint are given by:
\begin{equation}
\begin{pmatrix}
  \xi^0  \\
  \xi^1 \\
  \xi^2
 \end{pmatrix} =  \bm{R}^{-1} \begin{pmatrix}
  \rho \\
  u \\
  p
 \end{pmatrix},
\end{equation}
where $\bm{R}^{-1}$ is the matrix of the left eigenvectors (inverse of the matrix of the right eigenvectors $\bm{R}$) of the linearized Euler system given by:
\begin{equation}
\bm{R}^{-1} = \begin{pmatrix}
  0 & -\frac{{\rho} c}{2} & \frac{1}{2}  \\
  1 & 0 & -\frac{1}{c^2} \\
  0 &  \frac{{\rho} c}{2} & \frac{1}{2}
 \end{pmatrix},
\end{equation}
where $c = \sqrt{\gamma p / \rho}$ is the speed of sound in the medium. The expressions for $\bm{R}^{-1}$ in 3D problems are given in section 7.1 of \citet{wong2017high}.

At a midpoint $j+1/2$, the characteristic variables for all points in the stencil are computed using the same left eigenvector matrix $\bm{R}^{-1}_{j+\frac{1}{2}}$ to maintain consistency between the transforms to and back from the characteristic space. $\bm{R}^{-1}_{j+\frac{1}{2}}$ is computed using $\rho$ and $c$ values given by the Roe average or arithmetic average of nodes $j$ and $j+1$. The interpolation scheme for characteristic variables is given by:
\begin{align}
\alpha^l_{j+\frac{1}{2}} \tilde{\xi}^l_{j-\frac{1}{2}} + \beta^l_{j+\frac{1}{2}} \tilde{\xi}^l_{j+\frac{1}{2}} + \gamma^l_{j+\frac{1}{2}} \tilde{\xi}^l_{j+\frac{3}{2}} = a^l_{j+\frac{1}{2}} \xi^l_{j-2} + b^l_{j+\frac{1}{2}}\xi^l_{j-1} + c^l_{j+\frac{1}{2}}\xi^l_{j} \nonumber \\ + d^l_{j+\frac{1}{2}}\xi^l_{j+1} + e^l_{j+\frac{1}{2}} \xi^l_{j+2} + f^l_{j+\frac{1}{2}} \xi^l_{j+3}, \quad l = 0, 1, 2,
\end{align}
where $\alpha^l_{j+\frac{1}{2}}$, $\beta^l_{j+\frac{1}{2}}$, $\gamma^l_{j+\frac{1}{2}}$, $a^l_{j+\frac{1}{2}}$, $b^l_{j+\frac{1}{2}}$, $c^l_{j+\frac{1}{2}}$, $d^l_{j+\frac{1}{2}}$, $e^l_{j+\frac{1}{2}}$, and $f^l_{j+\frac{1}{2}}$ are the coefficients obtained from the nonlinear explicit-compact interpolation method described in equation~\eqref{eq:nonlinear_ECI}. However, the above equation cannot be solved in the form presented above as the interpolated characteristic variables are coupled across grid points due to the compact nature of the interpolation. Solving it in this form would introduce a consistency error since each edge interpolation equation uses a different characteristic matrix for the decomposition. A solution to this is to recast the above equation of scalars to an equation of vectors of the primitive variables at the cell nodes $\bm{V} = (\rho, u, p)^T$ and the unknown midpoint interpolated primitive variables $\tilde{\bm{V}} = (\tilde{\rho}, \tilde{u}, \tilde{p})^T$:
\begin{align}
   \left( \bm{\alpha}_{j+\frac{1}{2}} \bm{R}^{-1}_{j+\frac{1}{2}} \right) \tilde{\bm{V}}_{j-\frac{1}{2}}
+ &\left( \bm{\beta}_{j+\frac{1}{2}}  \bm{R}^{-1}_{j+\frac{1}{2}} \right) \tilde{\bm{V}}_{j+\frac{1}{2}}
+  \left( \bm{\gamma}_{j+\frac{1}{2}} \bm{R}^{-1}_{j+\frac{1}{2}} \right) \tilde{\bm{V}}_{j+\frac{3}{2}} = \nonumber \\
&   \left( \bm{a}_{j+\frac{1}{2}}     \bm{R}^{-1}_{j+\frac{1}{2}} \right) {\bm{V}}_{j-2}
+   \left( \bm{b}_{j+\frac{1}{2}}     \bm{R}^{-1}_{j+\frac{1}{2}} \right) {\bm{V}}_{j-1}
+   \left( \bm{c}_{j+\frac{1}{2}}     \bm{R}^{-1}_{j+\frac{1}{2}} \right) {\bm{V}}_{j} \nonumber \\
+ & \left( \bm{d}_{j+\frac{1}{2}}     \bm{R}^{-1}_{j+\frac{1}{2}} \right) {\bm{V}}_{j+1}
+   \left( \bm{e}_{j+\frac{1}{2}}     \bm{R}^{-1}_{j+\frac{1}{2}} \right) {\bm{V}}_{j+2}
+   \left( \bm{f}_{j+\frac{1}{2}}     \bm{R}^{-1}_{j+\frac{1}{2}} \right) {\bm{V}}_{j+3},
\end{align}

\noindent where $\bm{\alpha_{j+\frac{1}{2}}}$, $\bm{\beta_{j+\frac{1}{2}}}$, $\bm{\gamma_{j+\frac{1}{2}}}$, $\bm{a_{j+\frac{1}{2}}}$, $\bm{b_{j+\frac{1}{2}}}$, $\bm{c_{j+\frac{1}{2}}}$, $\bm{d_{j+\frac{1}{2}}}$, $\bm{e_{j+\frac{1}{2}}}$, and $\bm{f_{j+\frac{1}{2}}}$ are diagonal matrices with the diagonal entries representing the coefficents obtained using the nonlinear weighting procedure for the corresponding characteristic variable. With the characteristic decomposition, the interpolation reduces to one block tri-diagonal system of equations instead of three tri-diagonal systems of equations if only the primitive variables are interpolated. Note that we only use arithmetic average of node values for the matrix $\bm{R}^{-1}_{j+\frac{1}{2}}$ in this work. Section~\ref{sec:block_tridiag} details an efficient algorithm to solve the block-tridiagonal system resulting from this characteristic interpolation.

\begin{figure}[!ht]
\begin{center}
\subfigure[Full]{\includegraphics[width=0.35\textwidth]{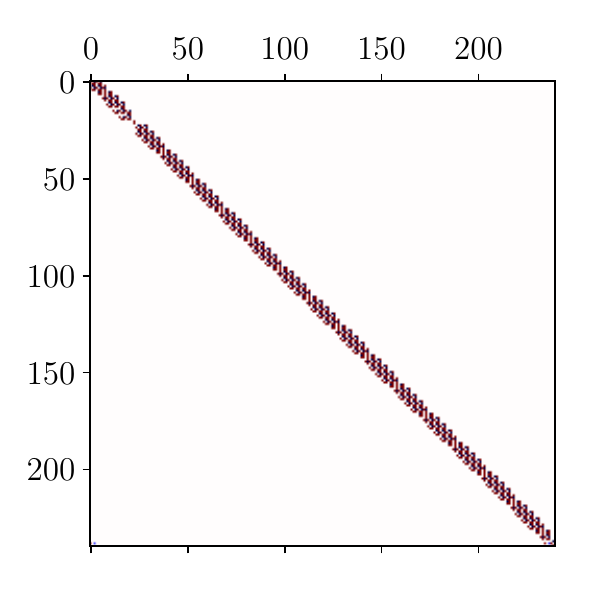}\label{fig:characteristic_matrix_a}}
\subfigure[Top $50\times50$ portion]{\includegraphics[width=0.35\textwidth]{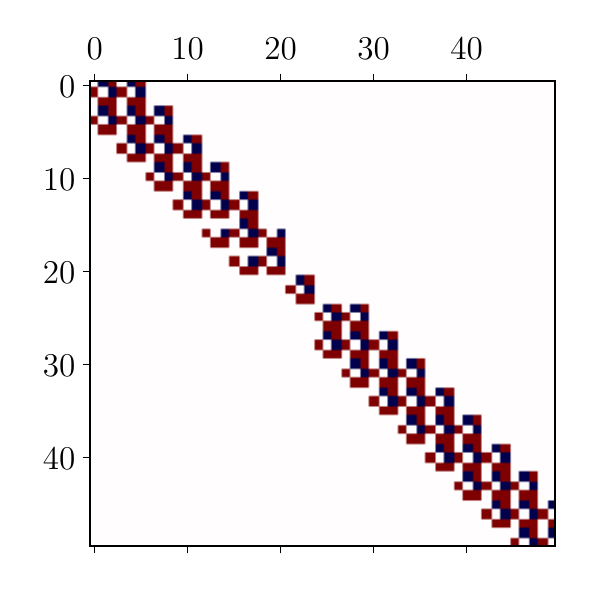}\label{fig:characteristic_matrix_b}}
\caption{Block tri-diagonal matrix structure for the characteristic decomposition in left-biased interpolation for initial condition of the Shu--Osher problem with 80 points. The top portion of the matrix shows that the interpolation is decoupled across the shock. Red indicates positive values, blue indicates negative values, and white indicates zero values.}
\label{fig:characteristic_matrix}
\end{center}
\end{figure}

\begin{figure}[!ht]
\begin{center}
\subfigure[Density]{\includegraphics[width=0.48\textwidth]{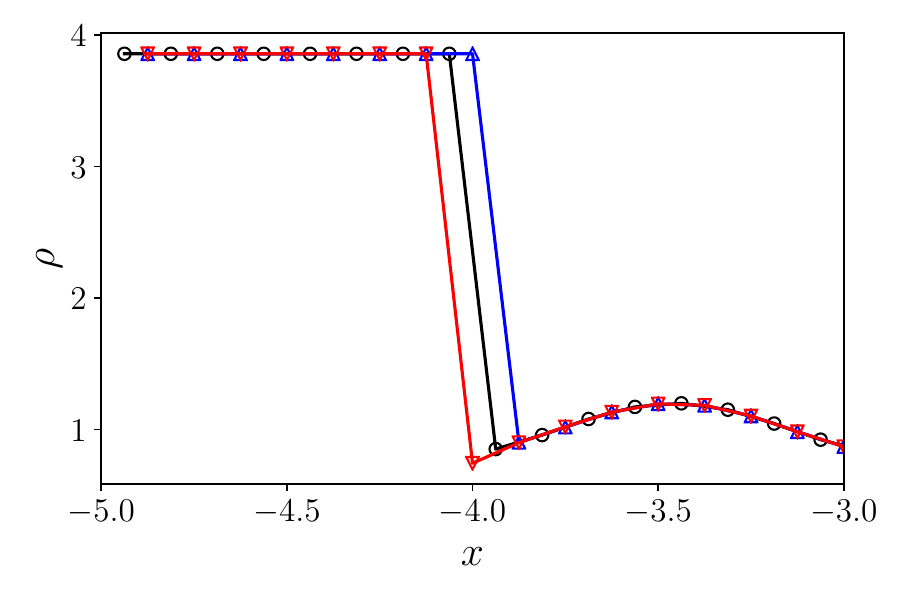}}
\subfigure[Velocity]{\includegraphics[width=0.48\textwidth]{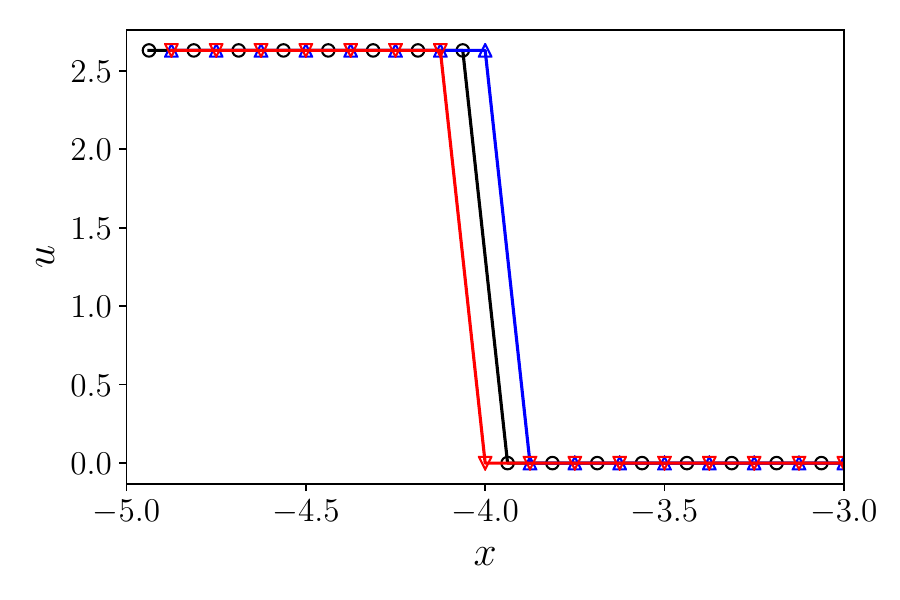}}
\subfigure[Pressure]{\includegraphics[width=0.48\textwidth]{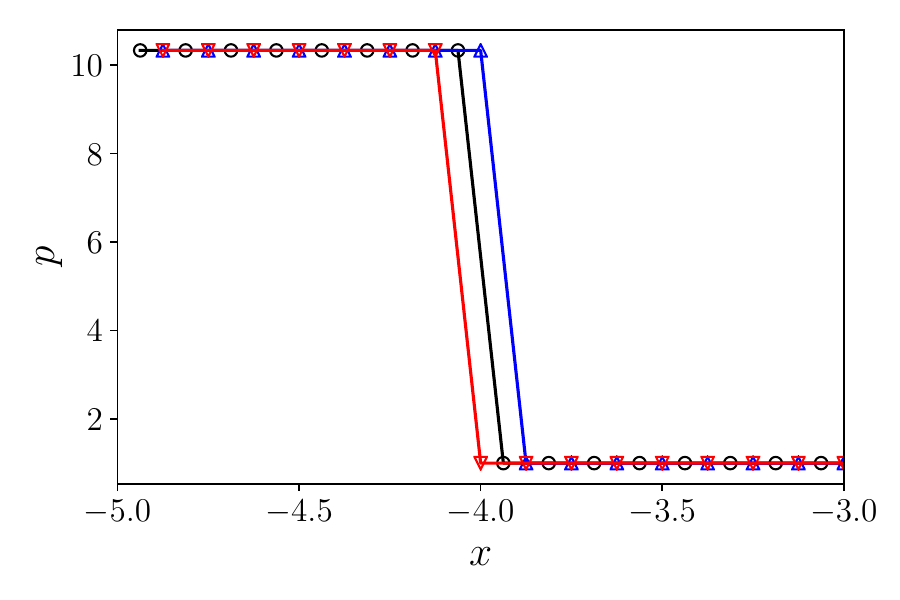}}
\caption{Interpolated values of primitive variables from initial conditions of the Shu--Osher problem with 80 points. Black circles: node values; blue upper triangles: left interpolated midpoint values; red lower triangles: right interpolated midpoint values}
\label{fig:characteristic_interpolation}
\end{center}
\end{figure}

Figure~\ref{fig:characteristic_matrix} shows the matrix structure for the left biased characteristic based weighted compact interpolation for the initial conditions of the Shu--Osher problem (section~\ref{sec:shuosher}) with 80 points in the domain. Since the matrix is a block tri-diagonal system, the size of the matrix is $240\times240$ and the full matrix structure is shown in figure~\ref{fig:characteristic_matrix_a}. Figure~\ref{fig:characteristic_matrix_b} shows the first $50\times50$ portion of the interpolation matrix. Here, we see that across the shock at index $\sim 24$, the matrix decouples. This means that the interpolation stencil never crosses the shock. Additionally, the point closest to the shock has just one block in it's row indicating that the nonlinear weighting procedure picked solely the most upwind stencil at the shock which is purely explicit. Figure~\ref{fig:characteristic_interpolation} shows the left and right interpolated density, velocity, and pressure. Since the interpolation stencil never crosses the shock, the interpolation is virtually perfect and no spurious oscillations are observed.

The method can be easily extended from 1D to multi-dimensional problems by applying the algorithm along each spatial dimension to get the flux derivatives in that direction. 

For the 3D Euler equations:
\begin{equation}
\pd{\bm{Q}}{t} + \pd{\bm{F}(\bm{Q})}{x} + \pd{\bm{G}(\bm{Q})}{y} + \pd{\bm{H}(\bm{Q})}{z} = 0,
\end{equation}
the flux derivatives $\partial \bm{F}(\bm{Q}) / \partial x$ are obtained using the algorithm outlined above in the $x$ direction and similarly for the flux derivatives $\partial \bm{G}(\bm{Q}) / \partial y$ and $\partial \bm{H}(\bm{Q}) / \partial z$ in the $y$ and $z$ directions using grid spacings $\Delta y$ and $\Delta z$ respectively.

\subsection{Cost estimate}
The cost estimates for a single left-biased interpolation for the 3D Euler equations using different interpolation schemes are shown in table~\ref{table:cost}. These are based on the operation count of each sub-algorithm per grid point. The LD nonlinear weights are used for all schemes in this comparison. Although the matrix solve portion of the interpolation algorithm for ECI on characteristic variables is approximately $20$ times more expensive than the corresponding EI, this difference is dwarfed by the large operation count of computing the smoothness indicators and nonlinear weights. In total, performing ECI on characteristic variables is $\approx 23\%$ more expensive than performing EI on characteristic variables in terms of the operation count.

\begin{table}[!ht]
  \begin{center}
  \begin{tabular}{| c | c | c | c | c |}
      \hline
      \textbf{Operation}     & \multicolumn{4}{c|}{\textbf{Interpolation methods}} \\
      \cline{2-5}
      \textbf{counts}              &    (a)  &     (b)   &    (c)   &    (d)    \\ \hline \hline
      Matrix solve                 &    $0$  &    $11$   &   $45$   &  $195$    \\ \hline
      R.H.S. interpolation         &   $55$  &    $55$   &   $55$   &   $55$    \\ \hline
      Characteristic decomposition &    $0$  &    $66$   &    $0$   &   $66$    \\ \hline
      Smoothness indicators        &  $440$  &   $440$   &  $440$   &  $440$    \\ \hline
      Nonlinear weights            &  $630$  &   $630$   &  $720$   &  $720$    \\ \hline \hline
      \textbf{Total}               & $1125$  &  $1202$   &  $1260$  & $1476$    \\ \hline
  \end{tabular}
  \caption{Operation counts per grid point for different interpolation methods with the LD nonlinear weights. (a) EI on primitive variables; (b) EI on characteristic variables;
  (c) ECI on primitive variables; (d) ECI on characteristic variables.}
  \label{table:cost}
  \end{center}
\end{table}

\subsection{Hybridization of Riemann solvers for multi-dimensional Euler equations}

\noindent The 3D Euler equations are given by:
\begin{align}
	\frac{\partial \rho}{\partial t} + \nabla \cdot \left( \rho \bm{u} \right) = 0, \\
    \frac{\partial \rho \bm{u}}{\partial t} + \nabla \cdot \left( \rho \bm{uu} + p \bm{\delta} \right) = 0, \\
    \frac{\partial E}{\partial t} + \nabla \cdot \left[ \left( E + p \right) \bm{u} \right] = 0,
\end{align}
where $\bm{u} = \left(u, v, w \right)^T = \left(u_1, u_2, u_3 \right)^T$ is the velocity vector.

In this work, we use the hybrid HLLC-HLL Riemann solver proposed by \citet{huang2011cures} (see \ref{appendix:HLLC_HLL} for details on the Riemann solver) when the Ducros-like shock sensor \cite{larsson2007effect} value, $s$, is greater than 0.65. $s$ is defined as:
\begin{equation} \label{eq:Larsson_switch}
	s = \frac{-\theta}{\left| \theta \right| + \left| \bm{\omega} \right| + \epsilon},
\end{equation}
where $\theta = \nabla \cdot \bm{u}$ is the rate of dilatation and $\bm{\omega} = \nabla \times \bm{u}$ is the vorticity. $\epsilon = 1.0\mathrm{e}{-15}$ is a small constant to prevent division by zero. If $s \leq 0.65$, the HLLC Riemann solver is used instead. The HLLC-HLL Riemann solver is a cure to the HLLC Riemann solver on the potential numerical instabilities near shocks for multi-dimensional problems when the shock normal direction does not align well with the grid normal surface direction.

\subsection{Positivity-preserving for Euler equations}
Negative density and pressure may arise during the nonlinear interpolation or the numerical time stepping processes to cause numerical failures for WCHR and WCNS's. While first order interpolation can be used instead to ensure that density and pressure are positive when it is detected that the nonlinearly interpolated density or pressure has become negative, a different positivity-preserving approach has to be considered regarding the positivity failures due to time stepping with the finite difference scheme. The positivity-preserving limiter designed by \citet{hu2013positivity} can be a cure for the positivity failures during the time stepping process for Euler problems but requires the use of reconstructed vector flux, $\widehat{\bm{F}}_{j+\frac{1}{2}}$, from the flux difference form given by equation~\eqref{eq:flux_reconstruction_form_derivative} in the vector form. During time stepping, the positivity-preserving method replaces $\widehat{\bm{F}}_{j+\frac{1}{2}}$ at any midpoint with a limited flux, $\widehat{\bm{F}}^{**}_{j+\frac{1}{2}}$, which is given by:
\begin{equation}
    \widehat{\bm{F}}^{**}_{j+\frac{1}{2}} = \left( 1 - \theta_{\rho, j+\frac{1}{2}} \theta_{p, j+\frac{1}{2}} \right) \widehat{\bm{F}}^{LF}_{j+\frac{1}{2}} + \theta_{\rho, j+\frac{1}{2}} \theta_{p, j+\frac{1}{2}} \widehat{\bm{F}}_{j+\frac{1}{2}},
\end{equation}
where $\theta_{\rho, j+\frac{1}{2}}$ and $\theta_{p, j+\frac{1}{2}}$ are blending functions between 0 and 1 to hybridize $\widehat{\bm{F}}_{j+\frac{1}{2}}$ with the Lax-Friederichs flux, $\widehat{\bm{F}}^{LF}_{j+\frac{1}{2}}$. $\widehat{\bm{F}}^{LF}_{j+\frac{1}{2}}$ for 1D Euler equations is given by:
\begin{equation}
    \widehat{\bm{F}}^{LF}_{j+\frac{1}{2}} = \frac{1}{2} \left[ \bm{F}_j + \bm{F}_{j+1} + \left( \left| u \right| + c \right)_{\mathrm{max}} \left( \bm{Q}_j - \bm{Q}_{j+1} \right) \right]
\end{equation}
The procedures to compute $\theta_{\rho, j+\frac{1}{2}}$ and $\theta_{p, j+\frac{1}{2}}$ are given by \citet{hu2013positivity}. The convex combination of the reconstructed flux and the positivity-preserving Lax-Friederichs flux ensures the density and pressure to remain positive for any time stepping method that is a convex combination of Euler-forward time steps under the condition that Courant--Friedrichs--Lewy number, $\textnormal{CFL}$, is smaller than 0.5. In this work, we suggest to use the five-stage fourth order strong stability preserving Runge--Kutta (SSP-RK54) scheme~\cite{spiteri2002new} which is a convex combination of Euler-forward steps.

The positivity-preserving flux limiters can be implemented in a dimension-by-dimension fashion for multi-dimensional Euler problems such as 3D problems if the time step size, $\Delta t$, is given by the following conditions:
\begin{equation}
    \Delta t = \frac{\textnormal{CFL}}{\tau_x + \tau_y + \tau_z},
\end{equation}
where
\begin{equation}
    \tau_x = \frac{\left( \left| u \right| + c \right)_{\mathrm{max}}}{\Delta x}, \quad
    \tau_y = \frac{\left( \left| v \right| + c \right)_{\mathrm{max}}}{\Delta y}, \quad
    \tau_z = \frac{\left( \left| w \right| + c \right)_{\mathrm{max}}}{\Delta z}.
\end{equation}

\subsection{Discretization of viscous and diffusive fluxes for Navier--Stokes equations}

\noindent The 3D compressible Navier--Stokes equations are given by:
\begin{align}
	\frac{\partial \rho}{\partial t} + \nabla \cdot \left( \rho \bm{u} \right) = 0, \\
    \frac{\partial \rho \bm{u}}{\partial t} + \nabla \cdot \left( \rho \bm{uu} + p \bm{\delta} \right) - \nabla \cdot \bm{\tau} = 0, \\
    \frac{\partial E}{\partial t} + \nabla \cdot \left[ \left( E + p \right) \bm{u} \right] - \nabla \cdot \left( \bm{\tau} \cdot \bm{u} - \bm{q_c} \right) = 0.
\end{align}
$\bm{\tau}$ and $\bm{q_c}$ are viscous stress tensor and conductive heat flux respectively. $\bm{\delta}$ is the identity tensor.

The viscous stress tensor $\bm{\tau}$ for a Newtonian fluid is given by:
\begin{equation}
	\bm{\tau} = 2 \mu \bm{S} + \left( \mu_v - \frac{2}{3} \mu \right) \bm{\delta} \left( \nabla \cdot \bm{u} \right),
\end{equation}
\noindent where $\mu$ and $\mu_v$ are the shear viscosity and bulk viscosity respectively. $\bm{S}$ is the strain-rate tensor given by:
\begin{equation}
    \bm{S} = \frac{1}{2} \left[ \nabla \bm{u} + \left( \nabla \bm{u} \right) ^{T} \right].
\end{equation}

The conductive flux $\bm{q_c}$ is given by:
\begin{equation}
	\bm{q_c} = - \kappa \nabla T,
\end{equation}
where $\kappa$ is the thermal conductivity. $T$ is the temperature given by the equation of state for ideal gas:
\begin{equation}
    T = \frac{p}{\rho R},
\end{equation}
where $R$ is the gas constant.

All the viscous and diffusive terms are discretized in their non-conservative forms by isolating the Laplacian operator as in \citet{nagarajan2003robust,pirozzoli2010generalized}. The viscous term in the momentum equation is split as:
\begin{equation}
    \nabla \cdot \bm{\tau} = \mu \left( \nabla^2 \bm{u} + \nabla \theta \right) + 2 \bm{S} \nabla \mu + \lambda \bm{\delta} \cdot \nabla \theta + \theta \bm{\delta} \cdot \nabla \lambda, \label{eq:viscous_nonconservative}
\end{equation}
where $\lambda = \mu_v - 2 \mu / 3$ and $\theta = \nabla \cdot \bm{u}$ is the dilatation. The second derivative terms in the gradient of $\theta$ are also isolated as:
\begin{equation}
    \pd{\theta}{x_i} = \frac{\partial^2 u_i}{\partial x_i^2} + \sum_{k \neq i} \frac{\partial^2 u_k}{\partial x_i \partial x_k}.
\end{equation}
Summation is not implied by repeating indices in the above equation.

The heat conduction term is split as:
\begin{equation}
    \nabla \cdot \bm{q_c} = - \kappa \nabla^2 T - \nabla T \cdot \nabla \kappa, \label{eq:conductive_nonconservative}
\end{equation}
and the viscous power term is also split in a non-conservative form as:
\begin{equation}
    \nabla \cdot \left( \bm{\tau} \cdot \bm{u} \right) = \bm{u} \cdot \left( \nabla \cdot \bm{\tau} \right) + \bm{\tau} : \nabla \bm{u}, \label{eq:viscous_power_nonconservative}
\end{equation}
where equation~\eqref{eq:viscous_nonconservative} is used for $\nabla \cdot \bm{\tau}$.

In equations~\eqref{eq:viscous_nonconservative}-\eqref{eq:viscous_power_nonconservative}, the Laplacian and second derivative terms are discretized directly using a sixth order accurate second derivative compact finite difference scheme~\cite{lele1992compact} given by:
\begin{equation}
\frac{2}{15} \widehat{f}_{j-1}^{\prime\prime} + \frac{11}{15} \widehat{f}_{j}^{\prime\prime} + \frac{2}{15} \widehat{f}_{j+1}^{\prime\prime} = \frac{1}{\Delta x^2} \left[ \frac{4}{5} \left( f_{j+1} -2f_j + f_{j-1} \right) + \frac{1}{20} \left( f_{j+2} -2f_j + f_{j-2} \right) \right], \label{eq:CND2}
\end{equation}
\noindent where $\widehat{f}_{j}^{\prime\prime}$ are numerically approximated second derivatives of any variables $f$ at cell nodes and $f_{j}$ are $f$ at cell nodes.

The other terms are discretized using successive applications of a sixth order accurate first derivative compact node-to-node finite difference scheme (CND)~\cite{lele1992compact} given by:
\begin{equation}
\frac{1}{5} \widehat{f}_{j-1}^\prime + \frac{3}{5} \widehat{f}_{j}^\prime + \frac{1}{5} \widehat{f}_{j+1}^\prime = \frac{1}{\Delta x} \left[ \frac{7}{15} \left( f_{j+1} - f_{j-1} \right) + \frac{1}{60} \left( f_{j+2} - f_{j-2} \right) \right], \label{eq:CND}
\end{equation}
\noindent where $\widehat{f}_{j}^{\prime}$ are numerically approximated first derivatives of any variables $f$ at cell nodes.

\section{Numerical results \label{sec:results}}

In this section, we present results using WCNS5-JS, WCNS5-Z, WCNS6-LD, and WCHR6 schemes in different test problems. All tests are inviscid except the compressible homogeneous isotropic turbulence case where the compressible Navier--Stokes equations are used. In all problems, the equations are integrated in time using the five-stage fourth order SSP-RK54 scheme~\cite{spiteri2002new}. Positivity-preserving limiter~\cite{hu2013positivity} is only used in the 1D planar Sedov blast wave problem and the 2D double Mach reflection problem to overcome the negative density and pressure issues encountered\footnote{The positivity-preserving limiter has no effect on problems that do not have occurrence of negative density and pressure.}.

\subsection{Convergence tests}
The formal order of accuracy of each scheme is verified and compared through 1D and 2D problems involving advection of an entropy wave. The initial conditions in a 1D periodic domain $\left[-1, 1 \right)$ and a 2D periodic domain $\left[-1, 1 \right) \times \left[-1, 1 \right)$ are respectively given by:
\begin{align}
	\left( \rho, u, p \right)  &= \left(1 + 0.5 \sin \left( \pi x \right), 1, 1 \right), \\
	\left( \rho, u, v, p \right)  &= \left(1 + 0.5 \sin \left[ \pi \left(x + y \right) \right], 1, 1, 1 \right).
\end{align}

\noindent Since the velocity and pressure are constant and only entropic disturbances are present, the problems reduce to linear advection of the entropy wave. Therefore, the exact solutions are given by:
\begin{align}
	\left( \rho_{\mathrm{exact}}, u_{\mathrm{exact}}, p_{\mathrm{exact}} \right) &= \left(1 + 0.5 \sin \left[ \pi \left( x - t \right) \right] , 1, 1 \right), \\
	\left( \rho_{\mathrm{exact}}, u_{\mathrm{exact}}, v_{\mathrm{exact}}, p_{\mathrm{exact}} \right) &= \left(1 + 0.5 \sin \left[ \pi \left( x + y - 2t \right) \right] , 1, 1, 1 \right).
\end{align}

\noindent The ratio of specific heats $\gamma$ is 1.4. The simulations using different schemes are conducted up to $t = 2$ with mesh refinements from $N = 8$ to $N = 128$ points in each direction. All simulations are run with very small constant time steps in order to isolate the spatial error and observe the order of accuracy of different numerical schemes. $\Delta t / \Delta x = 0.02$ is chosen for both 1D and 2D simulations. The $L_2$ errors for the 1D and 2D problems are computed as:
\begin{align}
    L_2\ \mathrm{error}\ (1D) &= \sqrt{ \sum_{j=0}^{N-1} \Delta x \left( \rho_j - \rho_{\mathrm{exact}}\left( x_j \right) \right)^2 / \sum_{j=0}^{N-1} \Delta x}, \\
    L_2\ \mathrm{error}\ (2D) &= \sqrt{ \sum_{i=0}^{N-1} \sum_{j=0}^{N-1} \Delta x \Delta y \left( \rho_{i, j} - \rho_{\mathrm{exact}}\left( x_i, y_j \right) \right)^2 / \sum_{i=0}^{N-1} \sum_{j=0}^{N-1} \Delta x \Delta y }.
\end{align}

\begin{figure}[!ht]
 \centering
 \subfigure[1D]{\includegraphics[width=0.48\textwidth]{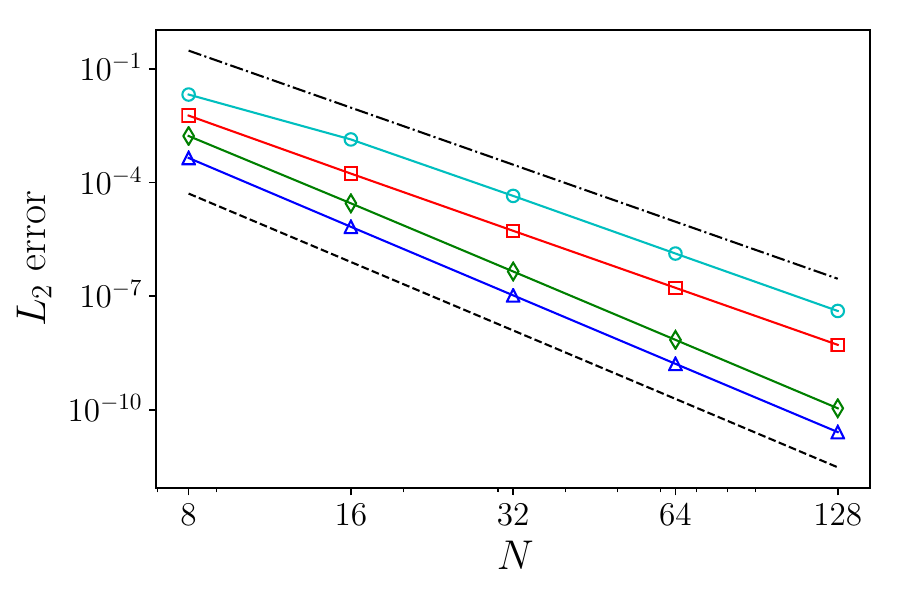} \label{fig:convergence_1D}}
 \subfigure[2D]{\includegraphics[width=0.48\textwidth]{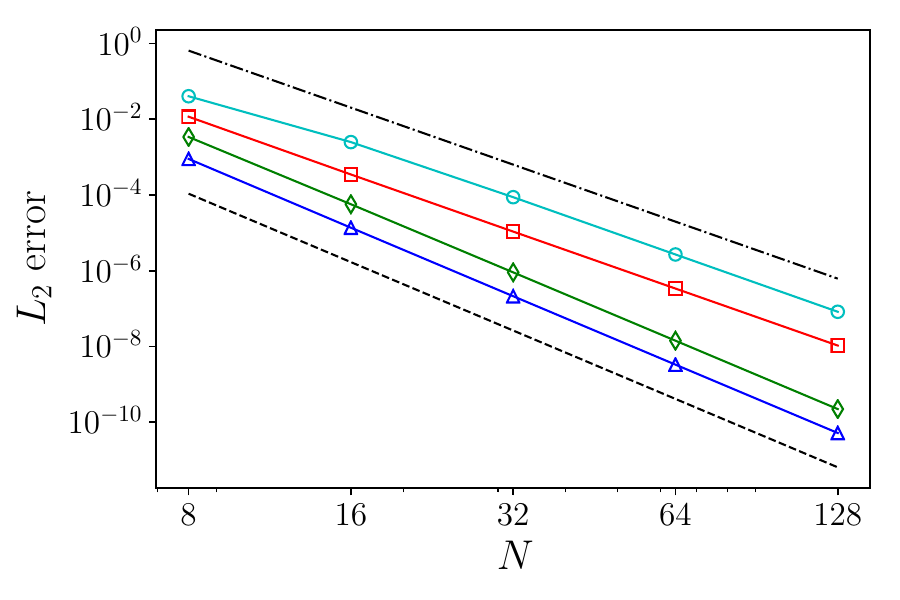} \label{fig:convergence_2D}}
\caption{$L_2$ errors against numbers of points, $N$, of different schemes for the 1D and 2D convergence tests. Cyan circles: WCNS5-JS; red squares: WCNS5-Z; green diamonds: WCNS6-LD; blue triangles: WCHR6; dashed-dotted line: fifth-order; dashed line: sixth-order}
\label{fig:convergence}
\end{figure}

From tables~\ref{table:L2_error_and_rate_of_convergence_1D} and \ref{table:L2_error_and_rate_of_convergence_2D} together with figure~\ref{fig:convergence}, we can see that all schemes can achieve their formal orders of accuracy when the number of points is large enough. Although both WCNS6-LD and WCHR6 are sixth order accurate, the latter scheme is more accurate than the former with errors that are $\approx 4-5$ times smaller. This is consistent with the ratio of their respective interpolation truncation errors which is $34/7 \approx 4.86$ since the major difference between the two schemes is the interpolation method.

\begin{table}[!ht]
\centering
\begin{tabular}{| c | c c | c c | c c | c c |}
	\hline
    Number & \multicolumn{2}{c|}{WCNS5-JS} & \multicolumn{2}{c|}{WCNS5-Z} & \multicolumn{2}{c|}{WCNS6-LD} & \multicolumn{2}{c|}{WCHR6} \\
    \cline{2-9}
    of points& error & order & error & order & error & order & error & order \\
    \hline
8   &   2.993e-02 &       &	8.328e-03 &	     &	2.410e-03 &	     &	6.339e-04 &      \\
16  &   1.954e-03 &  3.94 &	2.453e-04 &	5.09 &	4.028e-05 &	5.90 &	9.663e-06 &	6.04 \\
32  &   6.321e-05 &  4.95 &	7.579e-06 &	5.02 &	6.399e-07 &	5.98 &	1.500e-07 &	6.01 \\
64  &   1.905e-06 &  5.05 &	2.372e-07 &	5.00 &	1.004e-08 &	5.99 &	2.339e-09 &	6.00 \\
128 &   5.817e-08 &  5.03 &	7.416e-09 &	5.00 &	1.570e-10 &	6.00 &	3.697e-11 &	5.98 \\
    \hline
\end{tabular}
\caption{$L_2$ errors and orders of convergence of density for the 1D problem from different schemes at $t = 2$.}
\label{table:L2_error_and_rate_of_convergence_1D}
\end{table}

\begin{table}[!ht]
\centering
\begin{tabular}{| c | c c | c c | c c | c c |}
	\hline
    Number & \multicolumn{2}{c|}{WCNS5-JS} & \multicolumn{2}{c|}{WCNS5-Z} & \multicolumn{2}{c|}{WCNS6-LD} & \multicolumn{2}{c|}{WCHR6} \\
    \cline{2-9}
    of points& error & order & error & order & error & order & error & order \\
    \hline
    $8^2$   & 5.712e-02  &       & 1.647e-02  &       & 4.807e-03  & 	   & 1.265e-03  &      \\
    $16^2$  & 3.519e-03  & 4.02  & 4.915e-04  & 5.07  & 8.046e-05  & 5.90  & 1.930e-05  & 6.03 \\
    $32^2$  & 1.235e-04  & 4.83  & 1.526e-05  & 5.01  & 1.279e-06  & 5.98  & 2.999e-07  & 6.01 \\
    $64^2$  & 3.793e-06  & 5.02  & 4.778e-07  & 5.00  & 2.008e-08  & 5.99  & 4.683e-09  & 6.00 \\
    $128^2$ & 1.165e-07  & 5.03  & 1.494e-08  & 5.00  & 3.140e-10  & 6.00  & 7.332e-11  & 6.00 \\
    \hline
\end{tabular}
\caption{$L_2$ errors and orders of convergence of density for the 2D problem from different schemes at $t = 2$.}
\label{table:L2_error_and_rate_of_convergence_2D}
\end{table}

\subsection{Advection of broadband disturbances}
This problem is similar to the earlier one but with the density field of a uniform flow being disturbed by a broadband signal instead of a single mode. The initial conditions are given by:
\begin{equation}
	\begin{pmatrix}
		\rho \\
        u \\
        p \\
	\end{pmatrix}
    =
    \begin{pmatrix}
		1 + \delta \sum^{N/2}_{k=1} \left( E_{\rho}(k) \right) ^{1/2} \sin \left(2 \pi k \left( x +\psi_k \right) \right) \\
		1 \\
        1
	\end{pmatrix},
\end{equation}

\noindent where $\psi_k$ is a random number between 0 and 1 with uniform distribution, $\delta = 1.0\mathrm{e}{-2}$, and the ratio of specific heats $\gamma$ is 1.4. The density spectrum $E_{\rho}(k)$ is given by:
\begin{equation}
	E_{\rho}(k) = \left( \frac{k}{k_0} \right)^4 \exp \left( -2\left( \frac{k}{k_0} \right)^2 \right).
\end{equation}
\noindent We have chosen $k_0 = 12$. The computational domain is periodic on domain $x \in \left[0, 1\right)$. The simulations are run with $N = 128$ and $\Delta t = 0.002$ until $t=1$.

\begin{figure}[!ht]
\begin{center}
\subfigure[Global density profile]{\includegraphics[width=0.48\textwidth]{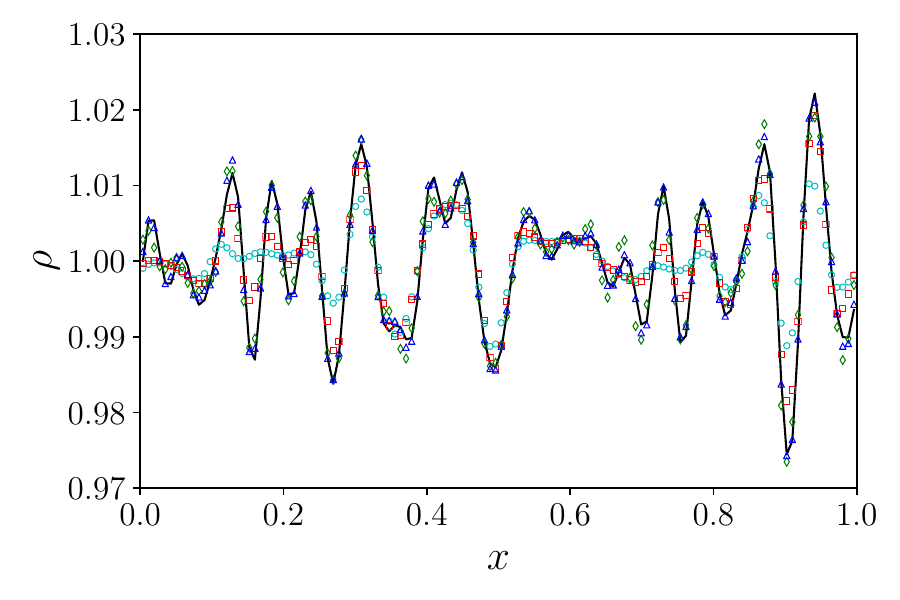} \label{fig:broadband_density_global}}
\subfigure[Spectrum of density disturbances]{\includegraphics[width=0.48\textwidth]{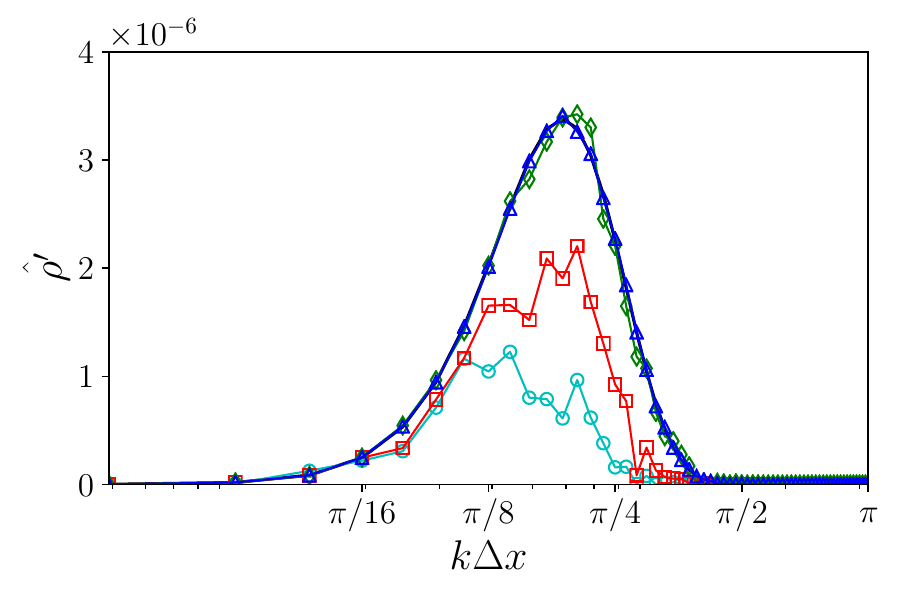} \label{fig:broadband_spectrum}}
\caption{Advection of broadband disturbances problem at $t = 1.0$ using different schemes. Black solid line: exact; cyan circles: WCNS5-JS; red squares: WCNS5-Z; green diamonds: WCNS6-LD; blue triangles: WCHR6.}
\label{fig:broadband_nonlinear}
\end{center}
\end{figure}

The density solutions from various schemes after one period are shown in figure~\ref{fig:broadband_density_global}. Since this problem reduces to linear advection, we should expect the initial density spectrum to be preserved without any corruption. However, the schemes themselves are nonlinear and would introduce some coupling between different modes. Figure~\ref{fig:broadband_spectrum} compares the spectra of the density disturbance from different schemes. We see that both WCNS5-JS and WCNS5-Z are too dissipative to preserve the initial spectrum due to their upwind nature. WCNS6-LD preserves the initial spectrum better, but still has some deviations from the prescribed spectrum. WCHR6 preserves the initial spectrum virtually perfectly. Unlike the WCNS's almost no errors due to the nonlinear nature of the scheme are seen. This is attributed to its higher resolution characteristics.

\subsection{Entropy wave leaving domain} \label{sec:Gaussain_boundary}

In this 1D inviscid problem, the advection of a Gaussian entropy wave leaving a domain $x \in \left[0, 1 \right]$ is simulated with the WCHR scheme and boundary closures. The initial conditions are given by:
\begin{equation}
	\left( \rho, u, p \right)
    =
    \left(1 + 0.1 \exp{ \left( - 400 \left( x - 0.5 \right)^{2} \right) }, 0.5, 1 \right).
\end{equation}

The ratio of specific heats $\gamma$ is 1.4. As the Gaussian pulse is being advected, it eventually reaches the right boundary and leaves the domain. Two boundary treatment methods to fill ghost cells at the boundaries are compared: (1) constant extrapolation from interior solutions and (2) sub-sonic inflow and outflow boundary conditions at the left and right boundaries respectively following the non-reflective characteristic ghost cell method in \citet{motheau2017navier}.

Primitive variables are used for the constant extrapolation method. For the non-reflective subsonic outflow method, $\sigma=0.005$, $l_x=0.1$, and $p_t=1$ are used. As for the non-reflective subsonic inflow method, $\eta_2 = \eta_3 = 0.005$, $l_x=0.1$, $u_t=0.5$, and $\left( p/\rho \right)_t=RT_t=1$ are set. The details of the implementation of the non-reflective characteristic method as well as interpretation of the parameters detailed above are explained in~\cite{motheau2017navier}. Simulations are performed with constant time steps $\Delta t = 0.002$ on a uniform grid composed of $N=128$ grid points.

\begin{figure}[!ht]
\begin{center}
\subfigure[$t=0.9$]{\includegraphics[width=0.31\textwidth]{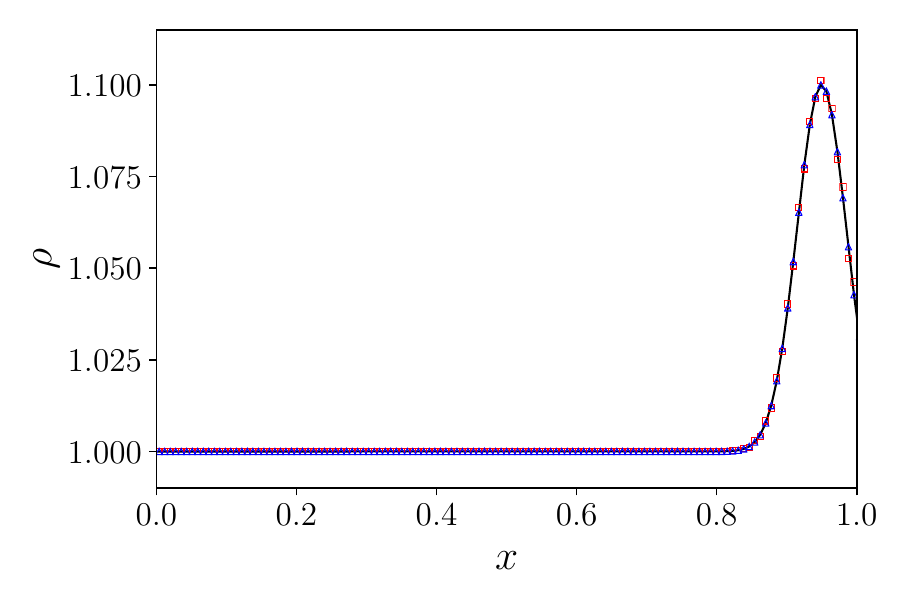} \label{fig:Gaussian_boundary_density_global_t1}}
%\subfigure[$t=1.0$]{\includegraphics[width=0.31\textwidth]{Gaussian_boundary/compare_Gaussian_boundary_density_global_N128_t2.pdf} \label{fig:Gaussian_boundary_density_global_t2}}
\subfigure[$t=1.1$]{\includegraphics[width=0.31\textwidth]{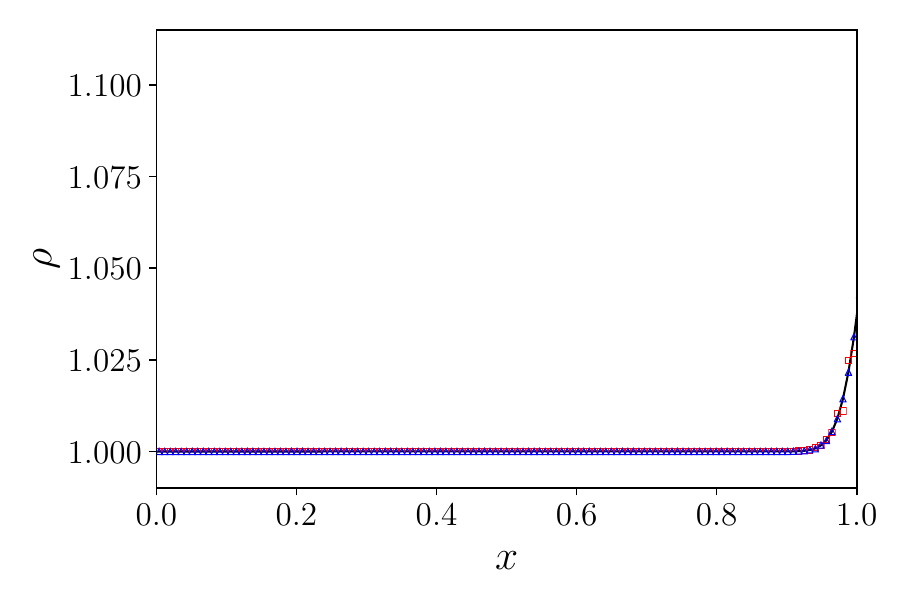} \label{fig:Gaussian_boundary_density_global_t3}}
\subfigure[$t=1.5$]{\includegraphics[width=0.31\textwidth]{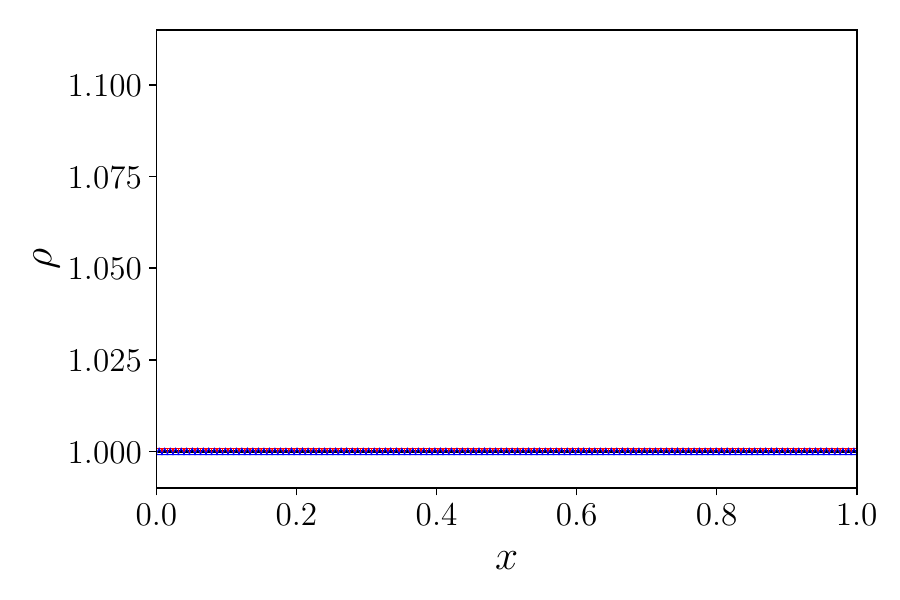} \label{fig:Gaussian_boundary_density_global_end}}
\caption{Global density profiles for the entropy wave leaving domain problem at different times. Black solid line: exact; red squares: extrapolation; blue triangles: non-reflective characteristic boundary conditions.}
\label{fig:Gaussian_boundary_density_global}
\end{center}
\end{figure}

\begin{figure}[!ht]
\begin{center}
\subfigure[$t=0.9$]{\includegraphics[width=0.48\textwidth]{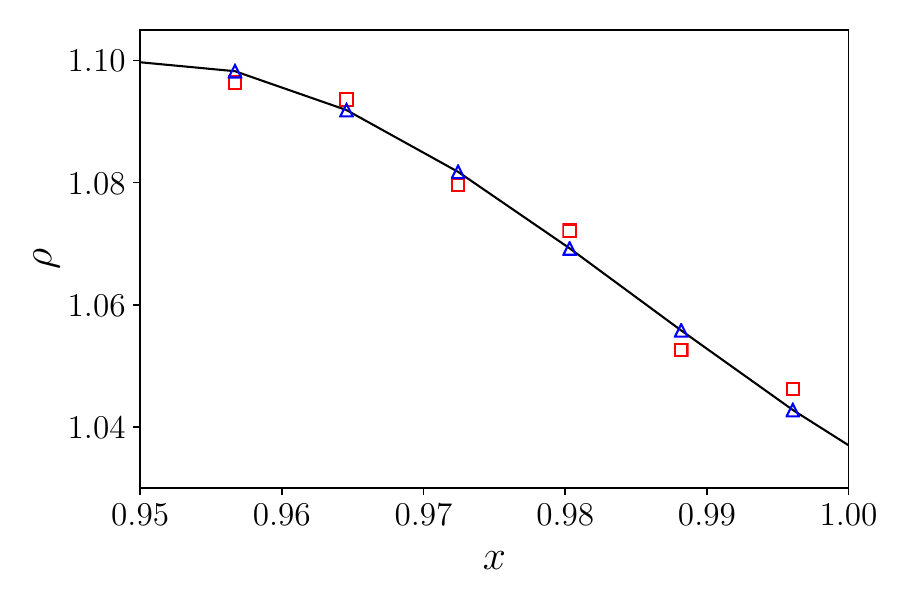} \label{fig:Gaussian_boundary_density_local_t1}}
%\subfigure[$t=1.0$]{\includegraphics[width=0.48\textwidth]{Gaussian_boundary/compare_Gaussian_boundary_density_local_N128_t2.pdf} \label{fig:Gaussian_boundary_density_local_t2}}
\subfigure[$t=1.1$]{\includegraphics[width=0.48\textwidth]{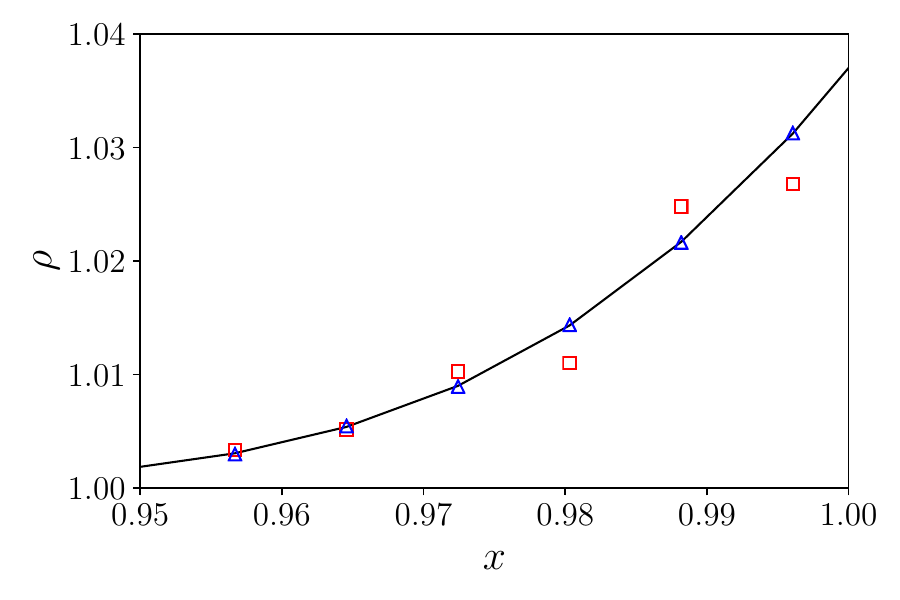} \label{fig:Gaussian_boundary_density_local_t3}}
\caption{Local density profiles for the entropy wave leaving domain problem at different times. Black solid line: exact; red squares: extrapolation; blue triangles: non-reflective characteristic boundary conditions.}
\label{fig:Gaussian_boundary_density_local}
\end{center}
\end{figure}

\begin{figure}[!ht]
\begin{center}
\includegraphics[width=0.5\textwidth]{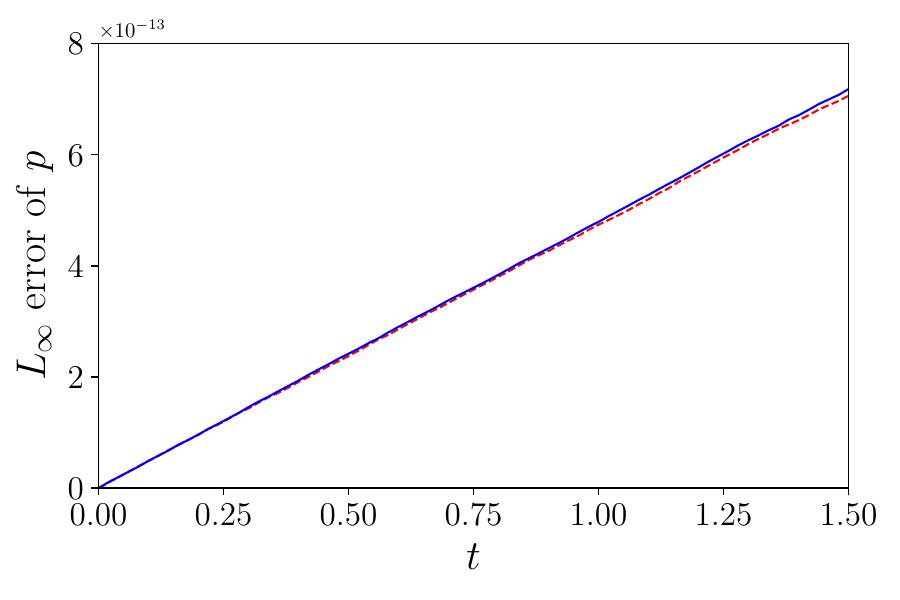}
\caption{$L_{\infty}$ errors of pressure against time for the entropy wave leaving domain problem. Red dashed line: extrapolation; blue solid line: non-reflective characteristic boundary conditions.}
\label{fig:Gaussian_boundary_pressure_error_over_time}
\end{center}
\end{figure}

From figures~\ref{fig:Gaussian_boundary_density_global} and \ref{fig:Gaussian_boundary_density_local}, it can be seen that both boundary methods allow the entropy wave to leave the domain when they are used with the boundary schemes in section~\ref{sec:boundary_closures}. Figure~\ref{fig:Gaussian_boundary_pressure_error_over_time} shows that the $L_{\infty}$ errors of pressure are very small for both methods. This indicates that acoustic components of any unphysical reflections at the outflow boundary are insignificant for both methods. However, the non-reflective characteristic method outperforms the extrapolation method in accuracy of the solution of density field at different times which shows the necessity of non-reflective characteristic method in the boundary treatment to properly treat the outgoing entropic wave.

\subsection{Sod shock tube problem}
This is a 1D shock tube problem introduced by \citet{sod1978survey}. The problem consists of the propagation of a shock wave, a contact discontinuity, and an expansion fan. The initial conditions are given by:
\begin{equation}
\begin{aligned}
	\left( \rho, u, p \right)
    =
    \begin{cases}
    	\left(1, 0, 1 \right), &\mbox{$x < 0$}, \\
    	\left(0.125, 0, 0.1 \right), &\mbox{$x \geq 0$}. \\
    \end{cases}
\end{aligned}
\end{equation}
The ratio of specific heats $\gamma$ is 1.4. The computational domain has size $x \in \left[-0.5, 0.5 \right]$. Simulations are performed with constant time steps $\Delta t = 0.002$ on a uniform grid composed of 100 grid points where $\Delta x = 0.01$.

\begin{figure}[!ht]
\begin{center}
\subfigure[Global density profile]{\includegraphics[width=0.48\textwidth]{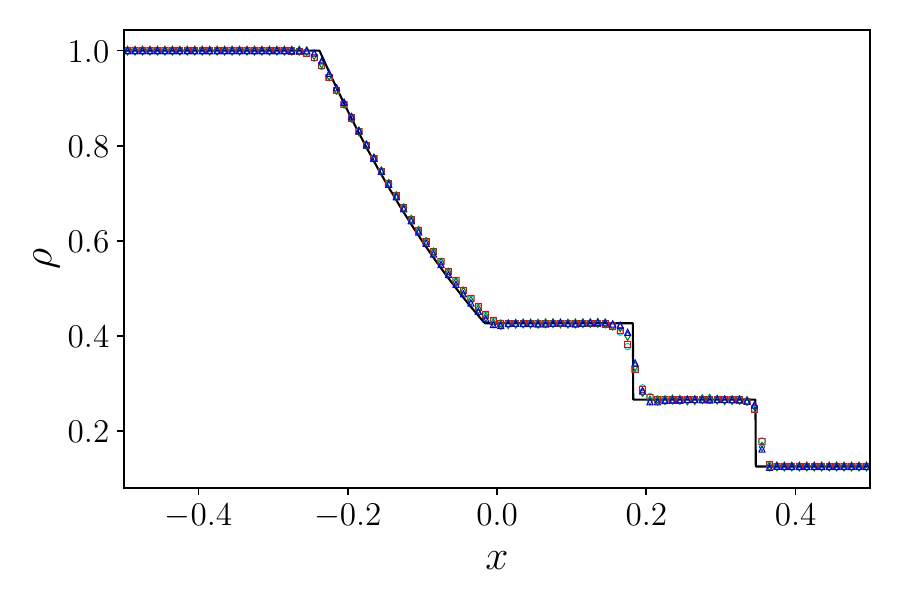} \label{fig:Sod_density_global}}
\subfigure[Local density profile]{\includegraphics[width=0.48\textwidth]{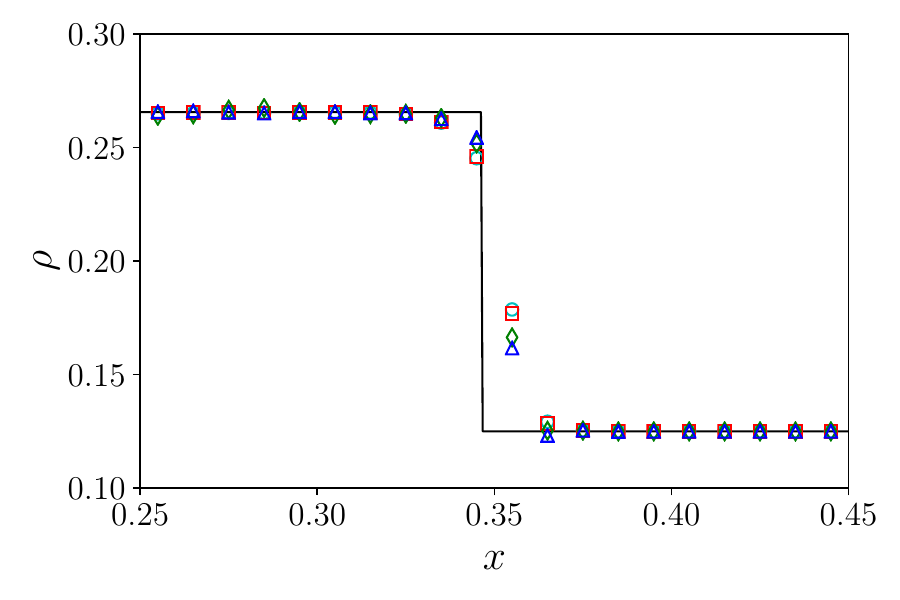} \label{fig:Sod_density_local}}
\caption{Sod shock tube problem at $t = 0.2$ using different schemes. Black solid line: exact; cyan circles: WCNS5-JS; red squares: WCNS5-Z; green diamonds: WCNS6-LD; blue triangles: WCHR6.}
\label{fig:Sod}
\end{center}
\end{figure}

Comparison between the exact solution and the numerical solution for the density at $t = 0.2$ is shown in figure~\ref{fig:Sod}. It can be seen that all of the schemes can capture the shock well. WCHR6 and WCNS6-LD have sharper profiles at the shock in comparison to WCNS5-JS and WCNS5-Z.

\subsection{Shu--Osher problem} \label{sec:shuosher}
This 1D problem first proposed by \citet{shu1988efficient} involves the interaction of a Mach 3 shock wave with an entropy wave. The interaction creates a high wavenumber entropy wave and a nonlinear acoustic wave that steepens and forms a shock train. This problem can hence assess the ability of a scheme to capture discontinuities well, while also retaining the smooth features of the solution. The initial conditions are given by:
\begin{equation}
\begin{aligned}
	\left( \rho, u, p \right)
    =
    \begin{cases}
    	\left(27/7, 4 \sqrt{35}/9, 31/3 \right), &\mbox{$x < -4$}, \\
    	\left(1 + 0.2 \sin{(5x)}, 0, 1 \right), &\mbox{$x \geq -4$}. \\
    \end{cases}
\end{aligned} 
\end{equation}

\noindent The ratio of specific heats $\gamma$ is 1.4. The spatial domain of the problem is $x \in \left[-5, 5 \right]$. Simulations are conducted with constant time steps $\Delta t = 0.005$ on a uniform grid with 150 grid points and also with constant time steps $\Delta t = 0.004$ on a uniform grid with 200 grid points. A reference solution is computed using the WCNS6-LD scheme with 2000 points and time step of $\Delta t = 0.0002$. All results shown here are at time $t = 1.8$.

\begin{figure}[!ht]
\begin{center}
\subfigure[Global density profile]{\includegraphics[width=0.48\textwidth]{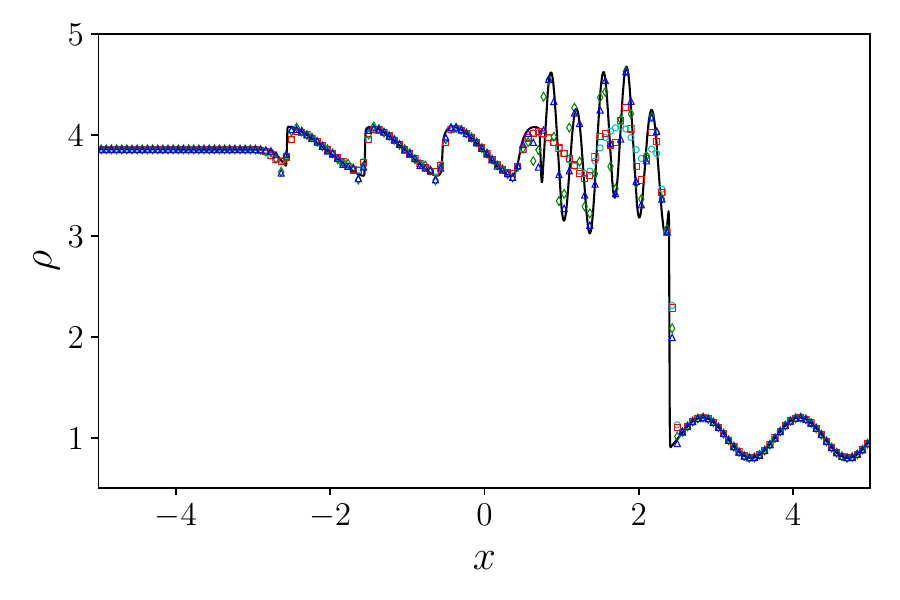} \label{fig:ShuOsher_density_global_coarse}}
\subfigure[Local density profile]{\includegraphics[width=0.48\textwidth]{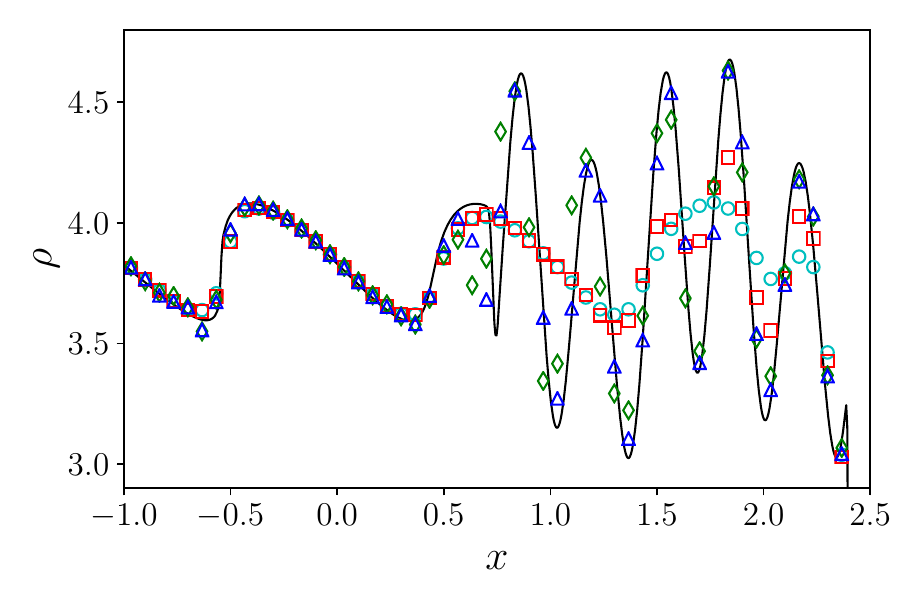} \label{fig:ShuOsher_density_local_coarse}}
\caption{Shu--Osher problem at $t = 0.2$ using different schemes with $\Delta x = 1/15$. Black solid line: reference; cyan circles: WCNS5-JS; red squares: WCNS5-Z; green diamonds: WCNS6-LD; blue triangles: WCHR6.}
\label{fig:ShuOsher_coarse}
\end{center}
\end{figure}

\begin{figure}[!ht]
\begin{center}
\subfigure[Global density profile]{\includegraphics[width=0.48\textwidth]{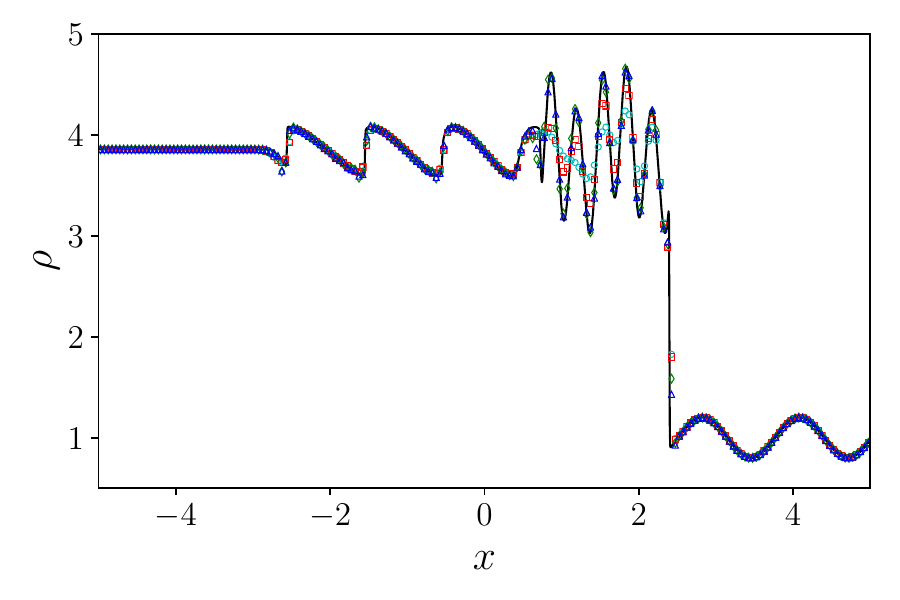} \label{fig:ShuOsher_density_global_fine}}
\subfigure[Local density profile]{\includegraphics[width=0.48\textwidth]{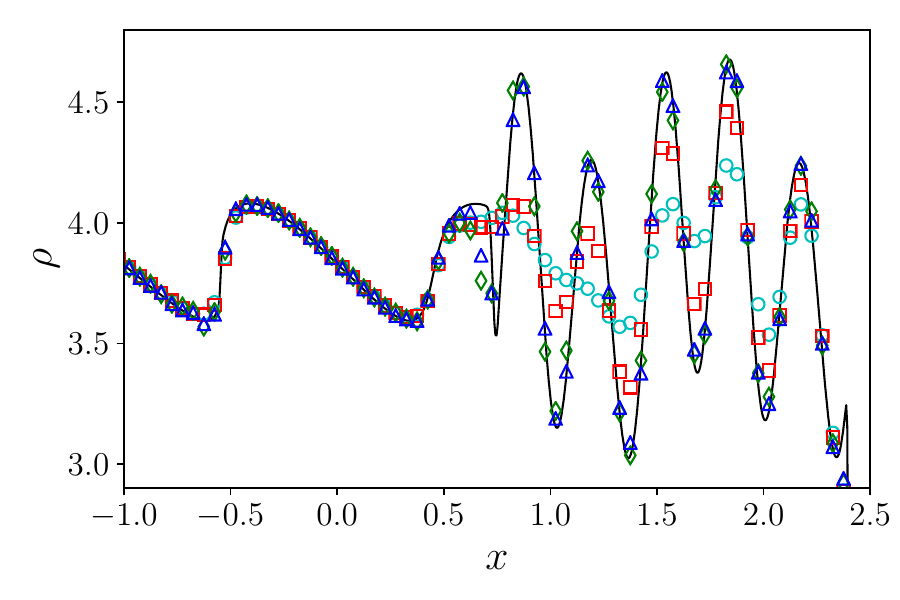} \label{fig:ShuOsher_density_local_fine}}
\caption{Shu--Osher problem at $t = 0.2$ using different schemes with $\Delta x = 0.05$. Black solid line: reference; cyan circles: WCNS5-JS; red squares: WCNS5-Z; green diamonds: WCNS6-LD; blue triangles: WCHR6.}
\label{fig:ShuOsher_fine}
\end{center}
\end{figure}

Figures~\ref{fig:ShuOsher_coarse} and \ref{fig:ShuOsher_fine} show the density profile at $t=1.8$ obtained using various schemes compared to the reference solution with the two different grid resolutions. Both WCNS5-JS and WCNS5-Z dissipate the high wavenumber entropy wave significantly which is not seen in the results from the WCNS6-LD and WCHR6 schemes. Figure~\ref{fig:ShuOsher_density_local_coarse} shows that WCHR6 has less dispersion error around the region where the entropy wave and weak shock interacts from the results with 150 points due to the higher resolution characteristics of WCHR6.

\subsection{One-dimensional planar Sedov blast wave problem}
This 1D planar Sedov blast wave problem \cite{sedov1993similarity, zhang2012positivity, hu2013positivity} is a near vacuum problem with the propagation of blast waves. The initial conditions are given by:
\begin{equation}
\begin{aligned}
	\left( \rho, u, p \right)
    =
    \begin{cases}
    	\left(1, 0, 4.0\mathrm{e}{-13} \right), &\mbox{$x < 2-0.5 \Delta x$, $x > 2+0.5\Delta x$}, \\
    	\left(1, 0, \frac{ 1.28\mathrm{e}{6} }{ \Delta x } \right), &\mbox{$2-0.5\Delta x \leq x \leq 2+0.5\Delta x$}. \\
    \end{cases}
\end{aligned} 
\end{equation}

\noindent The ratio of specific heats $\gamma$ is 1.4. The spatial domain of the problem is $x \in \left[0, 4 \right]$. Simulations are conducted with constant time steps $\Delta t = 1.0\mathrm{e}{-6}$ on a uniform grid with 201 grid points.

\begin{figure}[!ht]
\begin{center}
\subfigure[Global density profile]{\includegraphics[width=0.48\textwidth]{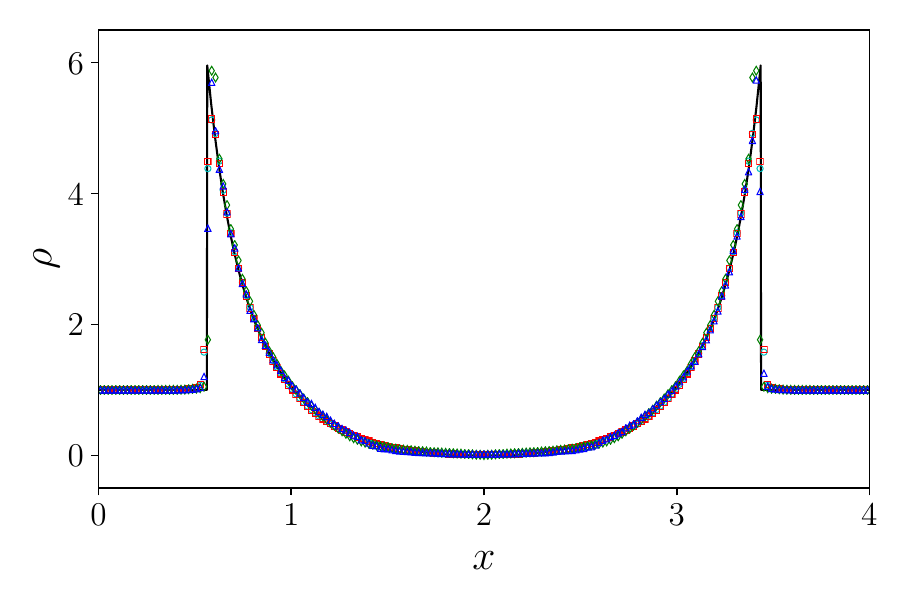} \label{fig:Sedov_density_global}}
\subfigure[Global pressure profile]{\includegraphics[width=0.48\textwidth]{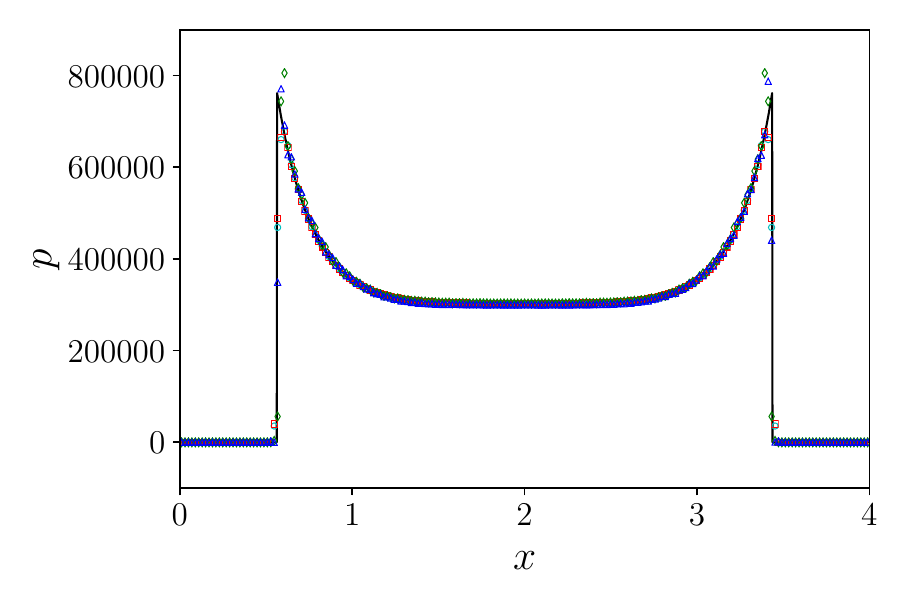} \label{fig:Sedov_pressure_global}}
\caption{1D planar Sedov blast wave problem at $t = 1.0\mathrm{e}{-3}$ using different schemes. Black solid line: exact; cyan circles: WCNS5-JS; red squares: WCNS5-Z; green diamonds: WCNS6-LD; blue triangles: WCHR6.}
\label{fig:Sedov}
\end{center}
\end{figure}

Figures~\ref{fig:Sedov_density_global} and \ref{fig:Sedov_pressure_global} show the density and pressure profiles respectively at $t = 1.0\mathrm{e}{-3}$ obtained using various schemes with the positivity limiter. It can be seen that all of the schemes can capture the blast waves. However, the pressure profiles computed with WCHR6 and WCNS6-LD have small overshoots at the peaks of the blast waves while density and pressure peaks obtained with WCNS5-JS and WCNS5-Z are damped.

\subsection{Two-dimensional vortex leaving domain}

\begin{figure}[!ht]
 \centering
	\includegraphics[width=0.5\textwidth]{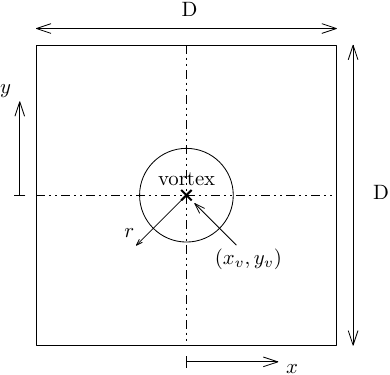}
	\caption{Schematic diagram of initial flow field and computational domain of the vortex leaving domain problem.}
    \label{fig:2D_vortex_boundary_IC_settings}
\end{figure}

This is a 2D test problem of the advection of an isothermal vortex out of a computation domain in a Mach number $M_{\infty} = 0.283$ uniform flow following case C in \citet{granet2010comparison} except that inviscid conditions are used here. The initial conditions of the vortex\footnote{This is actually a swirling flow with zero net circulation in the far field.} are given by:
\begin{equation}
	\begin{pmatrix}
		\rho \\
        p \\
        \delta u \\
        \delta v \\
	\end{pmatrix}
    =
    \begin{pmatrix}
		\rho_{\infty} \exp{ \left[ -\frac{\gamma}{2} \left( \frac{\Gamma_v}{c R_v} \right)^{2} \exp{ \left( - \left( \frac{r}{R_v} \right)^2 \right) } \right] } \\
        p_{\infty} \exp{ \left[ -\frac{\gamma}{2} \left( \frac{\Gamma_v}{c R_v} \right)^{2} \exp{ \left( - \left( \frac{r}{R_v} \right)^2 \right) } \right] } \\
        - \frac{\Gamma_v}{R_v^2} \exp{ \left[ - \frac{1}{2} \left( \frac{r}{R_v} \right)^2 \right] } (y - y_v) \\
          \frac{\Gamma_v}{R_v^2} \exp{ \left[ - \frac{1}{2} \left( \frac{r}{R_v} \right)^2 \right] } (x - x_v)
	\end{pmatrix},
\end{equation}

\noindent where $\Gamma_v = 0.024$, $R_v = 0.1$. The background flow has $\rho_{\infty} = 1$, $p_{\infty} = 1/\gamma$, $u_\infty = M_\infty c_\infty$, $v_\infty = 0$, and $c_\infty = 1$. $\delta u$ and $\delta v$ are the deviations of the $u$ and $v$ velocities from $u_\infty$ and $v_\infty$ respectively. The ratio of specific heats $\gamma = 1.4$ is used. The problem domain is chosen to be $\left[-D/2, D/2\right] \times \left[-D/2, D/2\right)$, where $D=1$ and the problem is periodic in the $y$ direction. The vortex is located at $(x_v,y_v) = (0,0)$ initially. Figure~\ref{fig:2D_vortex_boundary_IC_settings} shows the initial configuration and computation domain.

Similar to the 1D entropy wave leaving domain problem, the boundary schemes with two different ghost cell filling methods: (1) constant extrapolation of primitive variables from interior solutions and (2) sub-sonic inflow and outflow non-reflective boundary conditions at the left and right boundaries following \citet{motheau2017navier} are tested in this problem. When the non-reflective methods are used, $\sigma=0.005$, $l_x=R_v$, $\beta=M_{\infty}$, and $p_t=p_{\infty}$ are used for the non-reflective subsonic outflow method and $\eta_2=\eta_3=\eta_4=0.005$, $l_x=R_v$, $\beta=M_{\infty}$, $u_t=u_{\infty}$, and $\left( p/\rho \right)_t=RT_t=p_{\infty}/\rho_{\infty}$ are set for the non-reflective subsonic inflow method. All simulations in this section are run with $\textnormal{CFL} = 0.5$ and a grid with $64 \times 64$ points is used.

\begin{figure}[!ht]
\begin{center}
\subfigure[$t^{*}=0$]{\includegraphics[trim={0 0 3cm 0},clip,height=0.22\textwidth]{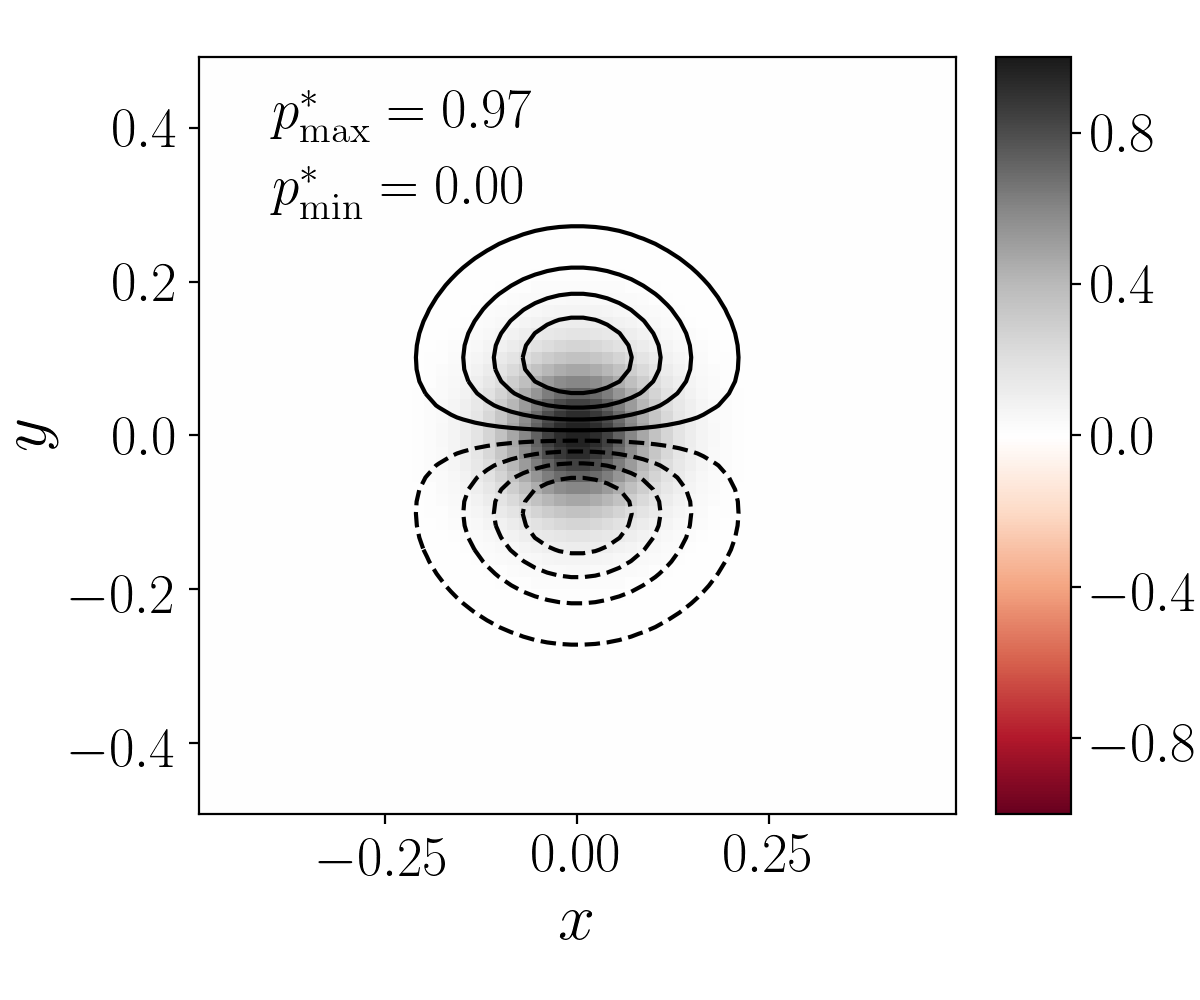}
\label{fig:vortex_boundary_extrapolation_t1}}
\subfigure[$t^{*}=0.6$]{\includegraphics[trim={0 0 3cm 0},clip,height=0.22\textwidth]{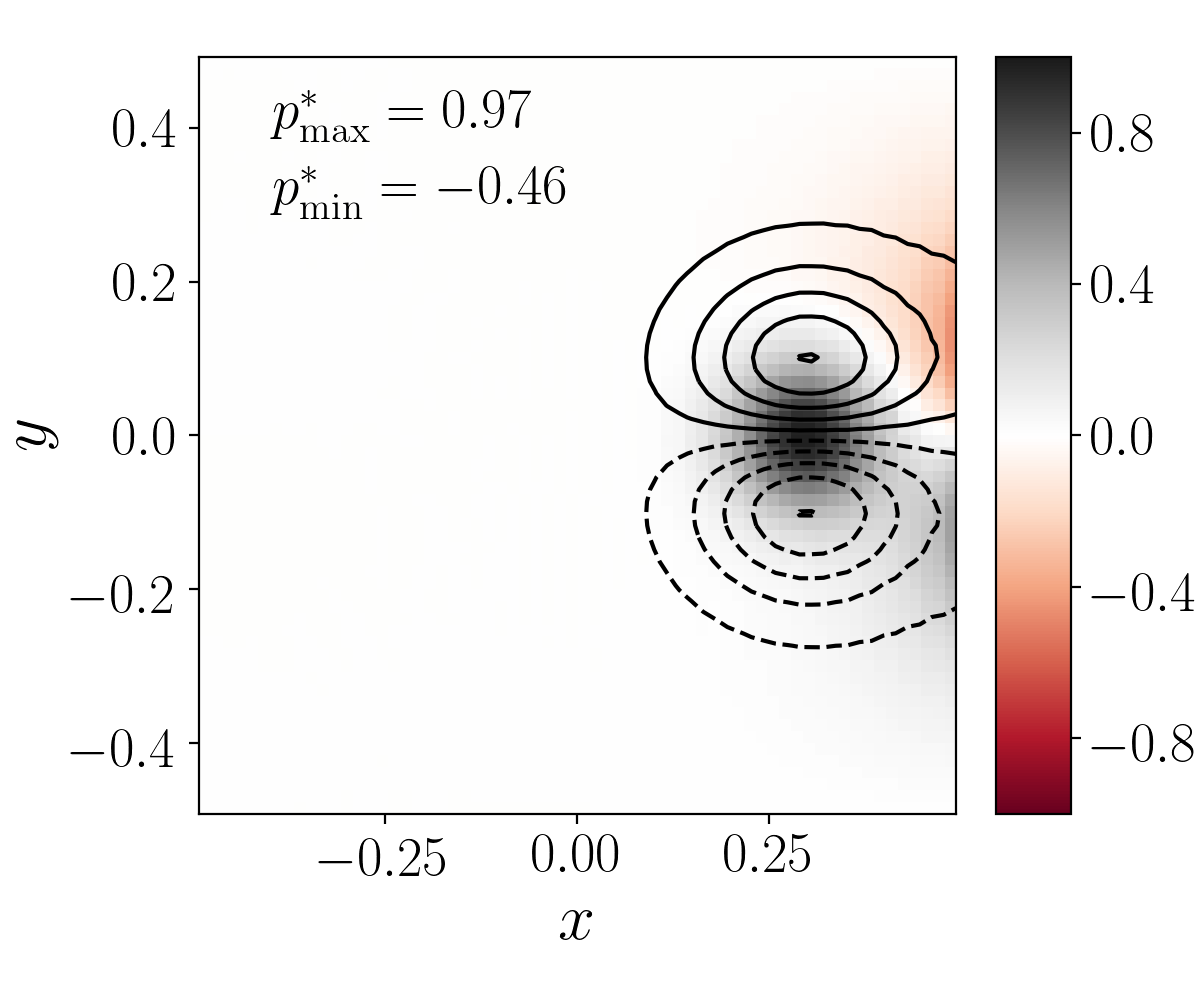}
\label{fig:vortex_boundary_extrapolation_t2}}
\subfigure[$t^{*}=1.2$]{\includegraphics[trim={0 0 3cm 0},clip,height=0.22\textwidth]{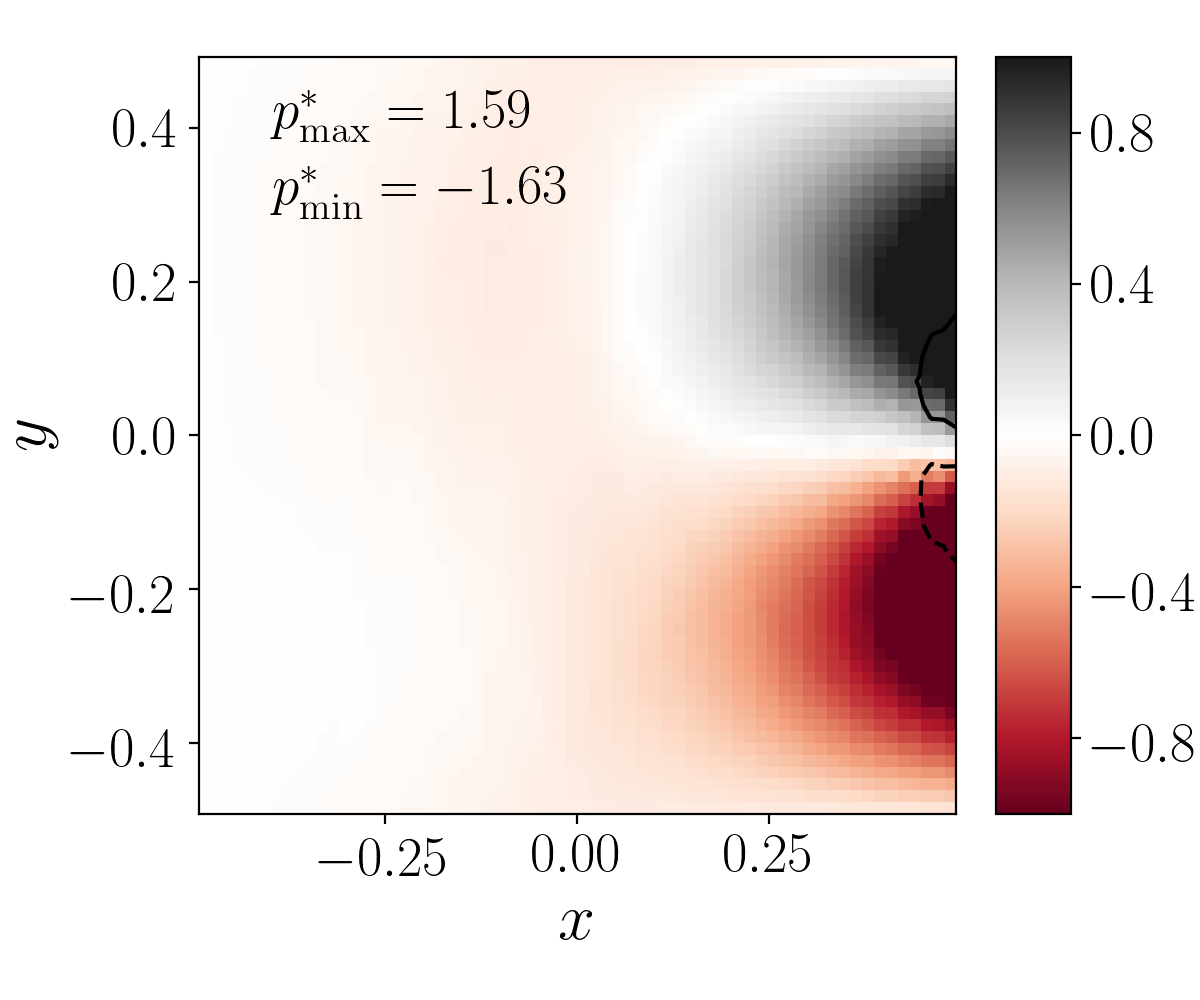}
\label{fig:vortex_boundary_extrapolation_t3}}
\subfigure[$t^{*}=1.6$]{\includegraphics[height=0.22\textwidth]{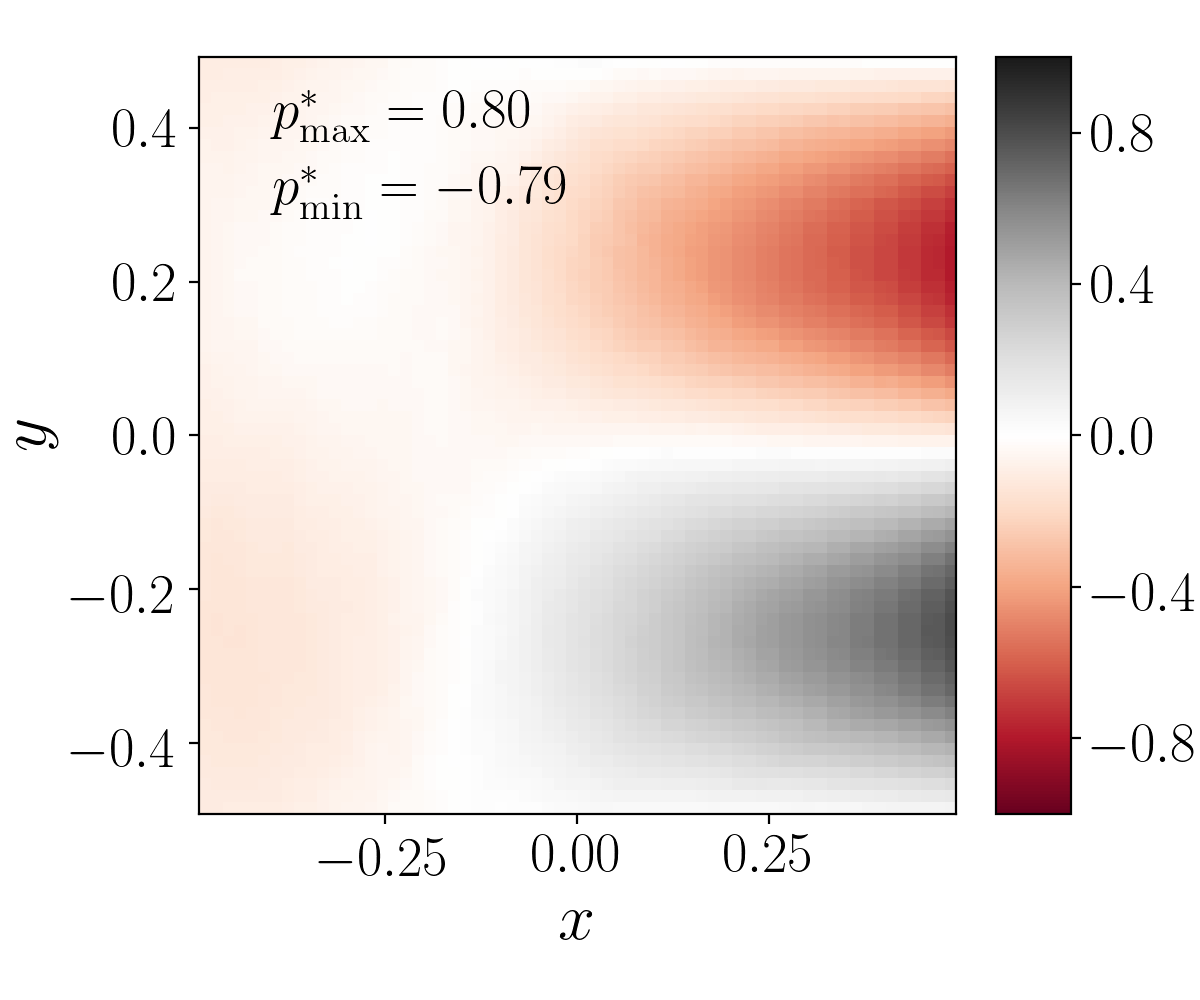}
\label{fig:vortex_boundary_extrapolation_t4}}
\end{center}
\begin{center}
\subfigure[$t^{*}=0$]{\includegraphics[trim={0 0 3cm 0},clip,height=0.22\textwidth]{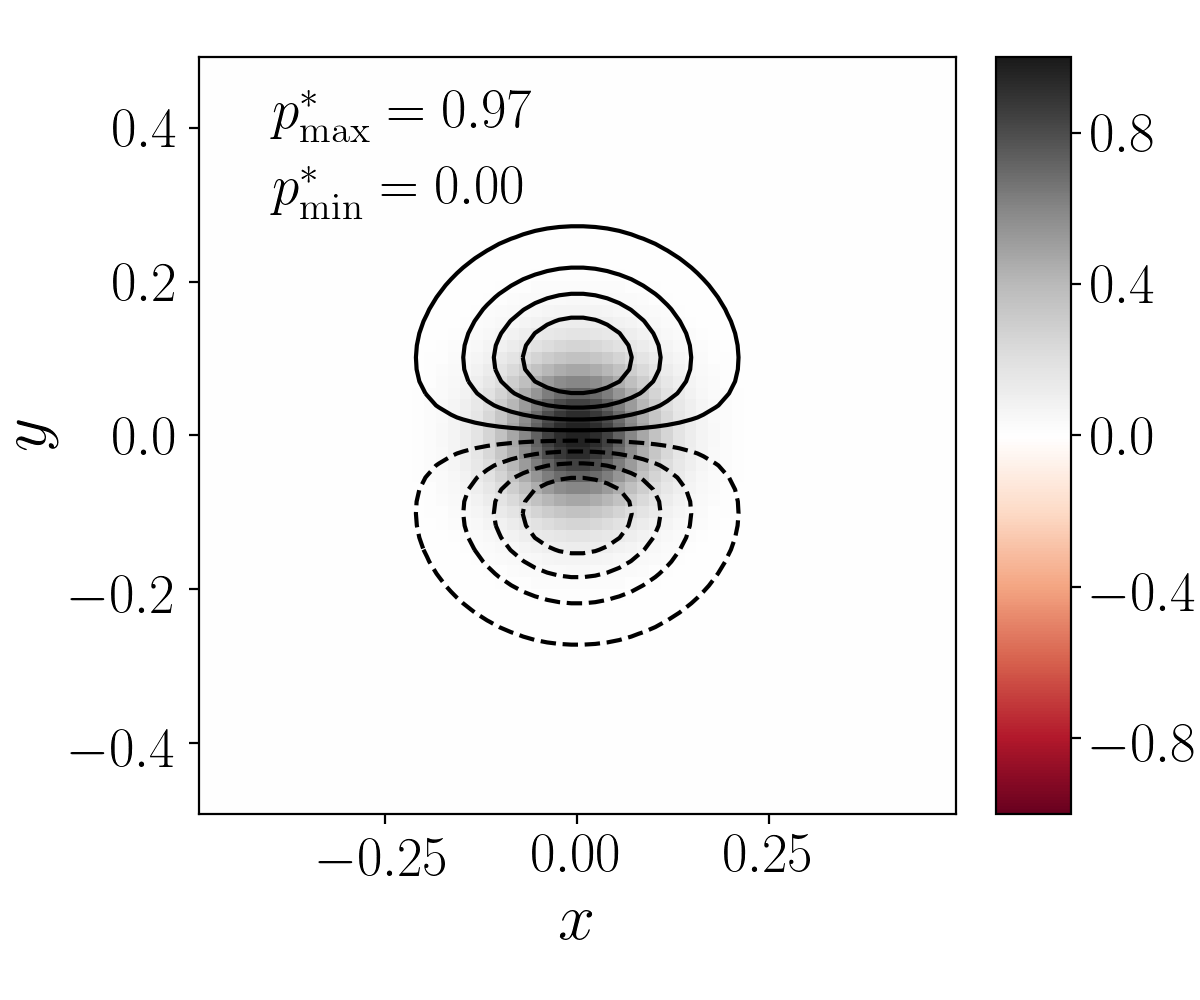}
\label{fig:vortex_boundary_NSCBC_t1}}
\subfigure[$t^{*}=0.6$]{\includegraphics[trim={0 0 3cm 0},clip,height=0.22\textwidth]{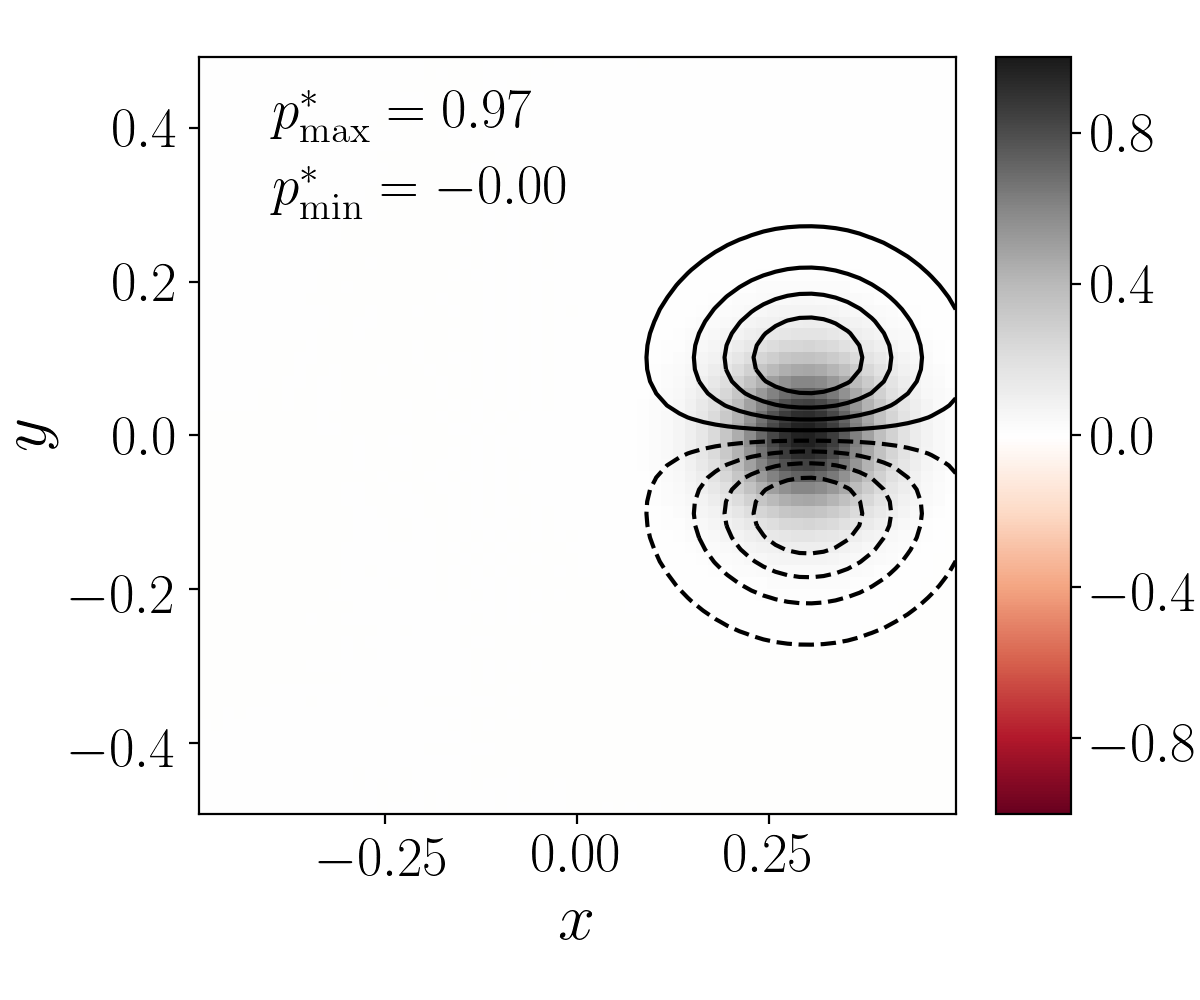}
\label{fig:vortex_boundary_NSCBC_t2}}
\subfigure[$t^{*}=1.2$]{\includegraphics[trim={0 0 3cm 0},clip,height=0.22\textwidth]{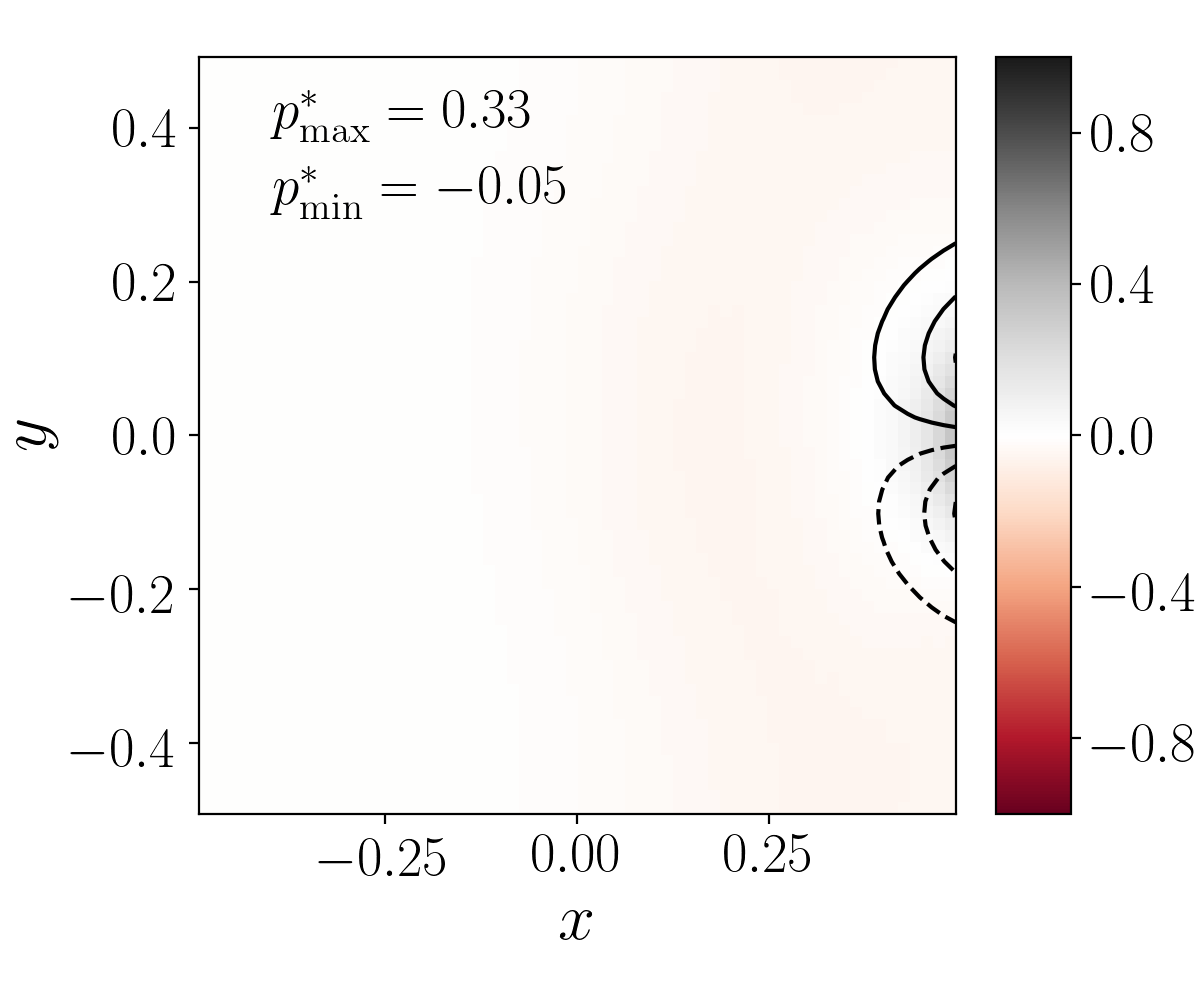}
\label{fig:vortex_boundary_NSCBC_t3}}
\subfigure[$t^{*}=1.6$]{\includegraphics[height=0.22\textwidth]{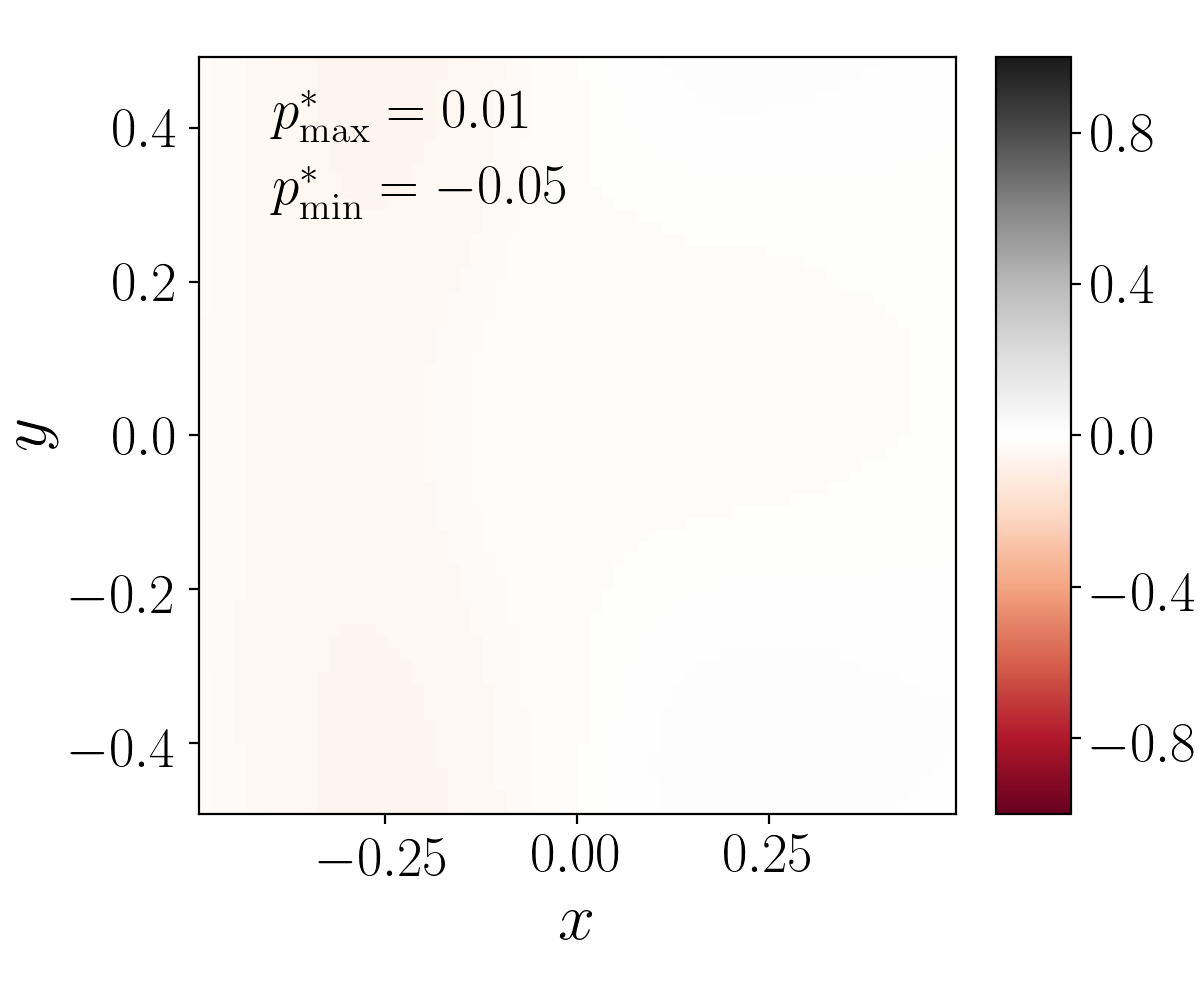}
\label{fig:vortex_boundary_NSCBC_t4}}
\end{center}
\caption{10 equally spaced streamwise velocity contours from 0.137 to 0.429 and normalized pressure fields, $p^{*}$, for the vortex leaving domain problem at different normalized times computed with the extrapolation boundary treatment (top row) and the non-reflecting boundary treatment (bottom row).}
\label{fig:vortex_boundary}
\end{figure}

Simulations computed with the boundary schemes and both ghost cell methods give stable results. Figure~\ref{fig:vortex_boundary} shows the streamwise velocity contours and the normalized pressure field at different normalized times computed with the two different boundary treatments. The pressure field and time are normalized as:
\begin{align}
    p^{*} &= \left( p_{\infty} - p \right) \frac{2 R_v^2}{\rho_{\infty} \Gamma_v^2}, \\
    t^{*} &= \frac{2 u_{\infty} t}{D}.
\end{align}

\noindent From the figures, it can be seen that the non-reflective boundary condition methods give accurate results, without any spurious waves reflected at the boundaries. However, in the solutions computed with the extrapolation method, spurious pressure waves are introduced at the right outflow boundary and the vortex is highly distorted as it crosses the domain boundary. These findings are similar to those observed in \citet{motheau2017navier}.

\subsection{Two-dimensional shock-vortex interaction}
\begin{figure}[!ht]
 \centering
	\includegraphics[width=0.5\textwidth]{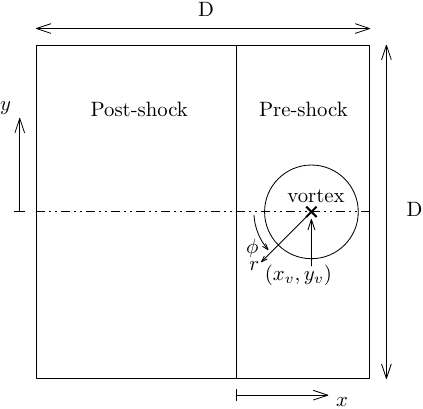}
	\caption{Schematic diagram of initial flow field and computational domain of the shock-vortex interaction problem.}
    \label{fig:2D_SVI_IC_settings}
\end{figure}

This 2D shock-vortex interaction problem was studied previously in several papers~\cite{inoue1999sound, zhang2005multistage, chatterjee2008multiple}. The inviscid version of this problem is studied here which consists of a stationary Mach $1.2$ shock and a strong isentropic vortex\footnote{Like in the previous problem, this is also actually a swirling flow with zero net circulation in the far field.} characterized by the vortex Mach number $M_v$ initially in the pre-shock region. The initial configuration and computation domain are shown in figure~\ref{fig:2D_SVI_IC_settings}. The shock is at $x = 0$ and the vortex is located upstream of the shock at $(x_v,y_v) = (D/10,0)$ initially. The initial conditions of the vortex are given by:
\begin{equation}
	\begin{pmatrix}
		\rho \\
        p \\
        \delta u \\
        \delta v \\
	\end{pmatrix}
    =
    \begin{pmatrix}
		\rho_{\infty} \left[ 1 - \frac{1}{2} \left( \gamma - 1 \right) M_v^2 \exp{ \left( 1 - \left( \frac{r}{R_v} \right)^2 \right)} \right]^{\frac{1}{\gamma-1}} \\
        p_{\infty} \left[ 1 - \frac{1}{2} \left( \gamma - 1 \right) M_v^2 \exp{ \left( 1 - \left( \frac{r}{R_v} \right)^2 \right)} \right]^{\frac{\gamma}{\gamma-1}} \\
        - \frac{M_v c_{\infty}}{R_v} \exp{\left[ \frac{1}{2} \left( 1 - \left( \frac{r}{R_v} \right)^2 \right) \right]}  (y - y_v) \\
          \frac{M_v c_{\infty}}{R_v} \exp{\left[ \frac{1}{2} \left( 1 - \left( \frac{r}{R_v} \right) ^2 \right) \right]} (x - x_v)
	\end{pmatrix}.
\end{equation}

\noindent where $\rho_{\infty}=1.0$, $p_{\infty}=1/\gamma$, $u_\infty = M_\infty c_\infty$, $v_\infty = 0$, $c_\infty = 1$, $M_v = 1.0$, and vortex radius $R_v = 1.0$ are chosen in this paper. $\delta u$ and $\delta v$ are the deviations of the $u$ and $v$ velocities from $u_\infty$ and $v_\infty$ respectively. The ratio of specific heats $\gamma = 1.4$ is used. The problem domain is chosen to be $[-3D/4, D/4] \times [-D/2, D/2)$, where $D = 40$ and the problem is periodic in the $y$ direction. The shock is initialized at $x = 0$.  Dirichlet post-shock and pre-shock boundary conditions are used to fill ghost cells for the boundary schemes at the left and right boundaries. A 2D grid with $512 \times 512$ points is used for all the schemes. All cases in this section are run with $\textnormal{CFL} = 0.5$.

\begin{figure}[!ht]
\begin{center}
\subfigure[WCNS5-JS]{
  \includegraphics[width=0.47\textwidth]{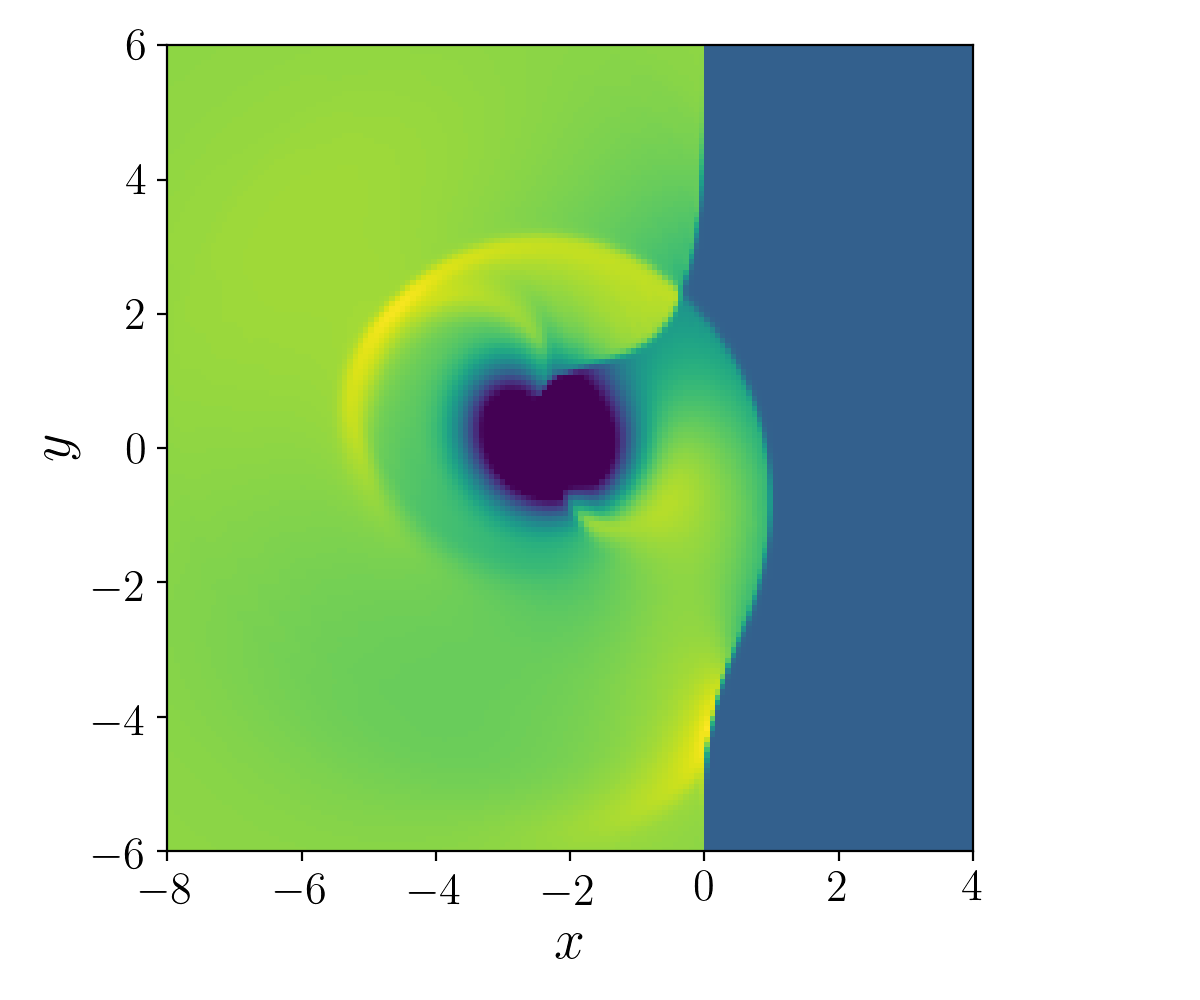}
  \label{fig:shock-vortex_N0512_p_contours_JS}
}
\subfigure[WCNS5-Z]{
  \includegraphics[width=0.47\textwidth]{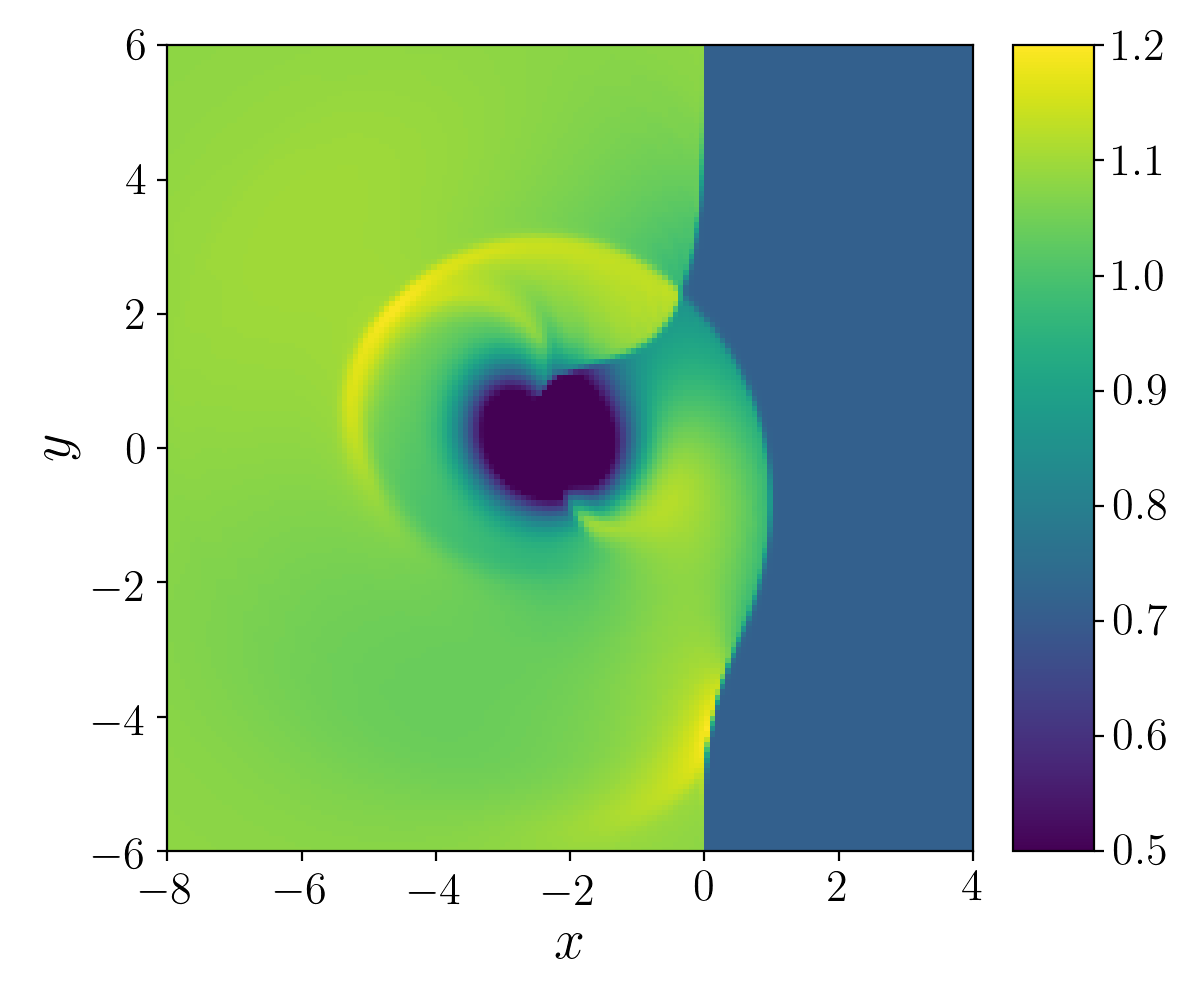}
  \label{fig:shock-vortex_N0512_p_contours_Z}
} \\
\subfigure[WCNS6-LD]{
  \includegraphics[width=0.47\textwidth]{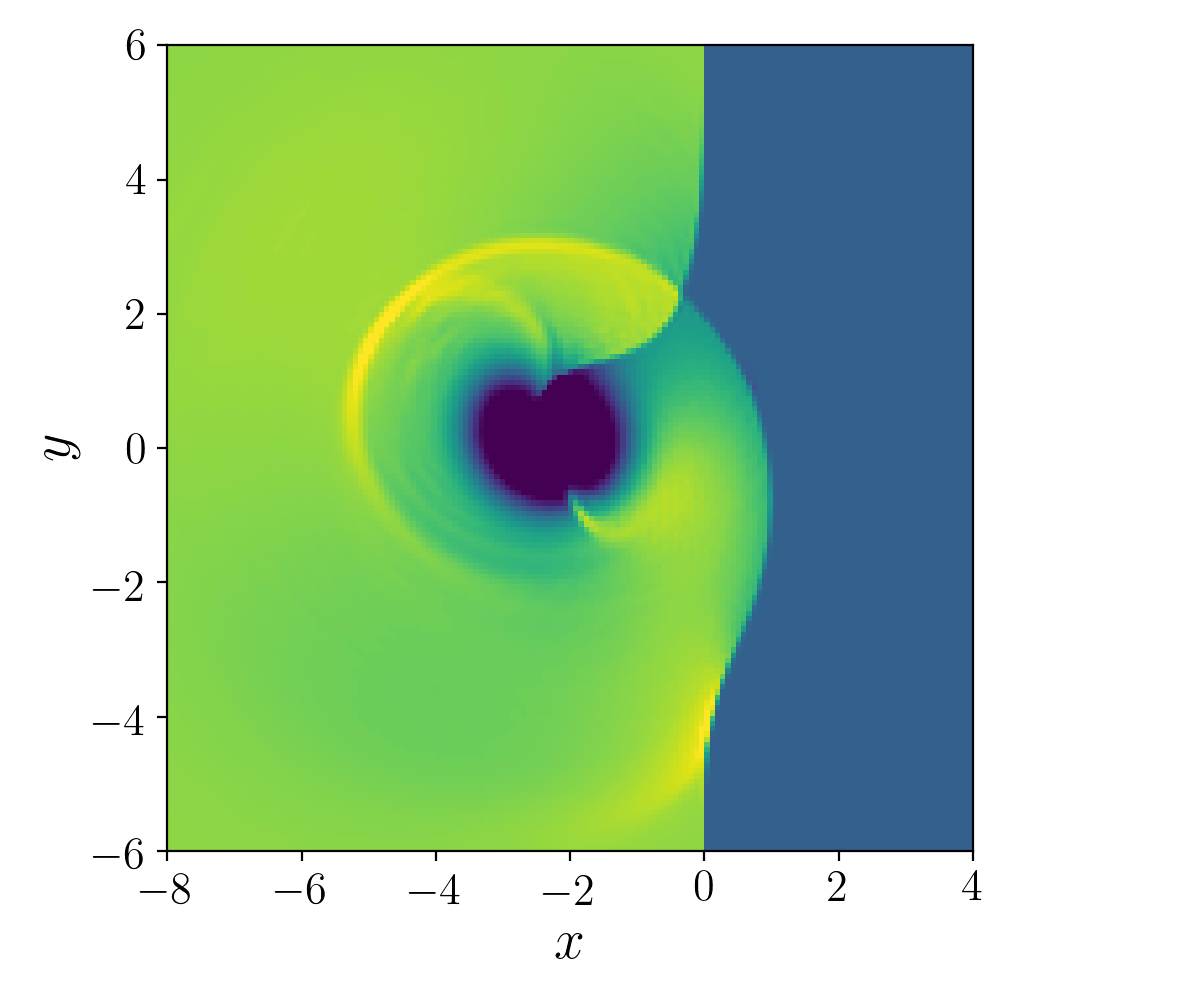}
  \label{fig:shock-vortex_N0512_p_contours_LD}
}
\subfigure[WCHR6]{
  \includegraphics[width=0.47\textwidth]{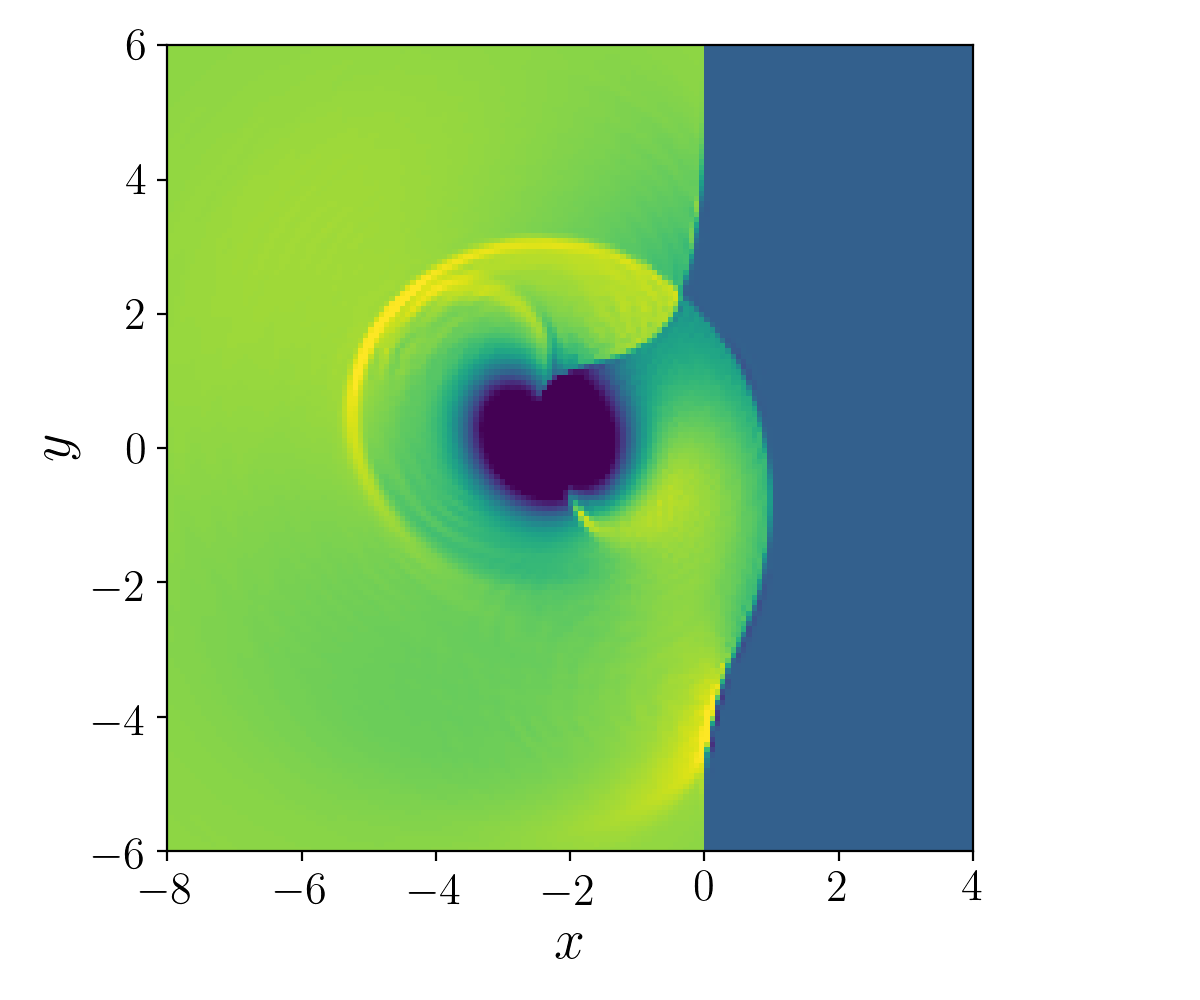}
  \label{fig:shock-vortex_N0512_p_contours_WCHR}
}
\caption{Pressure fields for the shock-vortex interaction problem on a $512\times512$ grid at $t = 6$.}
\label{fig:shock-vortex_N0512_p_contours}
\end{center}
\end{figure}

Figure~\ref{fig:shock-vortex_N0512_p_contours} shows the pressure fields at $t = 6$ for the four different schemes. At this time instant, the vortex has passed through the nominal shock line, but its interaction with the shock leads to several curved and highly deformed shock structures. WCNS5-JS and WCNS5-Z are dissipative but yield non-oscillatory solutions. WCNS6-LD and WCHR6 are less dissipative and have crispier features. However, they both have some mild oscillations at the radial shock front.

\begin{figure}[!ht]
\begin{center}
\subfigure[WCNS5-JS]{
  \includegraphics[width=0.47\textwidth]{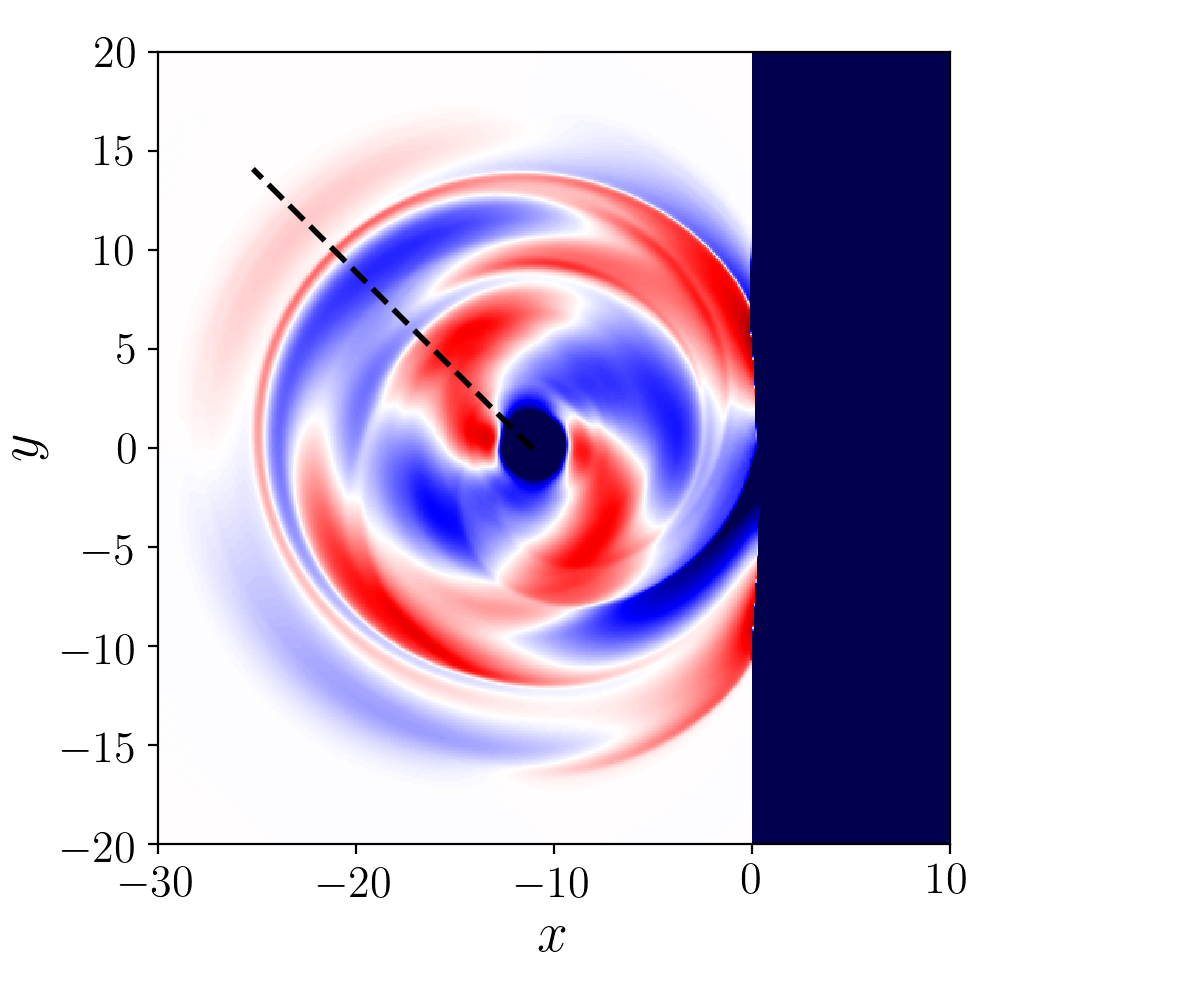}
  \label{fig:shock-vortex_N0512_psound_contours_JS}
}
\subfigure[WCNS5-Z]{
  \includegraphics[width=0.47\textwidth]{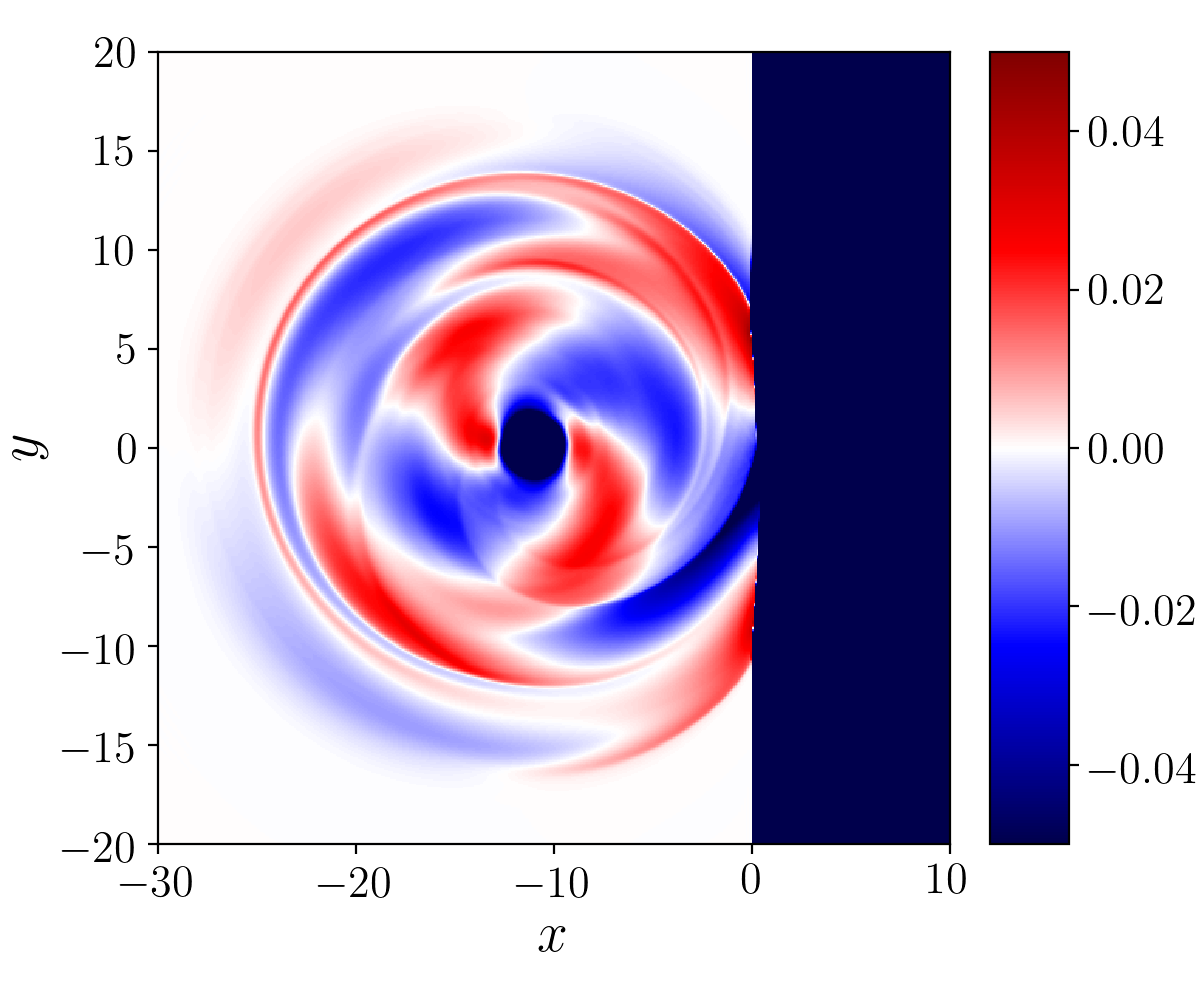}
  \label{fig:shock-vortex_N0512_psound_contours_Z}
} \\
\subfigure[WCNS6-LD]{
  \includegraphics[width=0.47\textwidth]{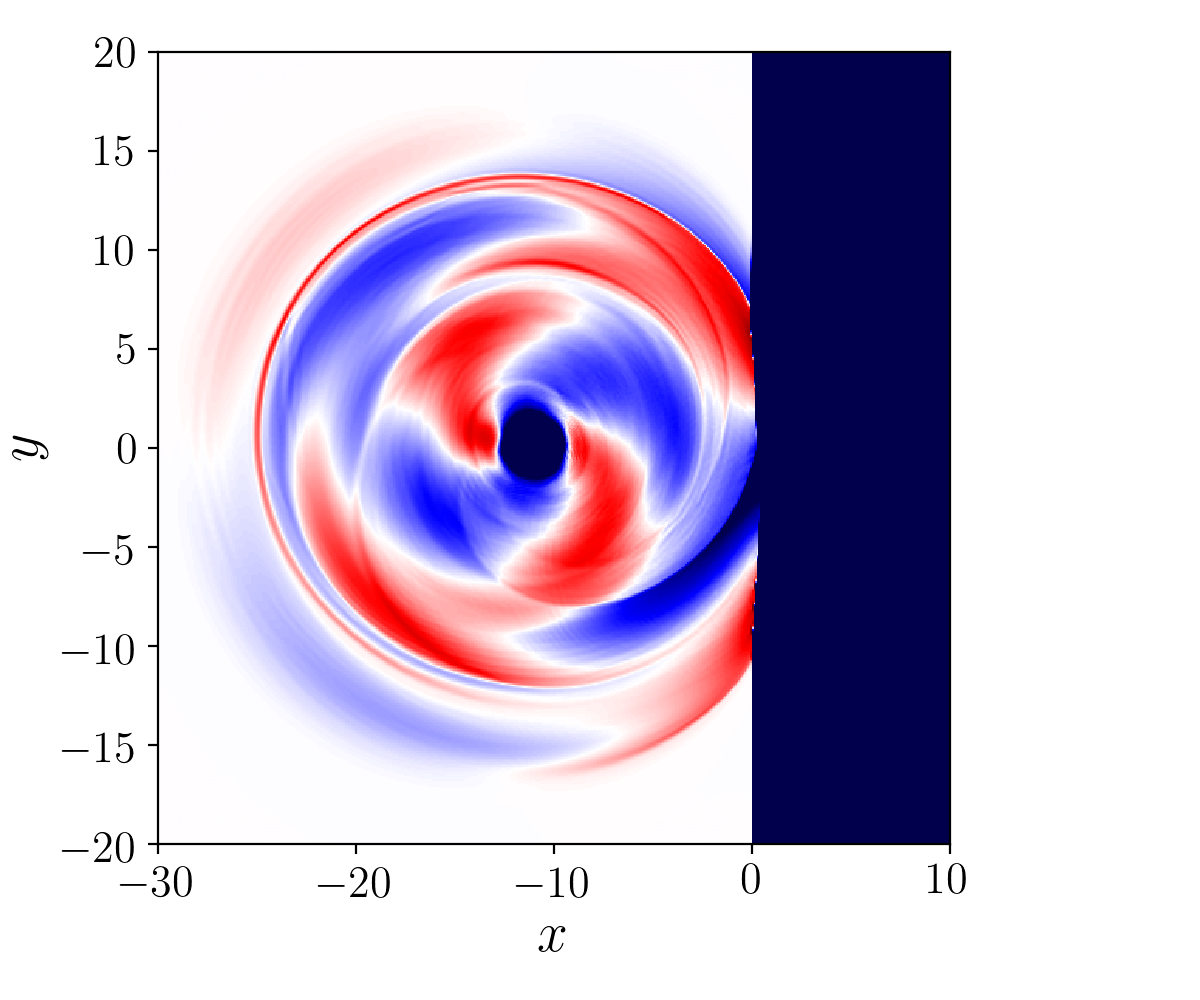}
  \label{fig:shock-vortex_N0512_psound_contours_LD}
}
\subfigure[WCHR6]{
  \includegraphics[width=0.47\textwidth]{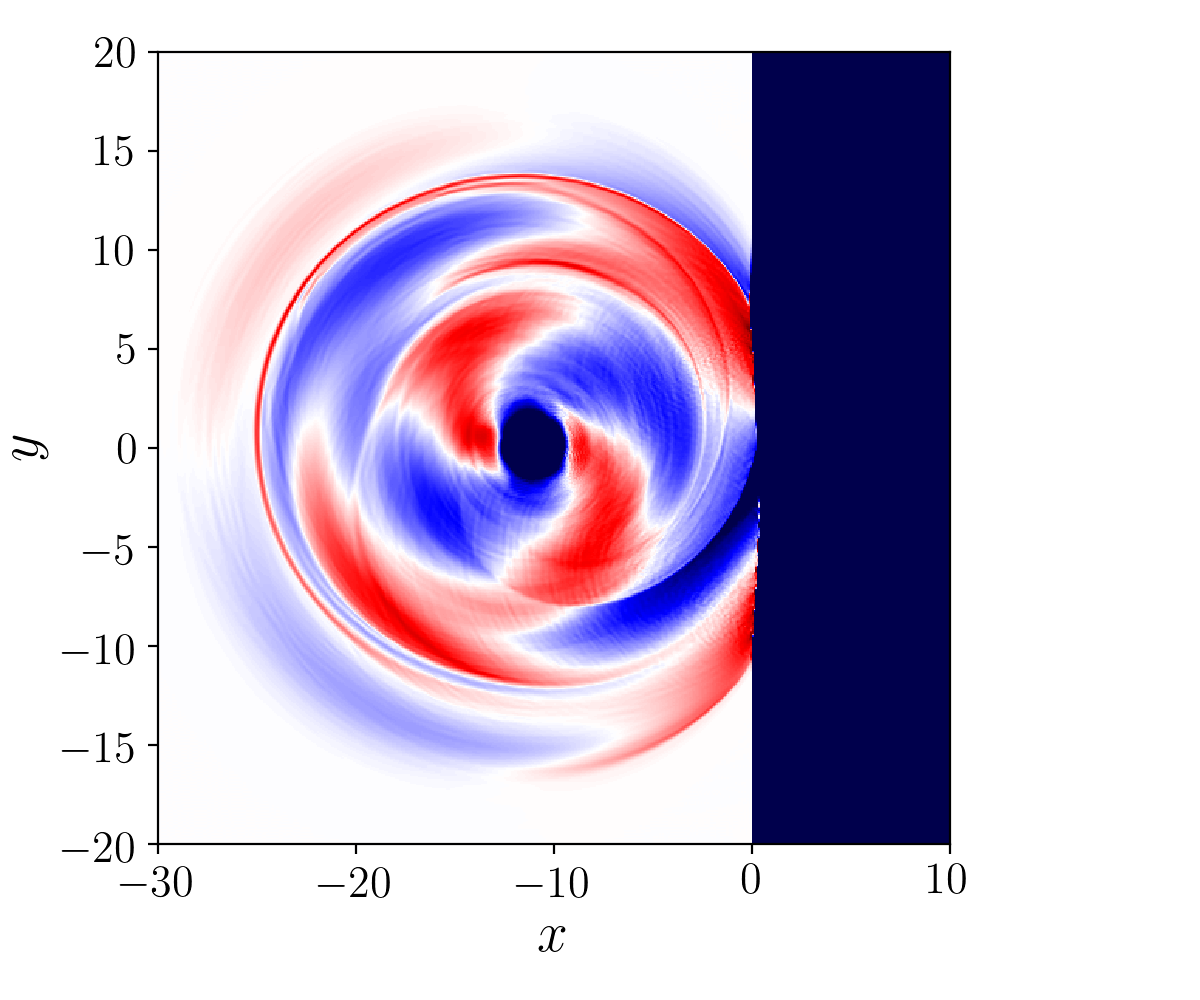}
  \label{fig:shock-vortex_N0512_psound_contours_WCHR}
}
\caption{Sound pressure fields defined by $\left( p - p_\infty \right)/\left( \rho_\infty c_{\infty}^2 \right)$ for the shock-vortex interaction problem on a $512\times512$ grid at $t = 16$. The black dashed line in (a) shows the line used to get the radial sound pressure profile plotted in figure~\ref{fig:shock-vortex_N0512_psound_line}. }
\label{fig:shock-vortex_N0512_psound_contours}
\end{center}
\end{figure}

\begin{figure}[!ht]
\begin{center}
\subfigure[Global profile]{
  \includegraphics[width=0.47\textwidth]{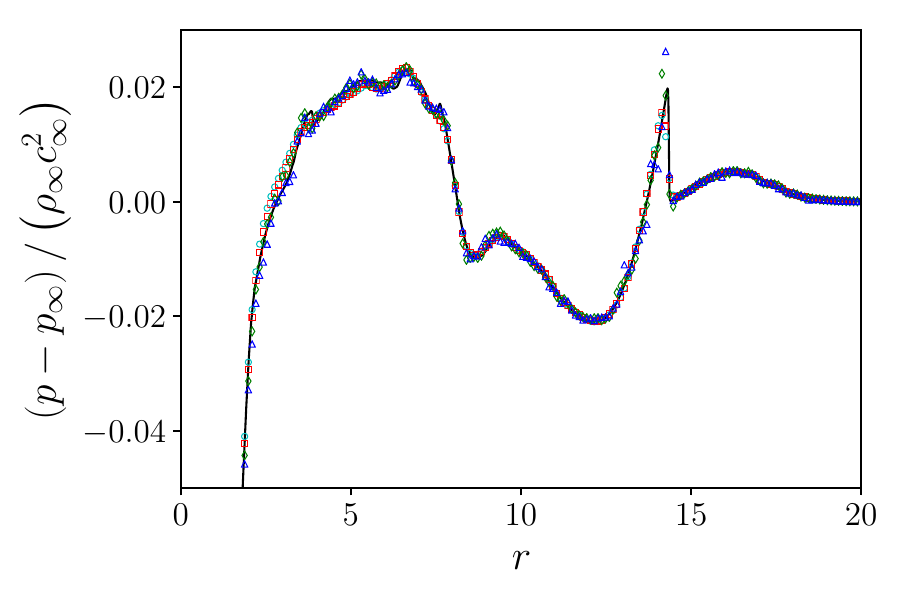}
  \label{fig:shock-vortex_N0512_psound_line_global}
}
\newline
\subfigure[Local profile 1]{
  \includegraphics[width=0.47\textwidth]{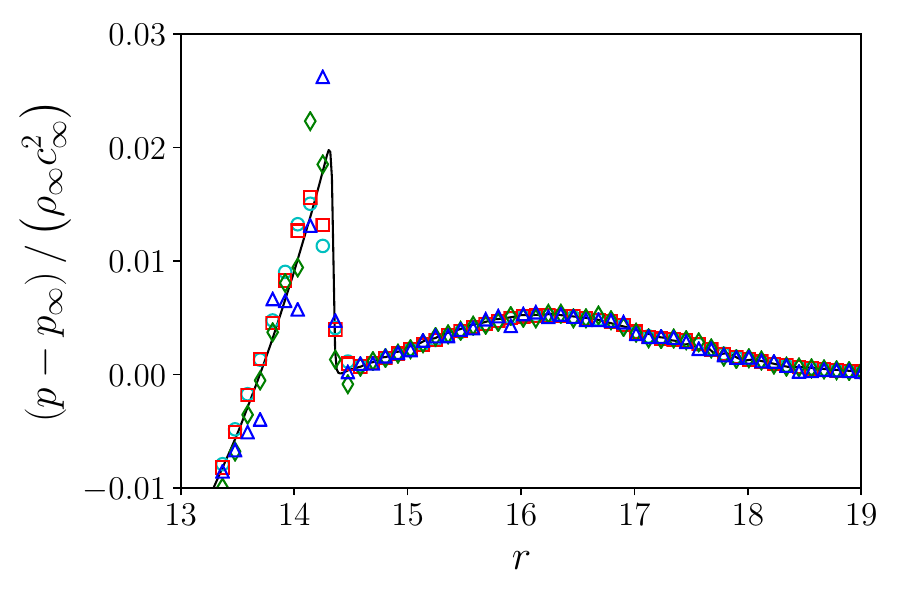}
  \label{fig:shock-vortex_N0512_psound_line_zoom1}
}
\subfigure[Local profile 2]{
  \includegraphics[width=0.47\textwidth]{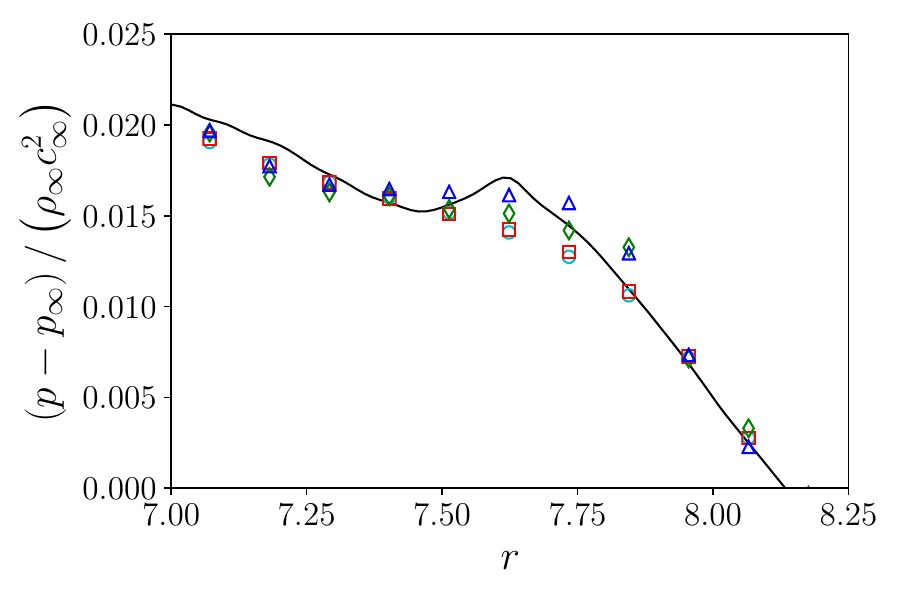}
  \label{fig:shock-vortex_N0512_psound_line_zoom2}
}
\caption{Sound pressure defined by $\left( p - p_\infty \right)/\left( \rho_\infty c_{\infty}^2 \right)$ for the shock vortex interaction problem at $t = 16$ on a radial line centered at the vortex core with $\theta = -45^\circ$ as indicated by the black dashed line in figure~\ref{fig:shock-vortex_N0512_psound_contours_JS}. Black solid line: reference; cyan circles: WCNS5-JS; red squares: WCNS5-Z; green diamonds: WCNS6-LD; blue triangles: WCHR6.}
\label{fig:shock-vortex_N0512_psound_line}
\end{center}
\end{figure}

Figure~\ref{fig:shock-vortex_N0512_psound_contours} shows the sound pressure fields defined as $\left( p - p_\infty \right)/\left( \rho_\infty c_{\infty}^2 \right)$ at $t = 16$ for the four different schemes. Here, the $(\cdot)_\infty$ quantities are all taken to be the post-shock values. The vortex, having passed through the shock gets deformed and as a result we see a quadrupole sound signature. However, since the vortex strength is very high, many weak shock waves are generated and propagate radially outward. Again, from figure~\ref{fig:shock-vortex_N0512_psound_contours}, we see that WCNS5-JS and WCNS5-Z are more dissipative and damp the fine-scale structures of the sound field. WCNS6-LD and WCHR6 are less dissipative and have more fine-scale features. Figure~\ref{fig:shock-vortex_N0512_psound_line} shows the sound pressure on a radial line from the center of the vortex with an angle of $\phi = -45^\circ$ (see figure~\ref{fig:2D_SVI_IC_settings}) for the four schemes considered here and a reference solution obtained using the WCNS5-Z on a grid with eight times the number of points in each direction. Figure~\ref{fig:shock-vortex_N0512_psound_line_global} plots a global view of the radial sound pressure profile and all schemes seem to overlap with the reference solution at this scale. Figure~\ref{fig:shock-vortex_N0512_psound_line_zoom1} shows a local view of the outgoing shock front at $r \approx 14.5$. Here we see that WCHR6 and WCNS6-LD overshoot the peak sound pressure while the WCNS5-JS and WCNS5-Z under-predict the peak sound pressure level. Figure~\ref{fig:shock-vortex_N0512_psound_line_zoom2} shows a local view of the radial sound pressure profile around $r = 7.5$. Here, the local peak of the sound pressure profile in the reference solution is not captured by any of the WCNS's while the WCHR6 scheme is able to capture the peak owing to its higher resolution property.

\subsection{Double Mach reflection}
This is a 2D problem with the domain size of $\left[ 0, 4 \right] \times \left[ 0, 1 \right]$ by \citet{woodward1984numerical}. The initial conditions are given by:
\begin{equation*}
\begin{aligned}
	\left( \rho, u, v, p \right)
    =
    \begin{cases}
    	\left(8, 8.25\cos{\left( \frac{\pi}{6} \right)}, -8.25\sin{\left( \frac{\pi}{6} \right)}, 116.5 \right), &\mbox{$x < \frac{1}{6} + \frac{y}{\sqrt{3}}$}, \\
    	\left(1.4, 0, 0, 1 \right), &\mbox{$x \geq \frac{1}{6} + \frac{y}{\sqrt{3}}$}. \\
    \end{cases}
\end{aligned}
\end{equation*}

\begin{figure}[!ht]
\begin{center}
\subfigure[WCNS5-JS]{
  \includegraphics[height=0.33\textwidth]{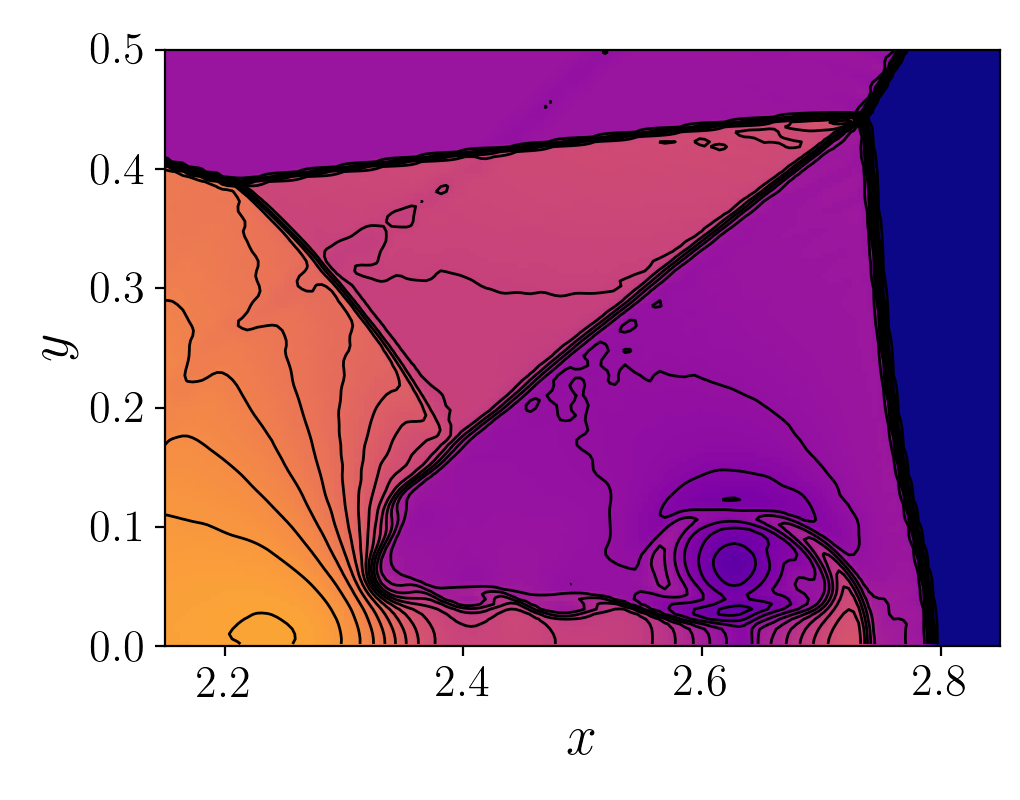}
  \label{fig:DMR_N0240_rho_JS_local}
}
\subfigure[WCNS5-Z]{
  \includegraphics[height=0.33\textwidth]{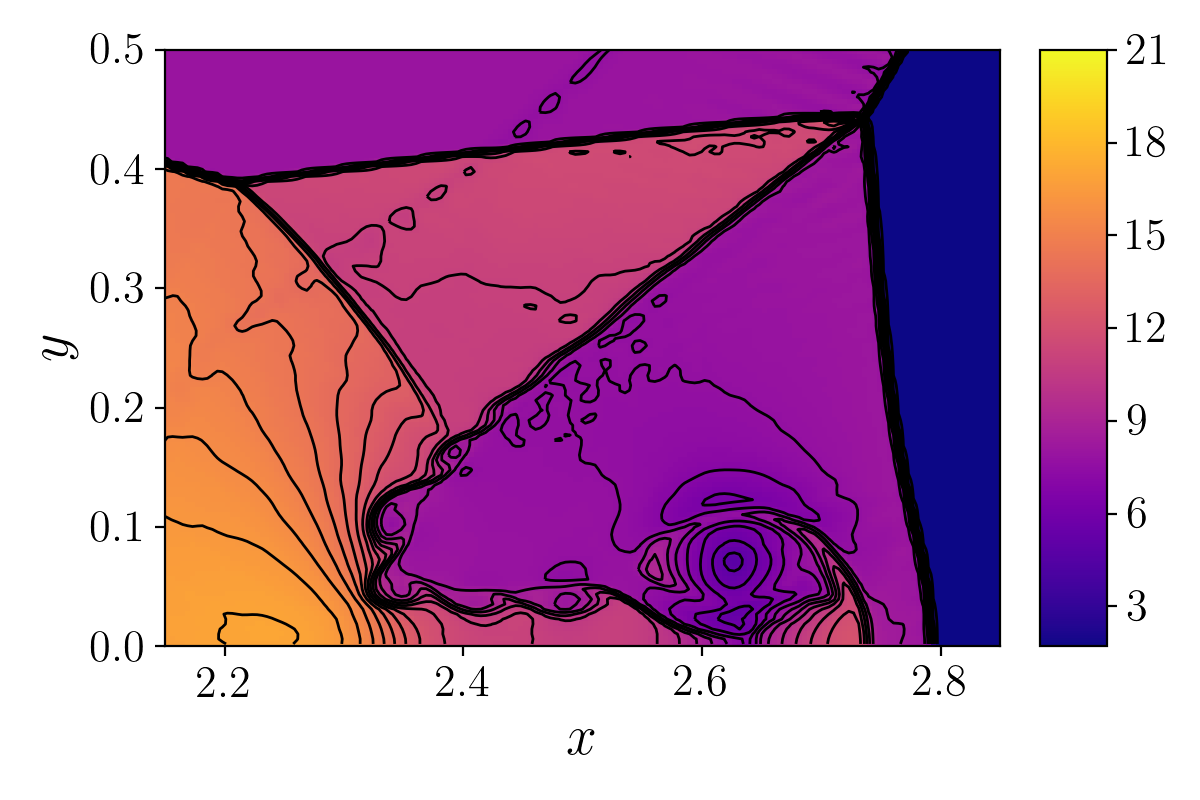}
  \label{fig:DMR_N0240_rho_Z_local}
} \\
\subfigure[WCNS6-LD]{
  \includegraphics[height=0.33\textwidth]{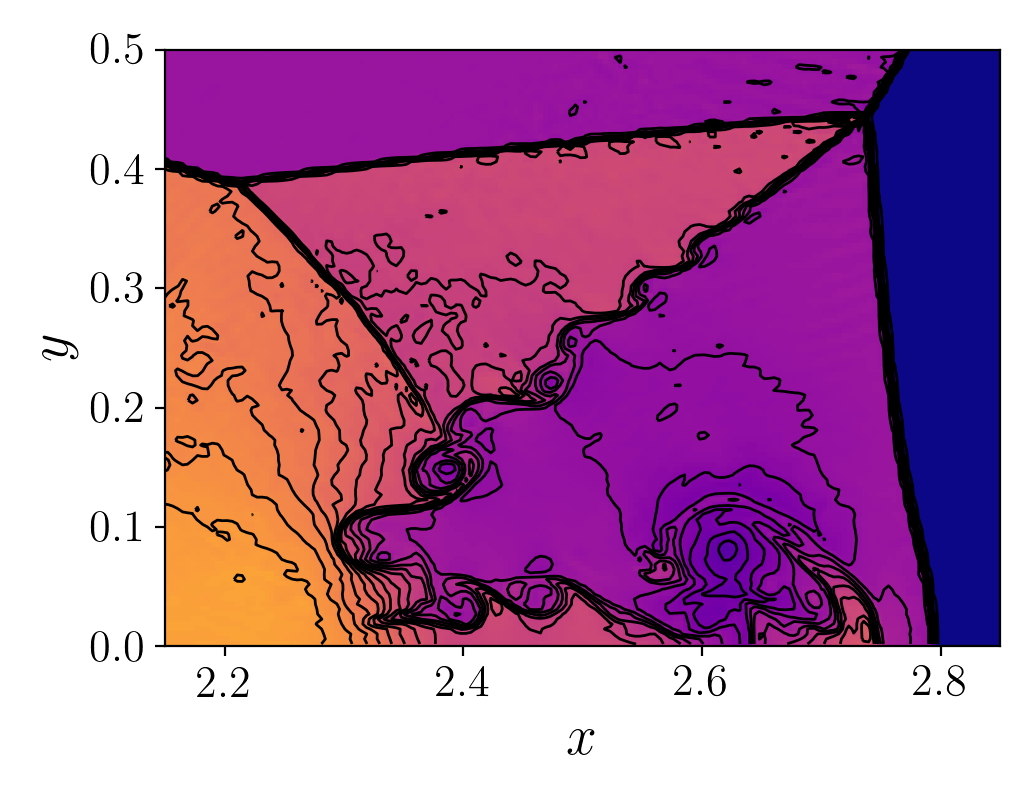}
  \label{fig:DMR_N0240_rho_LD_local}
}
\subfigure[WCHR6]{
  \includegraphics[height=0.33\textwidth]{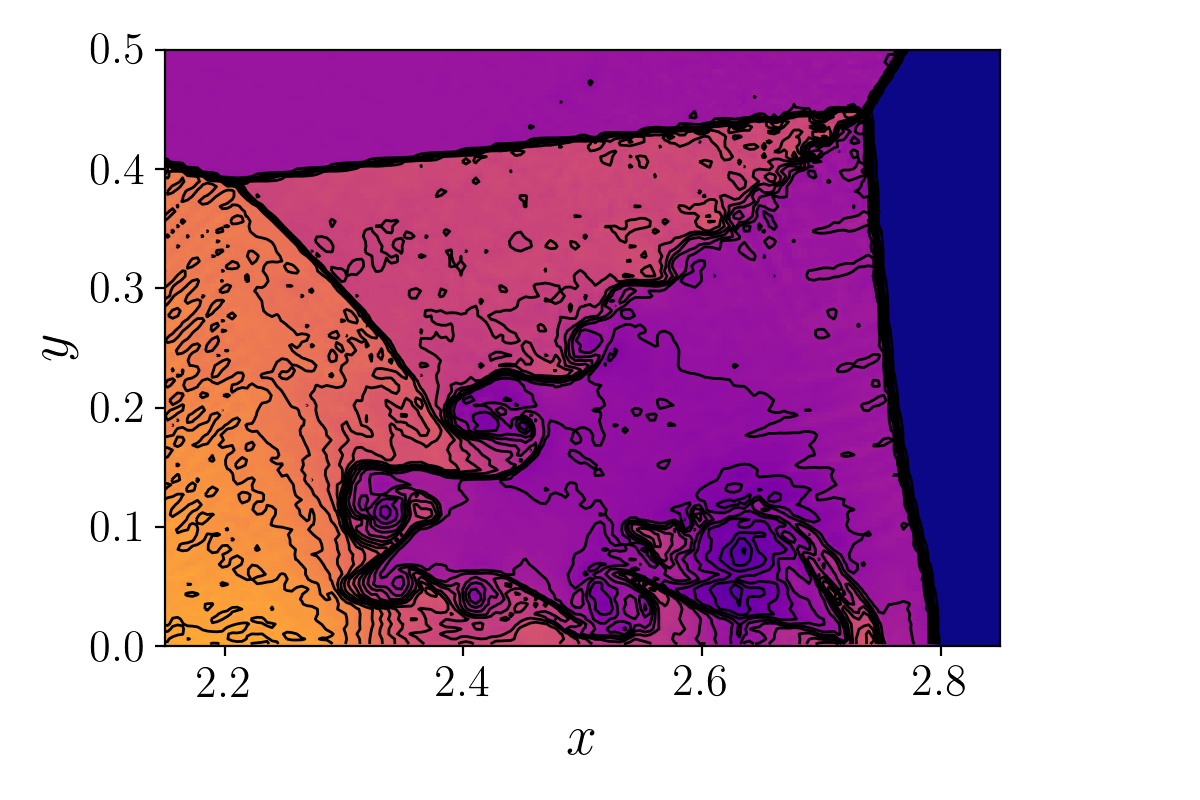}
  \label{fig:DMR_N0240_rho_WCHR_local}
}
\caption{30 equally spaced contours of density from 1.7 to 21 for the double Mach reflection problem at $t = 0.2$ using different schemes in the blown-up region around the Mach stem.}
\label{fig:DMR_N0240_rho_local}
\end{center}
\end{figure}

\noindent A Mach 10 strong shock initially makes a $60 ^{\circ}$ angle  with the horizontal wall at location $x = 1/6$ of the bottom boundary. As the shock moves and reflects on the wall, a complex shock structure with two triple points appears. The ratio of specific heats is $\gamma = 1.4$. The boundary conditions following those by \citet{woodward1984numerical} are used. At the bottom boundary, the conditions in the region $x \in \left[ 0, 1/6 \right]$ are fixed at Dirichlet boundary conditions with the post-shock flow conditions and reflecting boundary conditions are used for $x \geq 1/6$. Dirichlet boundary conditions with the post-shock flow conditions are set at the left boundary. Constant extrapolations of primitive variables are used to fill ghost cells at the right boundary to allow zero-gradient boundary conditions. Time-dependent conditions are applied on the top boundary to match the movement of the shock wave. The simulations are conducted with constant $\textnormal{CFL}=0.5$ until $t = 0.2$. All schemes can only provide stable results with the positivity limiter. The density fields for different schemes at $t = 0.2$ are shown in figure~\ref{fig:DMR_N0240_rho_local}.

At the shock triple point, a slip line is generated that is Kelvin--Helmholtz unstable. Since the inviscid Euler equations are solved, there is no physical dissipation in this test problem. The instability of the vortex sheet along the slip line is only damped by numerical dissipation. From figure~\ref{fig:DMR_N0240_rho_local}, we see that with the same mesh resolution of $960 \times 240$, both WCNS5-JS and WCNS5-Z are numerically too dissipative and completely inhibit the growth of Kelvin--Helmholtz vortices along the slip lines. On the other hand, both WCNS6-LD and WCHR6 can capture much more small-scale vortical structures along the slip lines as more localized dissipation is applied at the discontinuities. Since WCHR6 is the least dissipative, it exhibits the highest level of instability growth.

\begin{figure}[!ht]
\begin{center}
\subfigure[WCNS5-JS]{
  \includegraphics[width=0.8\textwidth]{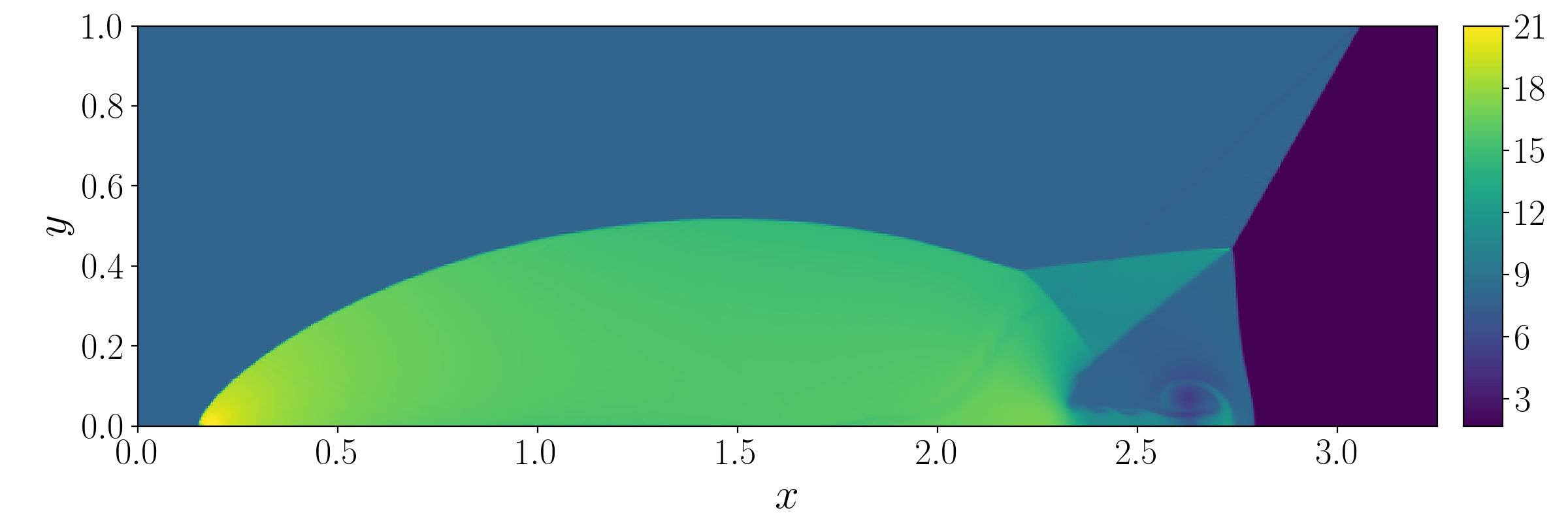}
  \label{fig:DMR_N0240_rho_JS}
}
\subfigure[WCNS5-Z]{
  \includegraphics[width=0.8\textwidth]{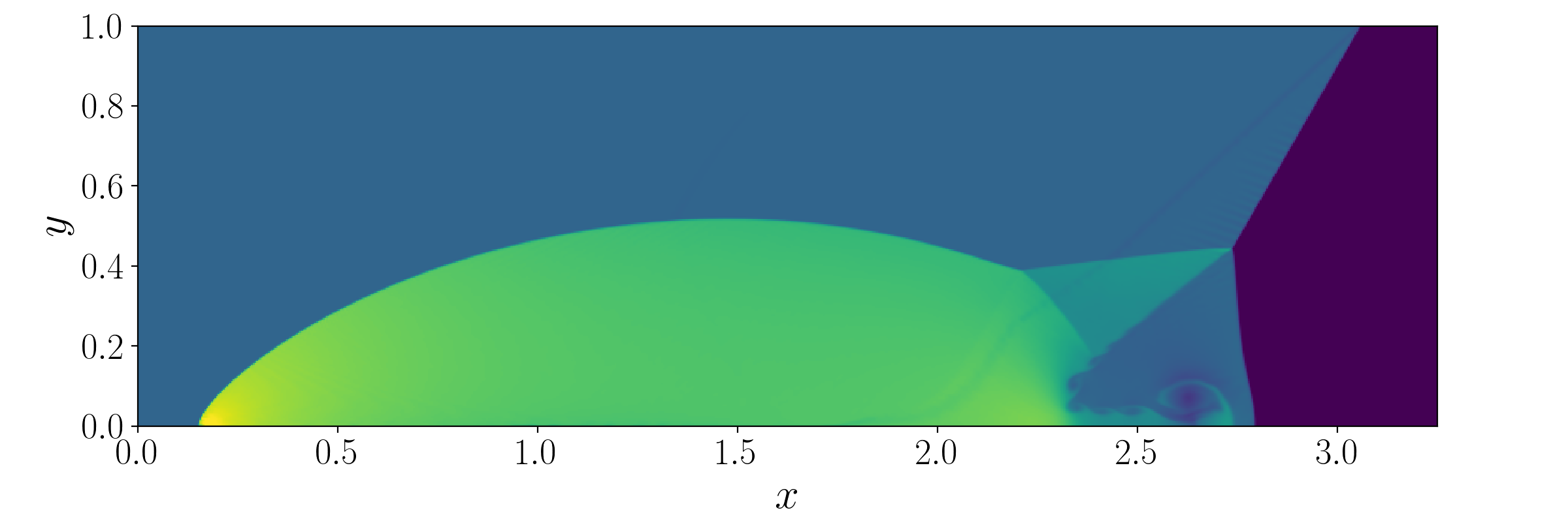}
  \label{fig:DMR_N0240_rho_Z}
} \\
\subfigure[WCNS6-LD]{
  \includegraphics[width=0.8\textwidth]{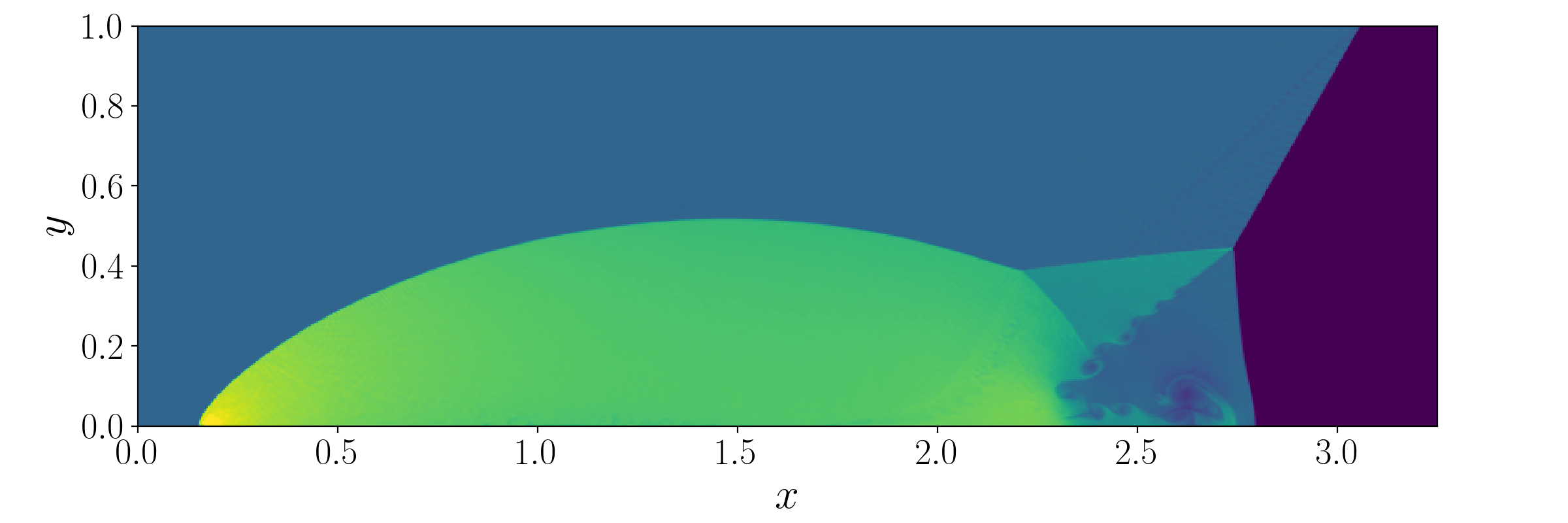}
  \label{fig:DMR_N0240_rho_LD}
}
\subfigure[WCHR6]{
  \includegraphics[width=0.8\textwidth]{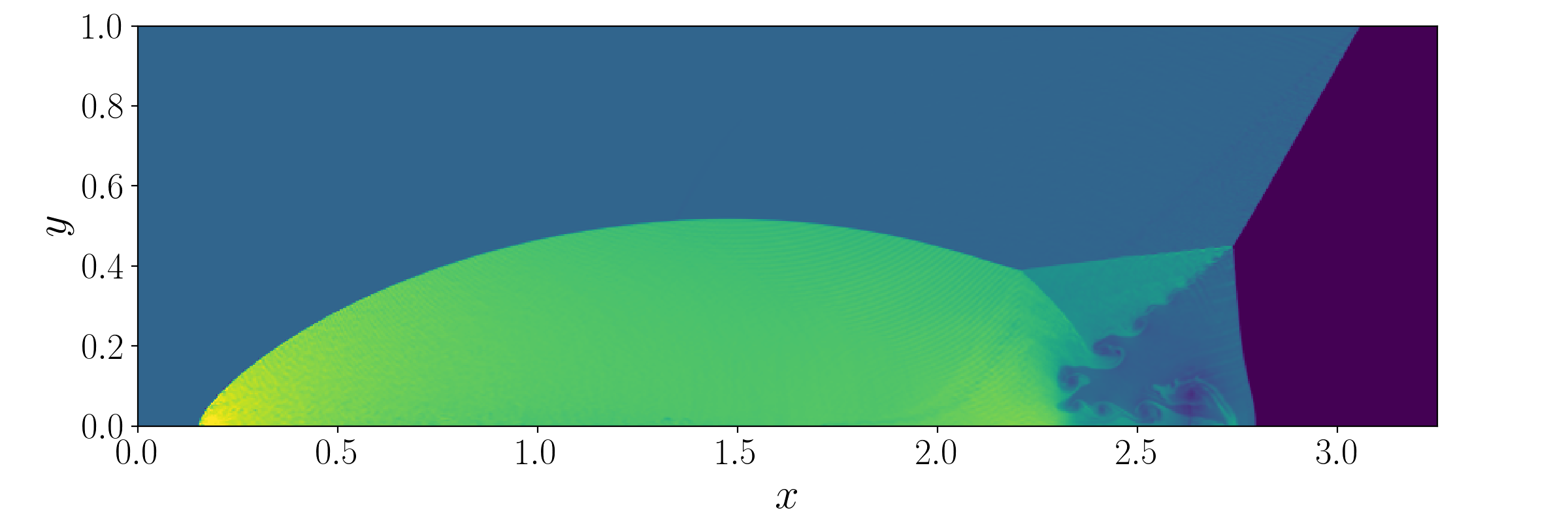}
  \label{fig:DMR_N0240_rho_WCHR}
}
\caption{Density fields for the double Mach reflection problem at $t = 0.2$ using different schemes.}
\label{fig:DMR_N0240_rho}
\end{center}
\end{figure}

\subsection{Taylor--Green vortex}
The 3D inviscid Taylor--Green vortex problem is a popular test case used to compare the numerical dissipation of different schemes and has been used widely in previous literature~\cite{johnsen2010assessment, hu2011scale}. The initial conditions of the problem are given by:
\begin{equation}
	\begin{pmatrix}
		\rho \\
        u \\
        v \\
        w \\
        p \\
	\end{pmatrix}
    =
    \begin{pmatrix}
		1 \\
        \sin{x} \cos{y} \cos{z} \\
        -\cos{x} \sin{y} \cos{z} \\
        0 \\
        100 + \frac{\left( \cos{(2z)} + 2 \right) \left( \cos{(2x)} + \cos{(2y)} \right) - 2}{16}
	\end{pmatrix}.
\end{equation}

\noindent The ratio of specific heats of the gas is $\gamma = 5/3$. The domain is periodic with size $\left[0, 2\pi \right)^3$. The problem is solved with the four schemes considered here on a $64^3$ grid. Simulations are conducted until $t = 10$ with a constant $\textnormal{CFL}=0.6$. 

As the mean pressure is chosen to be very large compared to the dynamic pressure, the flow problem is essentially incompressible. Thus, the kinetic energy of the flow is conserved in the inviscid limit and the problem can be used as a test to examine the dissipative property of different schemes. As time evolves, the initial flow gets stretched and energy is transferred from larger to finer scales.

\begin{figure}[!ht]
\begin{center}
\subfigure[Kinetic energy]{\includegraphics[width=0.48\textwidth]{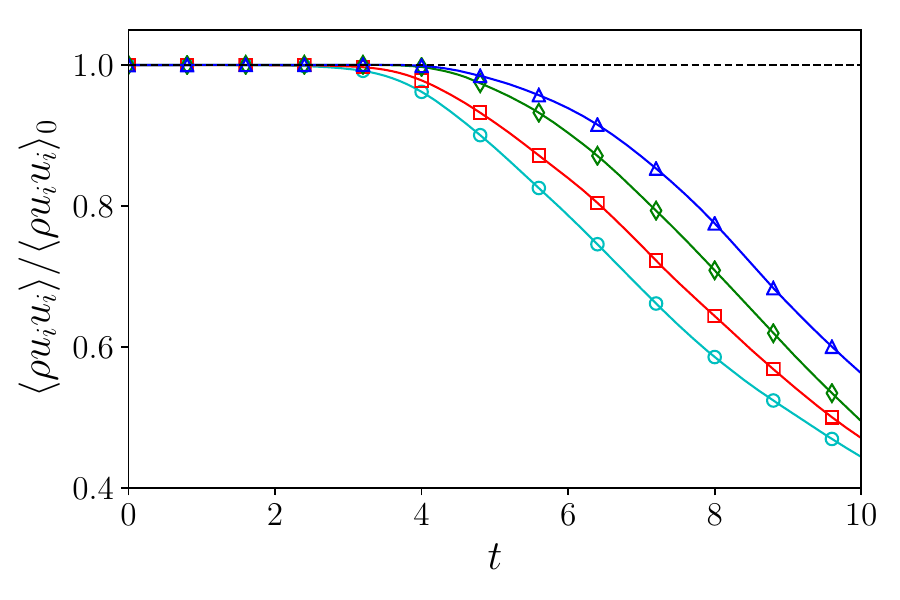} \label{fig:TGV_N0064_TKE}}
\subfigure[Enstrophy]{\includegraphics[width=0.48\textwidth]{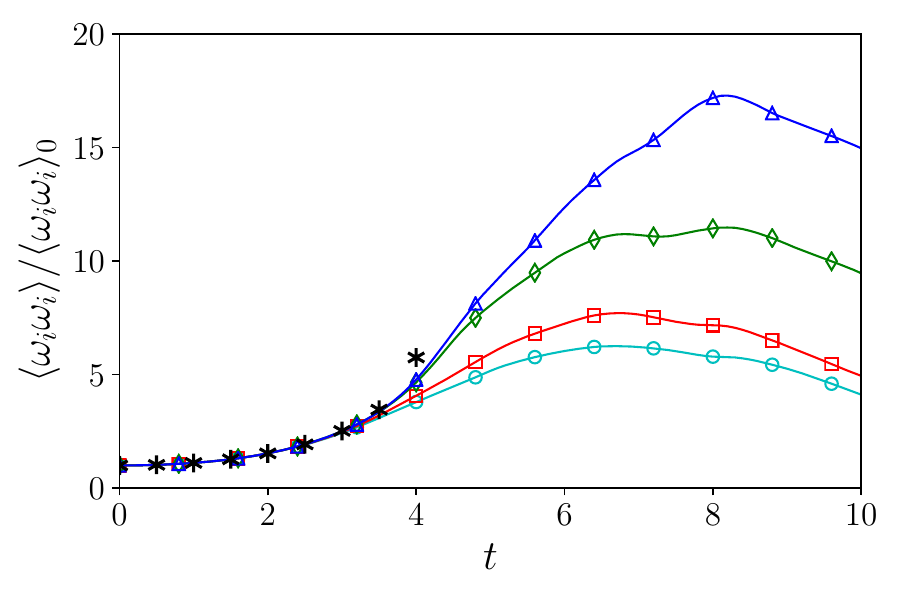} \label{fig:TGV_N0064_Enstrophy}}
\caption{Time evolution of statistical quantities for the Taylor--Green vortex problem on a $64^3$ grid. Black dashed line: exact for kinetic energy; black asterisks: semi-analytical result for enstrophy of \citet{brachet1983small}; cyan circles: WCNS5-JS; red squares: WCNS5-Z; green diamonds: WCNS6-LD; blue triangles: WCHR6.}
\label{fig:TGV_N0064}
\end{center}
\end{figure}

Figure~\ref{fig:TGV_N0064} plots the kinetic energy ($\left\langle \rho u_i u_i \right\rangle / 2$) and enstrophy ($\left\langle \omega_i \omega_i \right\rangle$) normalized by their respective initial values. The $\left\langle \cdot \right\rangle$ operator indicates averaging in space. Here, we see that WCHR6 is the least dissipative and retains the largest amount of the kinetic energy at $t = 10$. Both upwind biased schemes (WCNS5-JS and WCNS5-Z) are more dissipative than the hybrid central-upwind schemes (WCHR6 and WCNS6-LD). Similar trends can also been seen in the enstrophy plot. WCHR6 captures significantly larger amount of enstrophy compared to WCNS6-LD while the upwind biased schemes are very dissipative and deviate from the semi-analytical solution of \citet{brachet1983small} much earlier than the hybrid central-upwind schemes.

\begin{figure}[!ht]
\begin{center}
\subfigure[Velocity energy spectrum]{\includegraphics[width=0.48\textwidth]{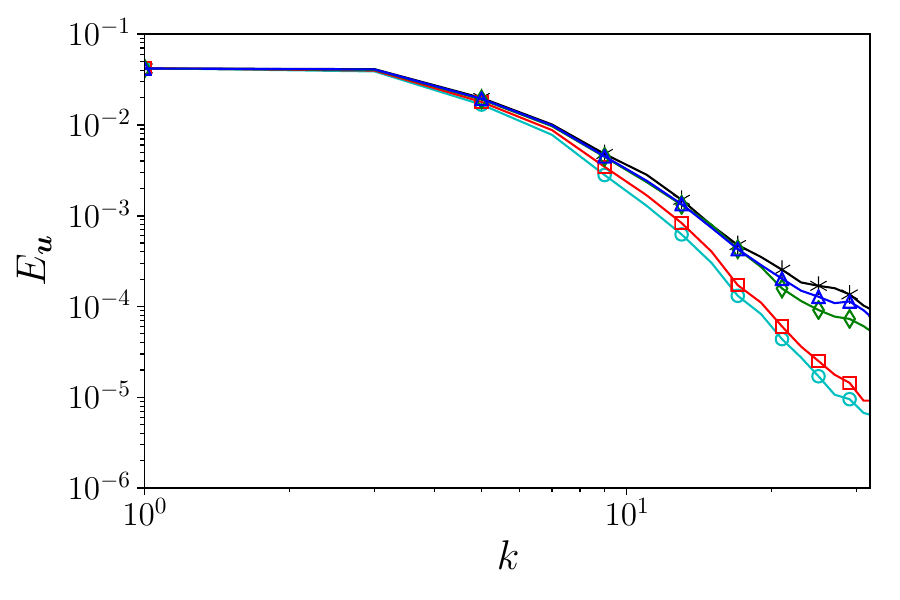} \label{fig:TGV_N0064_vel_spectrum_t_5}}
\subfigure[Vorticity energy spectrum]{\includegraphics[width=0.48\textwidth]{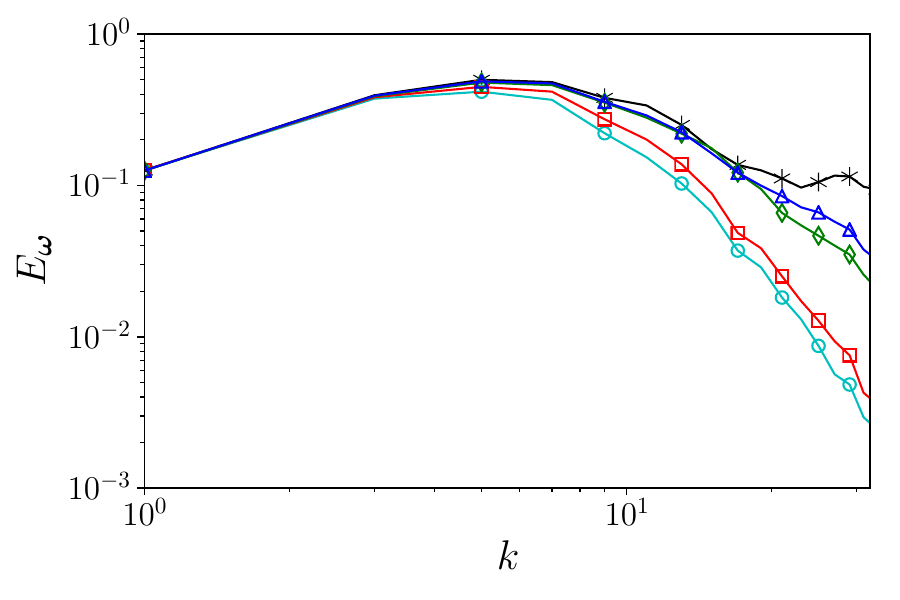} \label{fig:TGV_N0064_vor_spectrum_t_5}}
\caption{Spectra at $t = 5$ for the Taylor--Green vortex problem on a $64^3$ grid. Black asterisks: converged spectrum on a $256^3$ grid of tenth order compact scheme with localized artificial dissipation; cyan circles: WCNS5-JS; red squares: WCNS5-Z; green diamonds: WCNS6-LD; blue triangles: WCHR6.}
\label{fig:TGV_N0064_spectra_t_5}
\end{center}
\end{figure}

\begin{figure}[!ht]
\begin{center}
\subfigure[Velocity energy spectrum]{\includegraphics[width=0.48\textwidth]{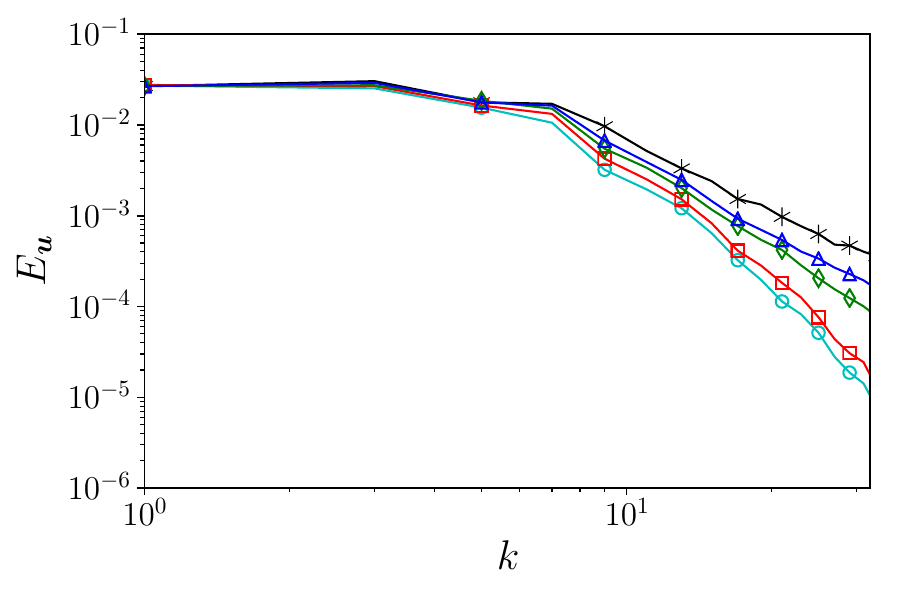} \label{fig:TGV_N0064_vel_spectrum_t_7}}
\subfigure[Vorticity energy spectrum]{\includegraphics[width=0.48\textwidth]{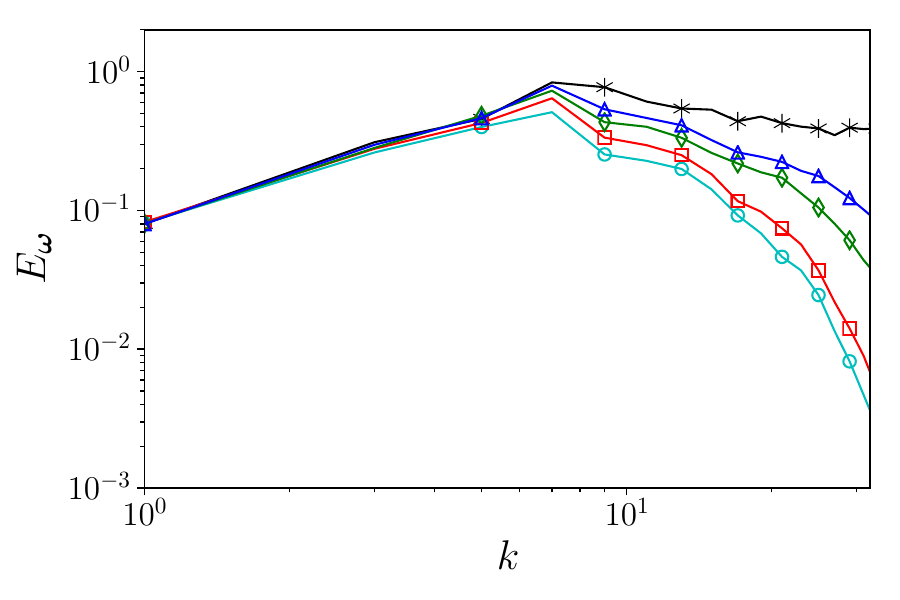} \label{fig:TGV_N0064_vor_spectrum_t_7}}
\caption{Spectra at $t = 7$ for the Taylor--Green vortex problem on a $64^3$ grid. Black asterisks: converged spectrum on a $256^3$ grid of tenth order compact scheme with localized artificial dissipation; cyan circles: WCNS5-JS; red squares: WCNS5-Z; green diamonds: WCNS6-LD; blue triangles: WCHR6.}
\label{fig:TGV_N0064_spectra_t_7}
\end{center}
\end{figure}

Figures~\ref{fig:TGV_N0064_spectra_t_5} and \ref{fig:TGV_N0064_spectra_t_7} compare the velocity and vorticity spectra of various schemes at $t=5$ and $t=7$ respectively. These spectra are also compared to a higher resolution simulation with $256^3$ grid points performed using a tenth order compact finite difference scheme~\cite{lele1992compact} with localized artificial dissipation. The velocity spectra are much more revealing than the kinetic energy plot. WCHR6 is the least dissipative since it is able to preserve more high wavenumber features while the other schemes dissipate the high wavenumber content more aggressively. WCHR6 agrees well with the higher resolution case until the Nyquist limit ($k=32$) at $t=5$ while WCNS6-LD agrees well till $k \approx 17$ after which it starts becoming more dissipative. WCNS5-JS and WCNS5-Z start adding dissipation from $k \approx 5$. The vorticity spectrum highlights the high wavenumber content more. From the vorticity spectrum, we again see that WCHR6 has much more energy in the high wavenumber region compared to the WCNS's. At $t=7$, the flow has much more fine scale features. At this time, all the schemes deviate from the high resolution case. The WCHR6 scheme has the highest energy content among all the other schemes and is closest to the high resolution case at all wavenumbers.

\subsection{Compressible homogeneous isotropic turbulence \label{sec:CHIT}}
A more realistic and pertinent test case for shock-capturing schemes than the Taylor--Green vortex problem is the decay of compressible homogeneous isotropic turbulence~\cite{lee1991eddy,johnsen2010assessment}. This is a viscous test case with the initial RMS velocity fluctuations being large enough to create eddy shocklets~\cite{lee1991eddy} and serves as a good problem to test the ability of numerical methods to capture shocks while also examine their dissipation characteristics for turbulence.

The initial velocity profile is a random solenoidal field that has an energy spectrum given by:
\begin{equation}
E(k) \propto k^4 \exp\left( -2 \left(\frac{k}{k_0}\right)^2 \right),
\end{equation}
where $k$ is the wavenumber and $k_0$ is the most energetic wavenumber. This gives an initial Taylor microscale, $\lambda = \lambda_0 = 2/k_0$. The RMS velocity fluctuation is given by $u_\mathrm{rms} = \sqrt{ \left\langle u_i u_i \right\rangle / 3 } = \sqrt{ (2/3) \int_0^\infty E(k) \mathrm{d}k }$. Details in obtaining the initial velocity profiles can be found in \citet{johnsen2010assessment}

The two important parameters in this problem are the turbulent Mach number, $M_t = \sqrt{3} u_\mathrm{rms} / \left\langle c \right\rangle$, and the Taylor scale Reynolds number, $\mathrm{Re}_\lambda = \left\langle \rho \right\rangle u_\mathrm{rms} \lambda / \left\langle \mu \right\rangle$. In this section, we consider the case with $M_t = M_{t,0} = 0.6$, $\mathrm{Re}_\lambda = \mathrm{Re}_{\lambda,0} = 100$, and $k_0 = 4$ initially. Ratio of specific heats, $\gamma = 1.4$, and the gas constant, $R=1$, are used. The density and pressure fields are taken to be constant at $\rho=1$ and $p=1/\gamma$ initially.

The shear viscosity is assumed to follow a power law temperature dependence given by:
\begin{equation}
\frac{\mu}{\mu_\mathrm{ref}} = \left(\frac{T}{T_\mathrm{ref}}\right)^{\frac{3}{4}},
\end{equation}
where $T_\mathrm{ref}=1/\gamma$ and $\mu_\mathrm{ref}=u_{\mathrm{rms},0} \lambda_0 / Re_{\lambda,0}$. $u_{\mathrm{rms},0}$ is the initial $u_\mathrm{rms}$. The bulk viscosity, $\mu_v$, is assumed to be zero. A constant Prandtl number, $\mathrm{Pr} = 0.7$, is used. The Prandtl number is defined as:
\begin{equation}
    \mathrm{Pr} = \frac{c_p \mu}{\kappa}, \quad c_p = \frac{\gamma R}{\gamma-1}
\end{equation}
The domain is periodic with size $\left[0, 2\pi \right)^3$. The problem is solved on a $64^3$ grid with the four schemes considered in this work. Reference solutions obtained from a direct numerical simulation (DNS) dataset spectrally filtered to a $64^3$ grid are used for comparison. The DNS dataset is obtained using a $512^3$ grid and a tenth order compact finite difference scheme. See section~\ref{sec:CHIT_postprocessing} for details on how the spectrally filtered DNS solutions are obtained. Simulations are run with a constant $\textnormal{CFL} = 0.5$ until $t/\tau = 4$ where $\tau$ is the eddy turnover time given by $\tau = \lambda_0 / u_\mathrm{rms,0}$. The simulations are also performed without the use of a subgrid-scale model in order to test the dissipation characteristics of the numerical scheme alone. Addition of subgrid-scale models in conjunction with this shock capturing scheme in a suitable and consistent way is left for future work.

\begin{figure}[!ht]
\begin{center}
\subfigure[]{\includegraphics[height=0.45\textwidth]{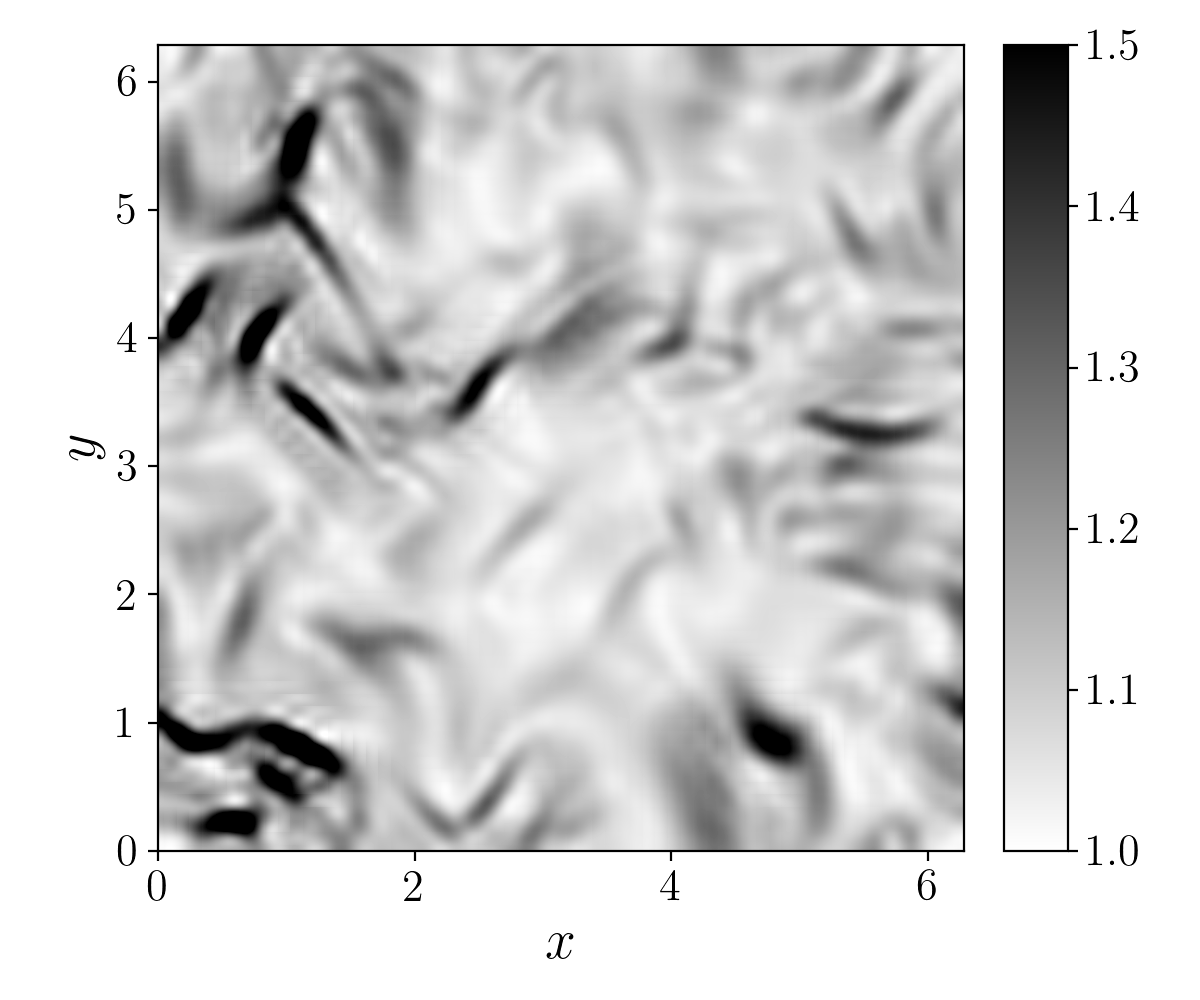} \label{fig:CHIT_schlieren}}
\subfigure[]{\includegraphics[height=0.45\textwidth]{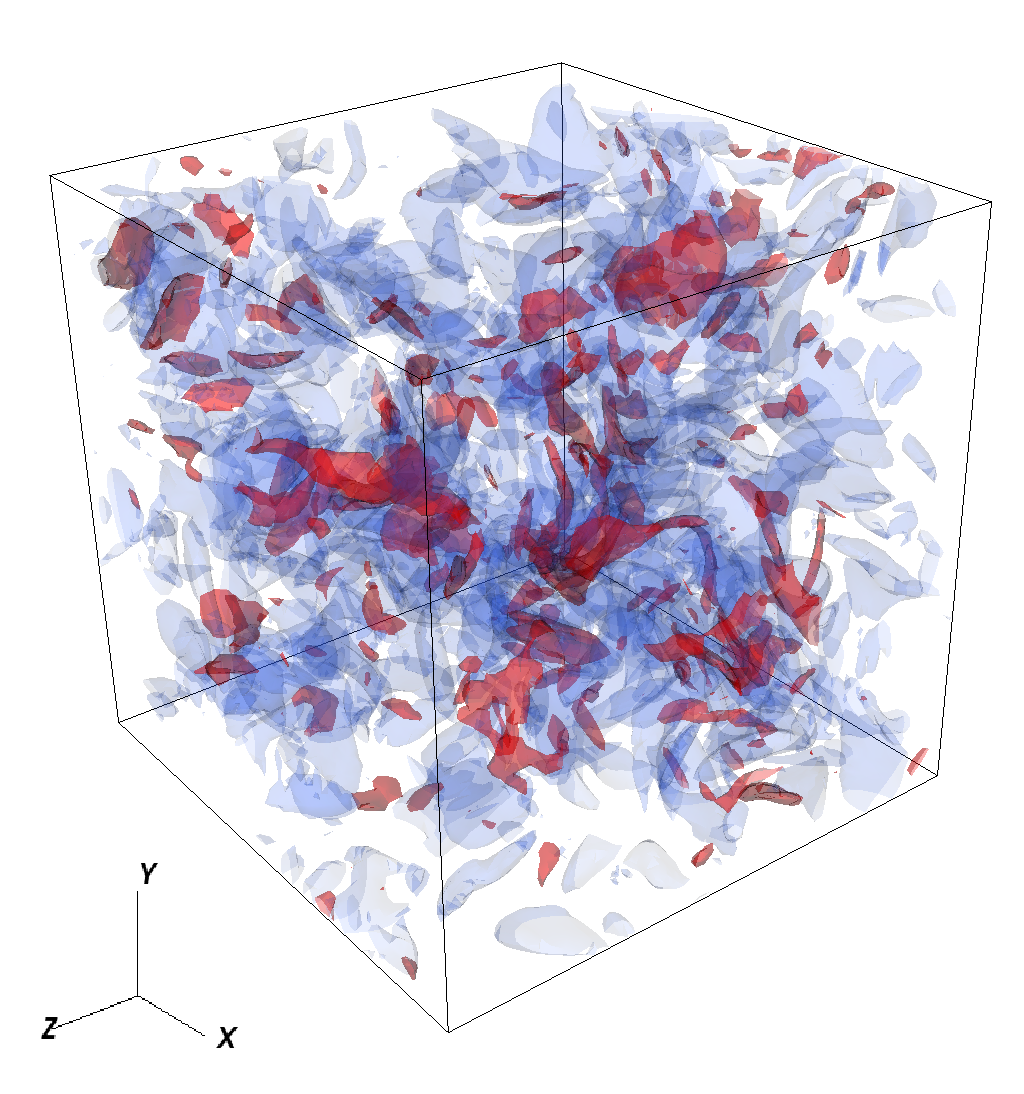} \label{fig:CHIT_3D}}
\caption{(a) Numerical schlieren visualized by $\exp \left( \frac{\| \nabla \rho \|}{\| \nabla \rho \|_{\textrm{max}}} \right)$ on a $z = 0$ slice for the compressible homogeneous isotropic turbulence problem on a $64^3$ grid at $t/\tau = 0.125$. (b) Isocontours of enstrophy at twice the mean (blue) and isocontours of dilatation at $3\sigma$ below the mean (red) for the same problem on the same grid at the same time, where $\sigma$ is the standard deviation.}
\label{fig:CHIT_shocklets}
\end{center}
\end{figure}

\begin{figure}[!ht]
\begin{center}
\subfigure[Velocity variance]{\includegraphics[width=0.48\textwidth]{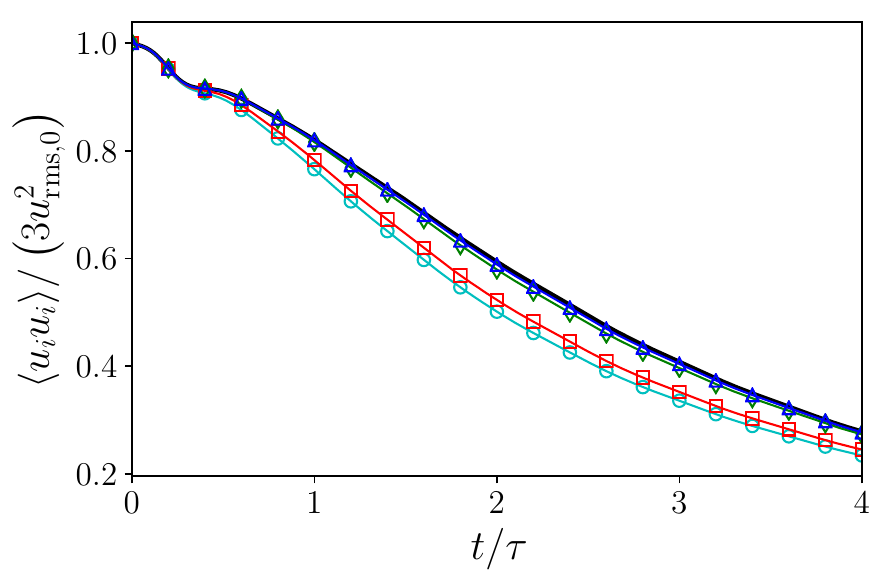} \label{fig:CHIT_N0064_TKE}}
\subfigure[Enstrophy]{\includegraphics[width=0.48\textwidth]{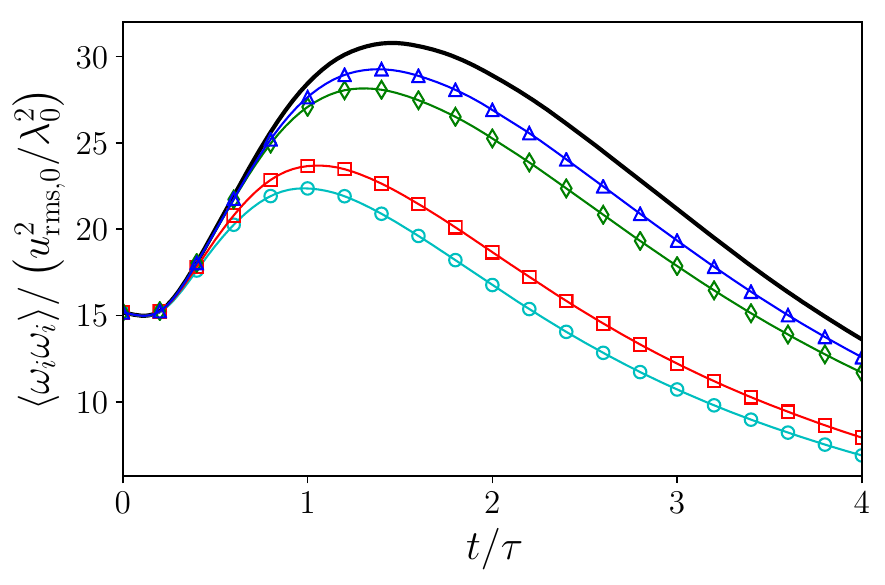} \label{fig:CHIT_N0064_Enstrophy}}
\subfigure[Dilatation variance]{\includegraphics[width=0.48\textwidth]{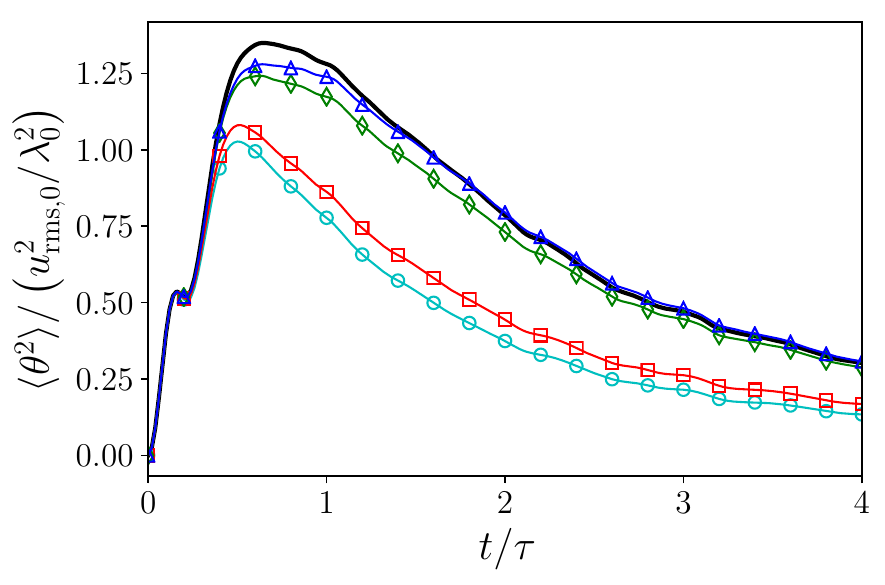} \label{fig:CHIT_N0064_dilatation_var}}
\caption{Time evolution of statistical quantities for the compressible homogeneous isotropic turbulence problem on a $64^3$ grid with $M_t = 0.6$. Black solid line : spectrally filtered DNS; cyan circles: WCNS5-JS; red squares: WCNS5-Z; green diamonds: WCNS6-LD; blue triangles: WCHR6.}
\label{fig:CHIT_N0064}
\end{center}
\end{figure}

Figure~\ref{fig:CHIT_schlieren} shows the numerical schlieren visualizing eddy shocklets in the domain. Figure~\ref{fig:CHIT_3D} shows contours of high enstrophy and high negative dilatation that visualizes the eddy shocklets. These distributed eddy shocklets make this test case challenging for numerical schemes and highlights the ability of schemes to capture turbulence structures as well as discontinuities. Figure~\ref{fig:CHIT_N0064} shows the velocity variance, enstrophy, and dilatation variance as a function of time for the four schemes and the filtered DNS solution. Here, we see that WCHR6 is the least dissipative and is the closest to the filtered DNS profiles for all the three statistics plotted. The enstrophy profiles highlight the difference between the schemes. WCNS5-JS and WCNS5-Z are excessively dissipative and capture very little amount of the enstrophy. WCHR6 agrees the best with the filtered DNS solution and shows that it is minimally dissipative even in the presence of eddy shocklets. Similar trends are seen in the plot of the dilatation variance. WCHR6 agrees very well with the filtered DNS solution while the other schemes dissipate dilatational motions more.

\begin{figure}[!ht]
\begin{center}
\subfigure[Velocity energy spectrum]{\includegraphics[width=0.48\textwidth]{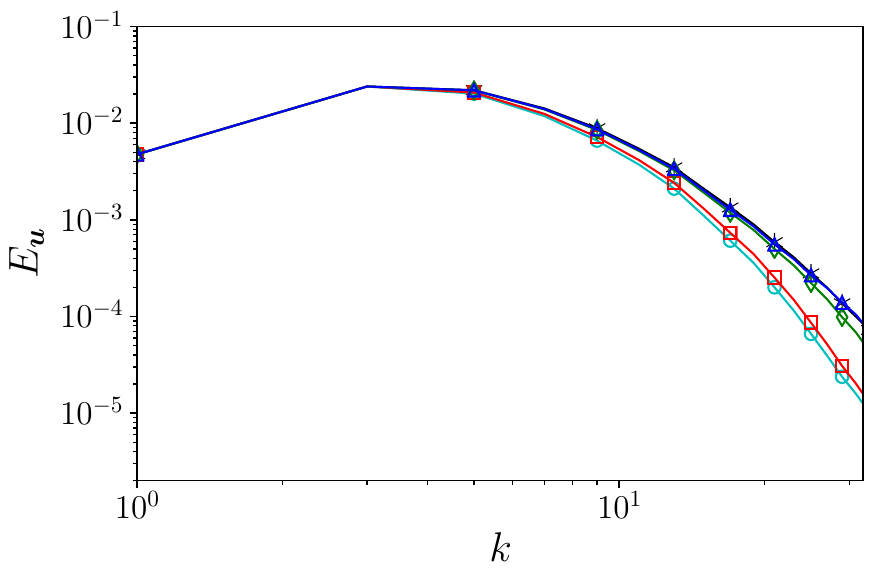} \label{fig:CHIT_VelocitySpectrum_t2}}
\subfigure[Vorticity energy spectrum]{\includegraphics[width=0.48\textwidth]{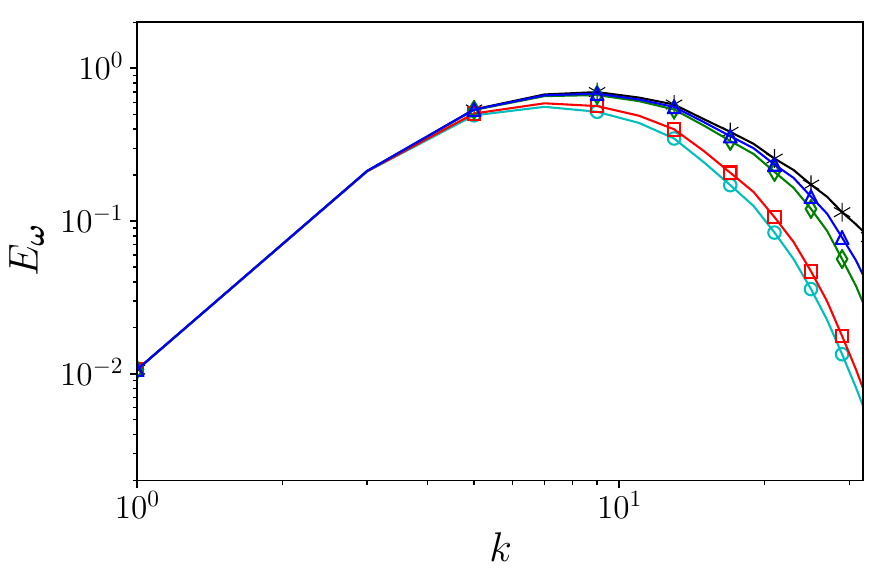} \label{fig:CHIT_VorticitySpectrum_t2}}
\subfigure[Dilatation energy spectrum]{\includegraphics[width=0.48\textwidth]{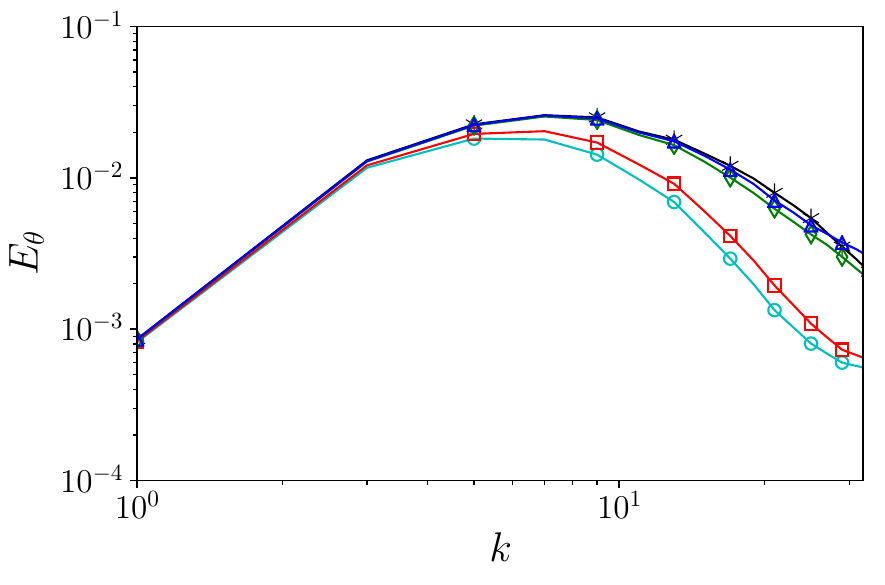} \label{fig:CHIT_DilatationSpectrum_t2}}
\subfigure[Density fluctuation energy spectrum]{\includegraphics[width=0.48\textwidth]{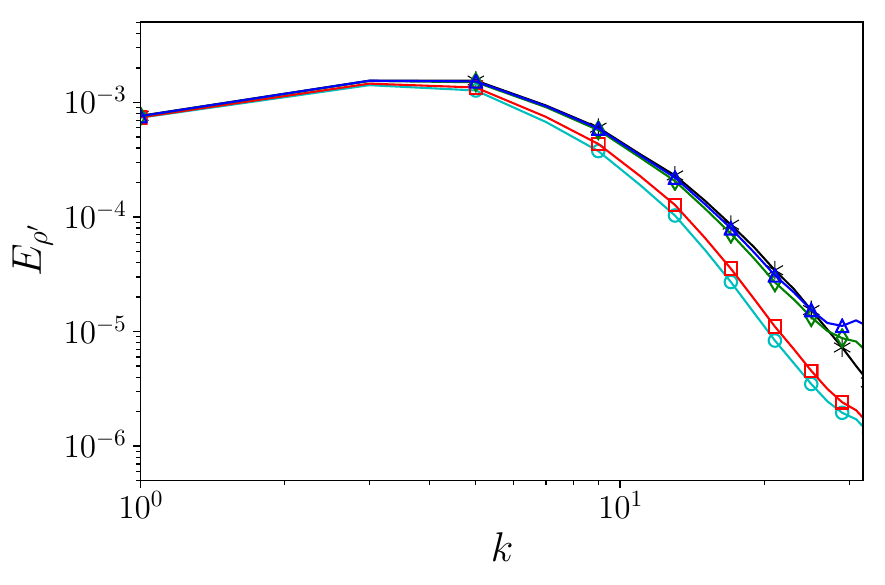} \label{fig:CHIT_DensitySpectrum_t2}}
\caption{Spectra at $t/\tau = 2$ for the compressible homogeneous isotropic turbulence problem on a $64^3$ grid with $M_t = 0.6$. Black asterisks: spectrally filtered DNS; cyan circles: WCNS5-JS; red squares: WCNS5-Z; green diamonds: WCNS6-LD; blue triangles: WCHR6.}
\label{fig:CHIT_spectra_t2}
\end{center}
\end{figure}

\begin{figure}[!ht]
\begin{center}
\subfigure[Velocity energy spectrum]{\includegraphics[width=0.48\textwidth]{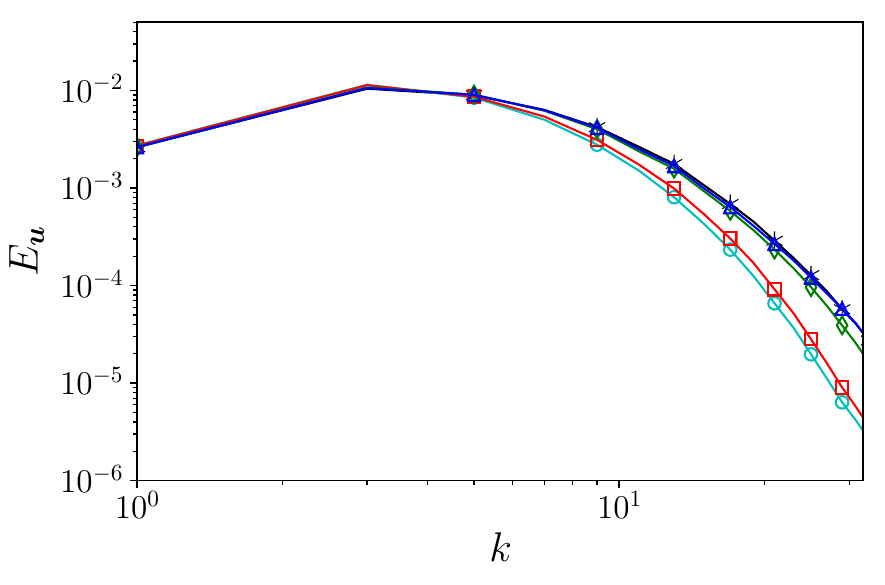} \label{fig:CHIT_VelocitySpectrum_t4}}
\subfigure[Vorticity energy spectrum]{\includegraphics[width=0.48\textwidth]{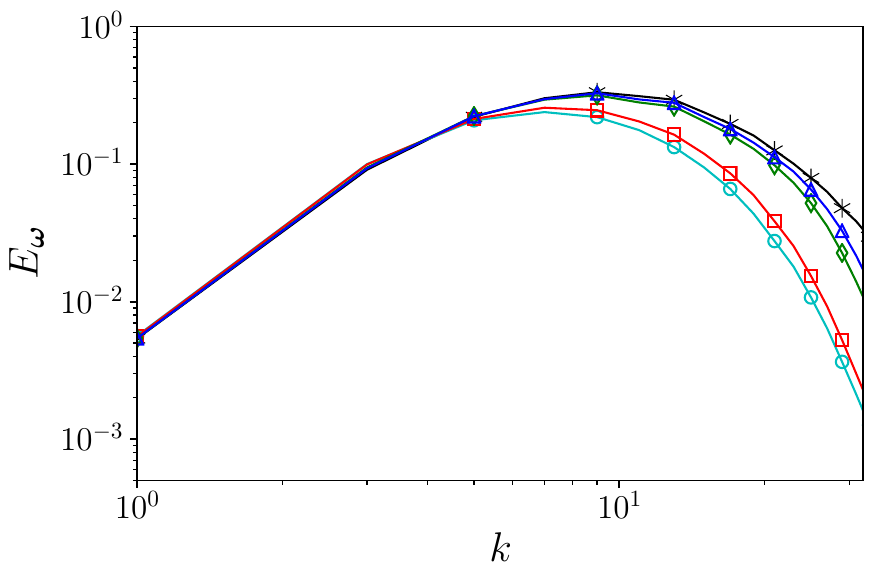} \label{fig:CHIT_VorticitySpectrum_t4}}
\subfigure[Dilatation energy spectrum]{\includegraphics[width=0.48\textwidth]{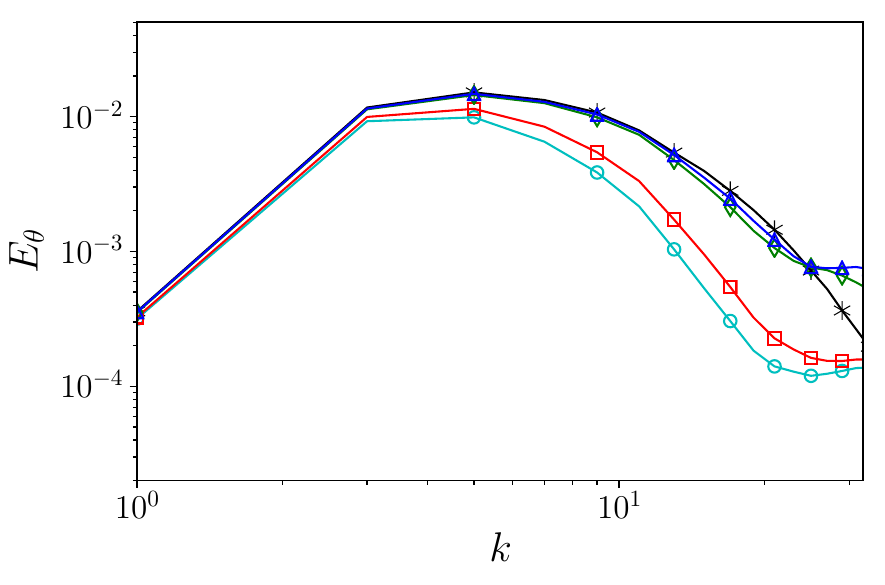} \label{fig:CHIT_DilatationSpectrum_t4}}
\subfigure[Density fluctuation energy spectrum]{\includegraphics[width=0.48\textwidth]{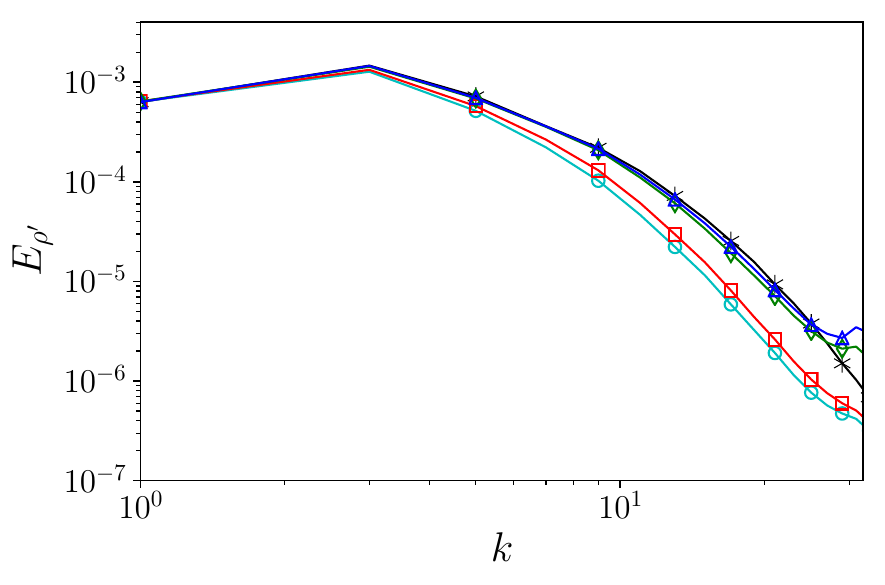} \label{fig:CHIT_DensitySpectrum_t4}}
\caption{Spectra at $t/\tau = 4$ for the compressible homogeneous isotropic turbulence problem on a $64^3$ grid with $M_t = 0.6$. Black asterisks: spectrally filtered DNS; cyan circles: WCNS5-JS; red squares: WCNS5-Z; green diamonds: WCNS6-LD; blue triangles: WCHR6.}
\label{fig:CHIT_spectra_t4}
\end{center}
\end{figure}

Figures~\ref{fig:CHIT_spectra_t2} and \ref{fig:CHIT_spectra_t4} show the velocity, vorticity, dilatation, and density spectra for the four different schemes. At this $\mathrm{Re}_{\lambda,0}$ of $100$, the peak of the vorticity energy spectrum is at $k \approx 9$ which is well below the maximum resolvable wavenumber of $32$. The two fifth order schemes don't capture this peak well but the two sixth order schemes do. Similar to the Taylor--Green vortex case, it can be seen that WCHR6 is the best at capturing fine scale features in both vorticity and dilatation while this advantage is less pronounced in this lower Reynolds number test case.

\section{Conclusions}

In summary, we have developed a new sixth order accurate weighted compact high resolution (WCHR6) scheme that has higher resolution and more localized dissipation than previous WCNS's. The high resolution property primarily comes from incorporating compact interpolation schemes directly into the WCNS interpolation mechanism. The scheme is presented for use with conservation equations such as the Euler equations and compressible Navier--Stokes equations in one, two, and three dimensions. The block tri-diagonal characteristic decomposition method is shown to be effective at interpolating primitive variables across shocks. Approximate dispersion relation (ADR) analysis of the scheme shows the superior resolution ability of the scheme compared to other WCNS's of similar orders of accuracy that use only explicit interpolations. Appropriate boundary schemes are also developed for non-periodic problems. Further, a conservative flux-difference form of compact finite difference schemes was derived for the first time and this allowed the use of central compact finite difference schemes with positivity-preserving limiters. Sixth order of accuracy of the scheme was demonstrated for the advection of an entropy wave in 1D and multi-dimensional settings. For all the test problems, the WCHR6 scheme was compared with WCNS's that utilized the same compact finite difference scheme but different interpolation methods to highlight the benefit of the new compact nonlinear interpolation method. Since all WCNS's in this paper use the same compact finite difference scheme as WCHR6, the advantage of the WCHR6 scheme might be expected to be larger when compared to the versions of the WCNS's which use explicit finite difference schemes. The advection of a broadband entropy wave showed that the WCHR6 scheme was better than the WCNS's at preserving the spectral content of the solution. The 1D Sod shock tube problem, the Shu--Osher problem, and the Sedov blast wave problem showed the ability of the method to capture shocks robustly while localizing the dissipation to regions near shocks. The WCHR6 scheme was shown to have much better dispersion and dissipation characteristics compared to the other schemes considered. The boundary schemes were also shown to be stable and accurate with appropriate boundary treatments for problems having features leaving the computational domain. The 2D shock interaction with a strong vortex showed the ability of the scheme to capture shocks with complex structures and large pressure variations. The robustness of the scheme while still being minimally dissipative was demonstrated in the double Mach reflection problem where the strong Mach 10 shock is captured robustly while the Kelvin--Helmholtz instability is minimally dissipated. The 3D Taylor--Green vortex problem highlighted the minimal dissipation characteristic of the scheme for a 3D problem with a large range of scales. Finally, the compressible homogeneous isotropic turbulence test case showed that the WCHR6 scheme was capable of capturing eddy shocklets randomly distributed in the turbulent field while still being minimally dissipative for both the solenoidal and dilatational motions.

\section*{Acknowledgments}

The code for all the 2D and 3D simulations in this paper is written in the high-level programming language Regent that uses the Legion tasking model developed at Stanford University. We acknowledge the support of Dr. Alex Aiken and Wonchan Lee with the task-based parallel programming in Regent and the Legion runtime system. It is our pleasure to acknowledge the benefits to this paper resulting from the comments of an anonymous referee who insisted on the Sedov blast wave and double Mach reflection test cases. Additional robust treatment was required to improve the scheme for these strong shock cases and resulted in an improved scheme, but this has an insignificant effect on the performance for Taylor--Green vortex and compressible homogeneous isotropic turbulence test cases.

\appendix

\section{Nonlinear weights} \label{appendix:nonlinear_weights}

Nonlinear weights are essential for nonlinear schemes such as WENO, WCNS, and WCHR schemes to capture discontinuities without spurious oscillations. Different forms of linear weights are discussed in this section.

\subsection{Classical upwind-biased (JS) nonlinear weights} \label{appendix:JS_nonlinear_weights}

For a weighted scheme with four sub-stencils, the classical JS nonlinear weighting method designed by \citet{jiang1995efficient} only assigns weights to the three upwind stencils and are therefore upwind-biased. The JS nonlinear weights $\omega_k^{\mathrm{JS}}$ are given by:

\begin{equation} \label{eq:JS_nonlinear_weights}
	\omega_k^{\mathrm{JS}} = \frac{\alpha_k^{\mathrm{JS}}}{\sum\limits_{k=0}^{2}\alpha_k^{\mathrm{JS}}}, \quad
    \alpha_k^{\mathrm{JS}} = \frac{d_k^{\mathrm{upwind}}}{ \left( \beta_k + \epsilon \right)^{p} }, \quad k = 0, 1, 2,
\end{equation}
\begin{equation}
	\omega^{\mathrm{JS}}_3 = 0,
\end{equation}

\noindent where $p$ is a positive integer and $\epsilon=1.0\mathrm{e}{-15}$ is a very small number to prevent division by zero. $\beta_k$ are smoothness indicators and are defined as:
\begin{equation}
	\beta_k = \sum^{2}_{l=1} \int^{x_{j+\frac{1}{2}}}_{x_{j-\frac{1}{2}}} \Delta x^{2l-1} \left( \frac{\partial^{l}}{\partial x^l} \tilde{u}^{(k)}(x) \right)^2 dx, \quad k = 0, 1, 2,
\end{equation}

\noindent where $\tilde{u}^{(k)}(x)$ is the Lagrange interpolating polynomial from stencil $S_k$ in figure~\ref{fig:stencil_WCNS} for WCNS and WCHR schemes. The integrated forms of the smoothness indicators are given by~\cite{zhang2008development}:
\begin{align}
	\beta_0 &= \frac{1}{3} \left[u_{j-2}\left(4u_{j-2} - 19u_{j-1} + 11u_j\right) + u_{j-1}\left(25u_{j-1} - 31u_j\right) + 10u_j^2\right], \\
    \beta_1 &= \frac{1}{3} \left[u_{j-1}\left(4u_{j-1} - 13u_j + 5u_{j+1}\right) + 13u_j\left(u_j - u_{j+1}\right) + 4u_{j+1}^2 \right], \\
    \beta_2 &= \frac{1}{3} \left[u_j\left(10u_j - 31u_{j+1} + 11u_{j+2}\right) + u_{j+1}\left(25u_{j+1}  - 19u_{j+2}\right) + 4u_{j+2}^2 \right].
\end{align}

\subsection{Improved upwind-biased (Z) nonlinear weights} \label{appendix:Z_nonlinear_weights}

The upwind-biased Z nonlinear weights designed by \citet{borges2008improved} improves the excessive dissipative nature of the JS nonlinear weights. The Z nonlinear weights $\omega_k^{\mathrm{Z}}$ are given by:

\begin{equation} \label{eq:Z5_nonlinear_weights}
	\omega_k^{\mathrm{Z}} = \frac{\alpha_k^{\mathrm{Z}}}{\sum\limits_{k=0}^{2}\alpha_k^{\mathrm{Z}}}, \quad
    \alpha_k^{\mathrm{Z}} = d_k^{\mathrm{upwind}} \left(1 + \left( \frac{\tau_5}{\beta_k + \epsilon} \right)^p \right), \quad k = 0, 1, 2,
\end{equation}
\begin{equation}
	\omega^{\mathrm{Z}}_3 = 0,
\end{equation}

\noindent where $\tau_5 = \left| \beta_2 - \beta_0 \right|$ is a reference smoothness indicator and $p$ is a positive integer.

\subsection{Localized dissipation (LD) nonlinear weigths} \label{appendix:LD_nonlinear_weights}

\noindent The nonlinear LD interpolation designed by \citet{wong2017high} also assigns nonlinear weight to the downwind stencil besides the upwind ones that helps the nonlinear interpolation recovers the non-dissipative central interpolation in smooth regions of the solutions. The LD nonlinear weights $\omega_k^{\mathrm{LD}}$ are given by:
\begin{equation} \label{eq:LD_nonlinear_weights}
 \omega_k^{\mathrm{LD}} = \begin{cases} 
  \sigma \omega^{\mathrm{upwind}}_k + (1 - \sigma) \omega^{\mathrm{central}}_k,
   &\mbox{if } R_{\tau} > \alpha^{\tau}_{RL}, \\
  \omega^{\mathrm{central}}_k,
   & \mbox{otherwise },
 \end{cases}
 \quad k = 0, 1, 2, 3,
\end{equation}\\

\noindent where $\omega^{\mathrm{upwind}}_k = \omega_k^{\mathrm{Z}}$ and $\omega^{\mathrm{central}}_k$ is given by:

\begin{equation} \label{eq:central_nonlinear_weights}
	\omega_k^\mathrm{central} = \frac{\alpha_k^\mathrm{central}}{\sum\limits_{k=0}^{3}\alpha_k^\mathrm{central}}, \quad
    \alpha_k^\mathrm{central} = d_k^\mathrm{central} \left( C + \left( \frac{\tau_6}{\beta_k + \epsilon} \right)^{q} \right), \quad k = 0, 1, 2, 3,
\end{equation}

\noindent where $q$ is a positive integer, $C$ is a positive constant, and $\beta_3$ is defined as:
\begin{equation}
\begin{aligned}
	\beta_3 = \sum^{5}_{l=1} \int^{x_{j+\frac{1}{2}}}_{x_{j-\frac{1}{2}}} \Delta x^{2l-1} \left( \frac{\partial^{l}}{\partial x^l} \tilde{u}^{(6)}(x) \right)^2 dx,
\end{aligned}
\end{equation}

\noindent where $\tilde{u}^{(6)}(x)$ is the Lagrange interpolating polynomial from stencil $S_{\mathrm{central}}$ in figure~\ref{fig:stencil_WCNS}. The integrated form of $\beta_3$ is given by~\cite{liu2015new}:

\begin{align}
	\beta_3 &=  \frac{1}{232243200} \left[u_{j-2}\left(525910327u_{j-2} - 4562164630u_{j-1} + 7799501420u_j \right.\right. \\
    & \quad \left.\left. - 6610694540u_{j+1} + 2794296070u_{j+2} - 472758974u_{j+3}\right) \right. \nonumber \\ 
    & \quad \left. + 5u_{j-1}\left(2146987907u_{j-1} - 7722406988u_j + 6763559276u_{j+1} - 2926461814u_{j+2} + 503766638u_{j+3}\right) \right. \nonumber \\
    & \quad \left. + 20u_j\left(1833221603u_j-3358664662u_{j+1}+1495974539u_{j+2}-263126407u_{j+3}\right) \right. \nonumber \\
    & \quad \left. + 20u_{j+1} \left( 1607794163u_{j+1} - 1486026707u_{j+2} + 268747951u_{j+3} \right) \right. \nonumber \\
    & \quad \left. + 5u_{j+2} \left(1432381427u_{j+2}-536951582u_{j+3}\right) + 263126407u_{j+3}^2\right]. \nonumber 
\end{align}

\noindent $\tau_6$ is a reference smoothness indicator:
\begin{align}
    \tau_6 &= \left| \beta_3 - \beta_\mathrm{avg} \right|,
\end{align}

\noindent where
\begin{equation}
	\beta_\mathrm{avg} = \frac{1}{8} \left( \beta_0 + 6\beta_1 + \beta_2 \right).
\end{equation}

\noindent $R_{\tau}$ is a relative sensor to distinguish smooth and non-smooth regions and is defined as:
\begin{equation}
	R_{\tau} = \frac{\tau_6}{\beta_\mathrm{avg} + \epsilon}.
\end{equation}

\noindent $\alpha^{\tau}_{RL}$ is a constant to determine the cut-off for the hybridization between upwind-biased and central nonlinear weights. $0 \leq \sigma \leq 1$ is a blending function that is close to one in regions near discontinuities and high wavenumber features. In this paper, the following form of $\sigma$ is used:
\begin{equation}
	\sigma_{j+\frac{1}{2}} = \max \left( \sigma_{j}, \sigma_{j+1} \right),
\end{equation}
where $\sigma_j$ is defined as:
\begin{eqnarray}
    \sigma_j &=& \frac{\left| \Delta u _{j+\frac{1}{2}} - \Delta u _{j-\frac{1}{2}} \right|}{\left|\Delta u _{j+\frac{1}{2}} \right| + \left| \Delta u _{j-\frac{1}{2}} \right| + \epsilon}, \\
    \Delta u _{j+\frac{1}{2}} &=& u_{j+1} - u_{j}.
\end{eqnarray}

\section{Coefficients of explicit-compact interpolations (ECI)}

\subsection{Interior scheme} \label{appendix:ECI_interior_coeffs}

The coefficients of the linear interpolations from $S_{\mathrm{upwind}}$ (equation~\eqref{eq:ECI_upwind}) and $S_{\mathrm{central}}$ (equation~\eqref{eq:ECI_central}) are given by:

\begin{align}
	\alpha^\mathrm{upwind} &= -\frac{5\left(\xi - 1 \right) \left(13\xi - 7 \right)}{8 \left( \xi + 5 \right) \left( 2\xi - 1 \right) }, \: &
    \beta^\mathrm{upwind} &= \frac{53\xi - 5}{8 \left(\xi + 5 \right)}, \: &
    \gamma^\mathrm{upwind} &= -\frac{5\left(\xi - 1 \right) \left(5\xi - 2 \right)}{8 \left( \xi + 5 \right) \left( 2\xi - 1 \right)}, \nonumber \\
    a^\mathrm{upwind} &= \frac{3 \left( 8\xi - 5 \right)}{64 \left(\xi + 5 \right)}, \: &
    b^\mathrm{upwind} &= - \frac{5 \left( 84\xi^2 - 103\xi + 31 \right)}{64 \left( \xi + 5 \right) \left( 2\xi - 1 \right)}, \: &
    c^\mathrm{upwind} &= \frac{5 \left( 68\xi^2 + 11\xi - 25 \right)}{64 \left( \xi + 5 \right) \left( 2\xi - 1 \right)}, \nonumber \\
    d^\mathrm{upwind} &= \frac{5 \left( 52\xi^2 - 11\xi - 5\right)}{64 \left( \xi + 5 \right) \left( 2\xi - 1 \right)}, \: &
    e^\mathrm{upwind} &= -\frac{5 \left(4\xi - 3\right) \left(5\xi - 2\right)}{64 \left( \xi + 5 \right) \left( 2\xi - 1 \right)}, & & &
\end{align}

\noindent and
\begin{align}
	\alpha^\mathrm{central} &= -\frac{45 \left(\xi - 1\right) }{16 \left(\xi + 5 \right)}, \: &
    \beta^\mathrm{central} &= \frac{53\xi - 5}{8 \left(\xi + 5 \right)}, \: &
    \gamma^\mathrm{central} &= -\frac{45 \left(\xi - 1\right) }{16 \left(\xi + 5 \right)}, \nonumber \\
    a^\mathrm{central} &= \frac{3 \left( 8\xi - 5 \right)}{128 \left(\xi + 5 \right)}, \: &
    b^\mathrm{central} &= -\frac{5 \left( 52\xi - 37 \right)}{128 \left( \xi + 5 \right) }, \: &
    c^\mathrm{central} &= \frac{75 \left( 2 \xi+ 1 \right)}{64 \left( \xi + 5 \right)}, \nonumber \\
    d^\mathrm{central} &= \frac{75 \left( 2 \xi+ 1 \right)}{64 \left( \xi + 5 \right)}, \: &
    e^\mathrm{central} &= -\frac{5 \left( 52\xi - 37 \right)}{128 \left( \xi + 5 \right) }, \: &
    f^\mathrm{central} &= \frac{3 \left( 8\xi - 5 \right)}{128 \left(\xi + 5 \right)}.
\end{align}

\subsection{Boundary scheme} \label{appendix:ECI_boundary_coeffs}

The coefficients of the left-biased interpolations (equations~\eqref{eq:WCHR6_LB_S2} and \eqref{eq:WCHR6_LB_S3}) at the left boundary (LB) are given by:

\begin{align}
    a^{\mathrm{LB}} &= - \frac{ 8 \xi_{0} - 3}{4},                  \quad &
    b^{\mathrm{LB}} &= \frac{8 \xi_{0} + 1}{4},                     \quad &
    c^{\mathrm{LB}} &= - \frac{4 \xi_{0} - 1}{4},                   \nonumber \\
    d^{\mathrm{LB}} &= \frac{3}{4},                                 \quad &
    e^{\mathrm{LB}} &= \xi_{0},                                     \quad &
    f^{\mathrm{LB}} &= -\frac{8\xi_{1} - 15}{8},                    \nonumber \\
    g^{\mathrm{LB}} &= \frac{12\xi_{1} - 5}{4},                     \quad &
    h^{\mathrm{LB}} &= - \frac{3 \left( 8 \xi_{1} - 1 \right) }{8}, \quad &
    i^{\mathrm{LB}} &= \xi_{1},
\end{align}

\noindent where $\xi_{0}$ and $\xi_{1}$ are two free parameters and the linear weights in equations~\eqref{eq:ECI_LB_upwind} and \eqref{eq:ECI_LB_central} are given by:

\begin{align}
&d_{0}^{(5), \mathrm{LB}} = \frac{56\xi_{0} - 5}{24 \left( 24\xi_{0} - 5 \right)}, \quad
d_{1}^{(5), \mathrm{LB}} = \frac{5 \left(104\xi_{0} - 11 \right)}{24 \left(24\xi_{0} - 5 \right)}, \quad
&d_{2}^{(5), \mathrm{LB}} = - \frac{5}{2 \left( 24 \xi_{0} - 5 \right)},
\end{align}

\noindent and

\begin{align}
&d_{0}^{(6), \mathrm{LB}} = \frac{6560\xi_{0}\xi_{1} + 552\xi_{0} - 716\xi_{1} - 75}{24 \left( 3648\xi_{0}\xi_{1} + 376\xi_{0} - 1080\xi_{1} - 145 \right)}, \nonumber \\
&d_{1}^{(6), \mathrm{LB}} = \frac{161984\xi_{0}\xi_{1} + 15480\xi_{0} - 20456\xi_{1} - 2385}{48 \left( 3648\xi_{0}\xi_{1} + 376\xi_{0} - 1080\xi_{1} - 145 \right)}, \nonumber \\
&d_{2}^{(6), \mathrm{LB}} = -\frac{624\xi_{1} + 90}{3648\xi_{0}\xi_{1} + 376\xi_{0} - 1080\xi_{1} - 145}, \nonumber \\
&d_{3}^{(6), \mathrm{LB}} = \frac{488\xi_{0}-35}{16 \left( 3648\xi_{0}\xi_{1} + 376\xi_{0} - 1080\xi_{1} - 145 \right)}.
\end{align}

In the case of $\xi=\frac{2}{3}$, if the truncation errors of interpolations from stencils $S_5^{\mathrm{LB}}$ and $S_5^{\mathrm{LB}}$ are matched with those of $S_{\mathrm{upwind}}$ and $S_{\mathrm{central}}$ respectively, we will get:
\begin{equation}
    \xi_{0} = \frac{9}{152}, \quad \xi_{1} = - \frac{14445}{171608}.
\end{equation}

\noindent Therefore,

\begin{align}
    a^{\mathrm{LB}} &= \frac{12}{19},          \quad &
    b^{\mathrm{LB}} &= \frac{7}{19},           \quad &
    c^{\mathrm{LB}} &= \frac{29}{152},         \quad &
    d^{\mathrm{LB}} &= \frac{3}{4},            \nonumber \\
    e^{\mathrm{LB}} &= \frac{9}{152},          \quad &
    f^{\mathrm{LB}} &= \frac{168105}{85804},   \quad &
    g^{\mathrm{LB}} &= -\frac{257845}{171608}, \quad &
    h^{\mathrm{LB}} &= \frac{13461}{21451},    \nonumber \\
    i^{\mathrm{LB}} &= -\frac{14445}{171608},   &&&&&&
\end{align}

\noindent and

\begin{align}
d_{0}^{(5), \mathrm{LB}} = \frac{1}{51}, \quad
d_{1}^{(5), \mathrm{LB}} = \frac{115}{408}, \quad
d_{2}^{(5), \mathrm{LB}} = \frac{95}{136},
\end{align}

\noindent and

\begin{align}
d_{0}^{(6), \mathrm{LB}} = \frac{34531}{2811392}, \quad
d_{1}^{(6), \mathrm{LB}} = \frac{324345}{1405696}, \quad
d_{2}^{(6), \mathrm{LB}} = \frac{3465}{4624}, \quad
d_{3}^{(6), \mathrm{LB}} = \frac{1129}{147968}.
\end{align}

The coefficients of the left-biased interpolations (equations~\eqref{eq:WCHR6_RB_S0} and \eqref{eq:WCHR6_RB_S1}) at the right boundary (RB) are given by:

\begin{align}
    a^{\mathrm{RB}} &= \xi_{2},                                    \quad &
    b^{\mathrm{RB}} &= -\frac{3 \left( 8 \xi_{2} - 1 \right) }{8}, \quad &
    c^{\mathrm{RB}} &= \frac{12\xi_{2} - 5}{4},                    \nonumber \\
    d^{\mathrm{RB}} &= -\frac{8\xi_{2} - 15}{8},                   \quad &
    e^{\mathrm{RB}} &= \frac{8\xi_{3} + 1}{4},                     \quad &
    f^{\mathrm{RB}} &= -\frac{8\xi_{3} - 3}{4},                    \nonumber \\
    g^{\mathrm{RB}} &= \xi_{3},                                    \quad &
    h^{\mathrm{RB}} &= \frac{3}{4},                                \quad &
    i^{\mathrm{RB}} &= -\frac{4\xi_{3} - 1}{4},
\end{align}

\noindent where $\xi_{2}$ and $\xi_{3}$ are two free parameters and the linear weights in equations~\eqref{eq:ECI_RB_upwind} and \eqref{eq:ECI_RB_central} are given by:

\begin{align}
&d_{0}^{(5), \mathrm{RB}} = \frac{56 \xi_{3} - 5}{32 \left(48 \xi_{2} \xi_{3} - 26 \xi_{2} + 8 \xi_{3} - 5\right)}, \quad
&d_{1}^{(5), \mathrm{RB}} = - \frac{76 \xi_{2} + 15}{4 \left(48 \xi_{2} \xi_{3} - 26 \xi_{2} + 8 \xi_{3} - 5\right)}, \nonumber \\
&d_{2}^{(5), \mathrm{RB}} = \frac{1536 \xi_{2} \xi_{3} - 224 \xi_{2} + 200 \xi_{3} - 35}{32 \left(48 \xi_{2} \xi_{3} - 26 \xi_{2} + 8 \xi_{3} - 5\right)},
\end{align}

\noindent and

\begin{align}
&d_{0}^{(6), \mathrm{RB}} = \frac{488 \xi_{3} - 35}{16 \left(3648 \xi_{2} \xi_{3} - 1080 \xi_{2} + 376 \xi_{3} - 145\right)}, \nonumber \\
&d_{1}^{(6), \mathrm{RB}} = - \frac{624 \xi_{2}+ 90 }{3648 \xi_{2} \xi_{3} - 1080 \xi_{2} + 376 \xi_{3} - 145}, \nonumber \\
&d_{2}^{(6), \mathrm{RB}} = \frac{161984 \xi_{2} \xi_{3} - 20456 \xi_{2} + 15480 \xi_{3} - 2385}{{48 \left(3648 \xi_{2} \xi_{3} - 1080 \xi_{2} + 376 \xi_{3} - 145\right)}}, \nonumber \\
&d_{3}^{(6), \mathrm{RB}} = \frac{6560 \xi_{2} \xi_{3} - 716 \xi_{2} + 552 \xi_{3} - 75}{24 \left(3648 \xi_{2} \xi_{3} - 1080 \xi_{2} + 376 \xi_{3} - 145\right)}.
\end{align}

In the case of $\xi=\frac{2}{3}$, if the truncation errors of interpolations from stencils $S_5^{\mathrm{RB}}$ and $S_5^{\mathrm{RB}}$ are matched with those of $S_{\mathrm{upwind}}$ and $S_{\mathrm{central}}$ respectively, we will get:
\begin{equation}
    \xi_{2} = - \frac{3182085}{37433632} - \frac{45 \sqrt{723535913}}{37433632}, \quad \xi_{3} = - \frac{9 \sqrt{723535913}}{7659176} + \frac{96676}{957397}.
\end{equation}

\noindent Therefore,

\begin{align}
    a^{\mathrm{RB}} &= -\frac{3182085}{37433632} - \frac{45 \sqrt{723535913}}{37433632},   \quad & 
    b^{\mathrm{RB}} &= \frac{135 \sqrt{723535913}}{37433632} + \frac{23583867}{37433632},  \nonumber \\
    c^{\mathrm{RB}} &= -\frac{56338295}{37433632} - \frac{135 \sqrt{723535913}}{37433632}, \quad &
    d^{\mathrm{RB}} &= \frac{45 \sqrt{723535913}}{37433632} + \frac{73370145}{37433632},   \nonumber \\
    e^{\mathrm{RB}} &= -\frac{9 \sqrt{723535913}}{3829588} + \frac{1730805}{3829588},      \quad &
    f^{\mathrm{RB}} &= \frac{9 \sqrt{723535913}}{3829588} + \frac{2098783}{3829588},       \nonumber \\
    g^{\mathrm{RB}} &= -\frac{9 \sqrt{723535913}}{7659176} + \frac{96676}{957397},         \quad &
    h^{\mathrm{RB}} &= \frac{3}{4},                                                        \nonumber \\
    i^{\mathrm{RB}} &= \frac{9 \sqrt{723535913}}{7659176} + \frac{570693}{3829588},        &&
\end{align}

\noindent and

\begin{align}
&d_{0}^{(5), \mathrm{RB}} = -\frac{135353}{41283072} + \frac{35 \sqrt{723535913}}{41283072},    &\quad
&d_{1}^{(5), \mathrm{RB}} = -\frac{145 \sqrt{723535913}}{82566144} + \frac{74237155}{82566144}, &\nonumber \\
&d_{2}^{(5), \mathrm{RB}} = \frac{25 \sqrt{723535913}}{27522048} + \frac{2866565}{27522048},     & &&
\end{align}

\noindent and

\begin{align}
&d_{0}^{(6), \mathrm{RB}} = -\frac{2038531}{157733888} + \frac{95 \sqrt{723535913}}{157733888}, &\quad
&d_{1}^{(6), \mathrm{RB}} = -\frac{95 \sqrt{723535913}}{39433472} + \frac{32791565}{39433472},  &\nonumber \\
&d_{2}^{(6), \mathrm{RB}} = \frac{135 \sqrt{723535913}}{78866944} + \frac{13590345}{78866944},   &\quad
&d_{3}^{(6), \mathrm{RB}} = \frac{15 \sqrt{723535913}}{157733888} + \frac{1425469}{157733888}.    &
\end{align}

\section{Block-tridiagonal matrix solution algorithm \label{sec:block_tridiag}}
\subsection{Matrix solution algorithm \label{sec:matrix_solution}}
Consider a block tridiagonal matrix system $\bm{A} \bm{x} = \bm{b}$ given by:
\begin{equation}
    \bm{A} =
    \begin{pmatrix}
        \bm{\beta}_1  & \bm{\gamma}_1 &                   &                  &                   \\
        \bm{\alpha}_2 & \bm{\beta}_2  & \bm{\gamma}_2     &                  &                   \\
                      &               & \ddots            &                  &                   \\
                      &               & \bm{\alpha}_{N-1} & \bm{\beta}_{N-1} & \bm{\gamma}_{N-1} \\
                      &               &                   & \bm{\alpha}_{N}  & \bm{\beta}_{N}    \\
   \end{pmatrix},
\end{equation}
\begin{equation}
   \bm{x} =
   \begin{pmatrix}
       \bm{x}_1     \\
       \bm{x}_2     \\
       \vdots       \\
       \bm{x}_{N-1} \\
       \bm{x}_{N}   \\
   \end{pmatrix},
   \qquad
   \bm{b} =
   \begin{pmatrix}
       \bm{b}_1     \\
       \bm{b}_2     \\
       \vdots       \\
       \bm{b}_{N-1} \\
       \bm{b}_{N}   \\
   \end{pmatrix}, \label{eq:block_tridiag_nonperiodic}
\end{equation}
where $\bm{\alpha}_i$, $\bm{\beta}_i$, and $\bm{\gamma}_i$ are $B_s \times B_s$ matrix blocks. $\bm{x}_i$ and $\bm{b}_i$ are $B_s \times 1$ vector elements of the solution and the RHS vectors respectively.

For the resulting block-tridiagonal system, we use derive a block version of the Thomas algorithm with a forward elimination step:
\begin{align}
    \bm{\Delta}_i =& \left[ \bm{\beta}_i - \bm{\alpha}_i \Delta_{i-1} \bm{\gamma}_{i-1} \right]^{-1}, \\
    \hat{\bm{b}}_i =& - \bm{b}_i - \bm{\alpha}_i \bm{\Delta}_{i-1} \hat{\bm{b}}_{i-1},
\end{align}
and a back substitution step:
\begin{equation}
    \bm{x}_{i} = - \bm{\Delta}_i \left[ \hat{\bm{b}}_i + \bm{\gamma}_i \bm{x}_{i+1} \right].
\end{equation}
 
For periodic problems resulting in a cyclic block tridiagonal matrix $\bm{A}_p$, we use the Sherman-Morrison low rank correction given by:
\begin{equation}
    \bm{A}_p^{-1} = \left( \bm{A} + \bm{U}\bm{V}^T \right)^{-1} = \bm{A}^{-1} - \bm{A}^{-1}\bm{U} \left(\bm{I} + \bm{V}^T \bm{A}^{-1} \bm{U}\right)^{-1} \bm{V}^T \bm{A}^{-1},
\end{equation}
where
\begin{equation}
\bm{A}_p = \begin{pmatrix}
    \tilde{\bm{\beta}}_1    & \tilde{\bm{\gamma}}_1    &        &                          & \tilde{\bm{\alpha}}_1     \\
    \tilde{\bm{\alpha}}_2   & \tilde{\bm{\beta}}_2     & \tilde{\bm{\gamma}}_2 &           &                           \\
                            &                          & \ddots &                          &                           \\
                            &       & \tilde{\bm{\alpha}}_{N-1} & \tilde{\bm{\beta}}_{N-1} & \tilde{\bm{\gamma}}_{N-1} \\
    \tilde{\bm{\gamma}}_{N} &                          &        & \tilde{\bm{\alpha}}_{N}  & \tilde{\bm{\beta}}_{N}    \\
   \end{pmatrix}, \label{eq:block_tridiag_periodic}
\end{equation}
\begin{equation}
\bm{A} = \begin{pmatrix}
   \tilde{\bm{\alpha}}_1 + \tilde{\bm{\beta}}_1  & \tilde{\bm{\gamma}}_1    &        &                          &                           \\
    \tilde{\bm{\alpha}}_2 & \tilde{\bm{\beta}}_2     & \tilde{\bm{\gamma}}_2 &           &                           \\
                          &                          & \ddots &                          &                           \\
                          &       & \tilde{\bm{\alpha}}_{N-1} & \tilde{\bm{\beta}}_{N-1} & \tilde{\bm{\gamma}}_{N-1} \\
                          &                          &        & \tilde{\bm{\alpha}}_{N}  & \tilde{\bm{\beta}}_{N} + \tilde{\bm{\gamma}}_{N}    \\
   \end{pmatrix},
\end{equation}
and $\bm{U}$ and $\bm{V}$ are given by:
\begin{equation}
\bm{U} = \begin{pmatrix}
       -\tilde{\bm{\alpha}}_1     \\
       \bm{0} \\
       \vdots               \\
       \bm{0} \\
       \tilde{\bm{\gamma}}_{N}   \\
   \end{pmatrix},
\qquad
\bm{V} = \begin{pmatrix}
       \bm{I}     \\
       \bm{0} \\
       \vdots               \\
       \bm{0} \\
       -\bm{I}   \\
   \end{pmatrix}.
\end{equation}
The pseudo code for the block-tridiagonal algorithm including the Sherman-Morrison correction is given in algorithm~\ref{alg:block_tridiag}.

\begin{algorithm}
\small
 \caption{Pseudo code for the block-tridiagonal matrix solution algorithm. Internal variables required are $\bm{\Delta}_i$ (memory cost of $B_s^2 N$) and $\widehat{\bm{b}}_i$ (memory cost of $B_s N$) and periodic problems require $\widehat{\bm{U}}_i$ (memory cost of $B_s^2 N$) in addition. $B_s$ is the size of each block in the system. The solution can be computed in-place by replacing $\widehat{\bm{b}}_i$ and $\bm{x}_i$ by $\bm{b}_i$.} \label{alg:block_tridiag}
 \begin{algorithmic}
  \If {periodic}
      \State $\bm{\beta}_1 \gets \tilde{\bm{\alpha}}_1 + \tilde{\bm{\beta}}_1$
      \State $\bm{\gamma}_1 \gets \tilde{\bm{\gamma}}_1$
      \For{$i = 2,3,\dots,N-1$}
        \State $\bm{\alpha}_i \gets \tilde{\bm{\alpha}}_i$
        \State $\bm{\beta}_i \gets \tilde{\bm{\beta}}_i$
        \State $\bm{\gamma}_i \gets \tilde{\bm{\gamma}}_i$
      \EndFor
      \State $\bm{\alpha}_N \gets \tilde{\bm{\alpha}}_N$
      \State $\bm{\beta}_N \gets \tilde{\bm{\beta}}_N + \tilde{\bm{\gamma}}_N$
  \EndIf
  \State $\bm{\Delta}_1 =  \bm{\beta}_1^{-1}$
  \State $\widehat{\bm{b}}_1 = - \bm{b}_1$
  \If {periodic}
      \State $\widehat{\bm{U}}_1 = \tilde{\bm{\alpha}}_1$
  \EndIf
  \For{$i = 2,3,\dots,N$}
    \State $\bm{\Delta}_i = \left[ \bm{\beta}_i - \bm{\alpha}_i \Delta_{i-1} \bm{\gamma}_{i-1} \right]^{-1}$
    \State $\widehat{\bm{b}}_i = - \bm{b}_i - \bm{\alpha}_i \bm{\Delta}_{i-1} \widehat{\bm{b}}_{i-1}$
    \If {periodic}
        \State $\widehat{\bm{U}}_i = -\bm{\alpha}_i \bm{\Delta}_{i-1} \widehat{\bm{U}}_{i-1}$
    \EndIf
  \EndFor
  \If {periodic}
      \State $\widehat{\bm{U}}_N \gets -\tilde{\bm{\gamma}}_N + \widehat{\bm{U}}_{N}$
  \EndIf
  \State $\bm{x}_N = - \bm{\Delta}_N \widehat{\bm{b}}_N$
  \If {periodic}
      \State $\widehat{\bm{U}}_N \gets -\bm{\Delta}_N \widehat{\bm{U}}_N$
  \EndIf
  \For{$i = N-1,N-2,\dots,1$}
    \State $\bm{x}_{i} = - \bm{\Delta}_i \left[ \widehat{\bm{b}}_i + \bm{\gamma}_i \bm{x}_{i+1} \right]$
    \If {periodic}
        \State $\widehat{\bm{U}}_{i} \gets - \bm{\Delta}_i \left[ \widehat{\bm{U}}_i + \bm{\gamma}_i \widehat{\bm{U}}_{i+1} \right]$
    \EndIf
  \EndFor
  \If {periodic}
      \State $\bm{M} = \bm{I} + \widehat{\bm{U}}_1 - \widehat{\bm{U}}_N$
      \State $\bm{y} = \bm{M}^{-1} \left( \bm{x}_1 - \bm{x}_N \right)$
      \For{$i = 1,2,\dots,N$}
          \State $\bm{x}_i \gets \bm{x}_i - \widehat{\bm{U}}_i \bm{y}$
      \EndFor
  \EndIf
 \end{algorithmic}
\end{algorithm}

\subsection{Application to compact interpolation}
The compact interpolations with characteristic decomposition used in this paper result in block tridiagonal systems as in equation~\eqref{eq:block_tridiag_nonperiodic} for non-periodic problems or equation~\eqref{eq:block_tridiag_periodic} for periodic problems. In both cases, each block is a $5 \times 5$ matrix that is a scaled version of the Jacobian matrix of the fluxes with respect to primitive variables. The Jacobian matrix forming each block in the $x$ direction interpolation of a 3D problem is given by:
\begin{equation}
\begin{pmatrix}
  0 & -\frac{{\rho} c}{2} & 0 & 0 & \frac{1}{2}  \\
  1 & 0                   & 0 & 0 & -\frac{1}{c^2} \\
  0 & 0                   & 1 & 0 & 0 \\
  0 & 0                   & 0 & 1 & 0 \\
  0 &  \frac{{\rho} c}{2} & 0 & 0 & \frac{1}{2}
 \end{pmatrix},
\end{equation}
where the rows correspond to the primitive variables $\left( \rho, u, v, w, p \right)$.

Given this structure, we can decouple the third and fourth rows (corresponding to $v$ and $w$) and split the problem into a $3\times3$ block tridiagonal system corresponding to $\left(\rho, u, p\right)$ and separate independent tridiagonal systems for $v$ and $w$. In the $y$ direction, the $u$ and $w$ interpolations are independent and in the $z$ direction, the $u$ and $v$ interpolations are independent.

The reduced $3\times3$ block tridiagonal system may be solved using the algorithm described in section~\ref{sec:matrix_solution}. The cost of the block tridiagonal algorithm scales as $\mathcal{O}\left(B_s^3 N\right)$ where $B_s$ is the block size and $N$ is the number of diagonal blocks in the system. Reducing the block size from $5$ to $3$ would then reduce the operation count by a factor of $\approx 4.6$. The tridiagonal systems for the two transverse velocity components may be solved using the Thomas algorithm or a symbolic factorization based algorithm~\cite{nguetchue2008computational}.

\section{Relation between compact finite difference schemes and flux difference form for provable discrete conservation} \label{appendix:flux_difference}

\subsection{Formulation}

Given a scalar hyperbolic equation of conservative variable $u(x,t)$ of the form:
\begin{equation}
	\frac{\partial u}{\partial t} + \frac{\partial F(u)}{\partial x} = 0,
\end{equation}
defined in the domain $x \in [x_a, x_b]$.
We can get a semi-discretized form using the finite difference formalism as:
\begin{equation} \label{eq:semi_discrete_eqn}
	\frac{\partial u_j}{\partial t} + \left. \frac{\partial F}{\partial x} \right|_{x = x_j} = 0,
\end{equation}
where $u_j(t) = u(x_j, t)$ and is available at discrete points $x_j = x_a + (1/2 + j) \Delta x, \; \forall j \in \{0, \: 1, \: \dots, \: N-1 \}$.

Let us define $h(x, t; \Delta x)$ implicitly as:
\begin{equation}
	F(u(x,t)) = F(x,t) = \frac{1}{\Delta x} \int^{x + \frac{\Delta x}{2}}_{x - \frac{\Delta x}{2}} h(\xi, t; \Delta x) d\xi.
\end{equation}

\noindent Equation~\eqref{eq:semi_discrete_eqn} can then be rewritten as:
\begin{align}
	\frac{\mathrm{d} u_j}{\mathrm{d} t} + \frac{1}{\Delta x} \left[ h \left(x_{j+\frac{1}{2}}, t; \Delta x\right) - h \left(x_{j-\frac{1}{2}}, t; \Delta x\right) \right] = 0,
\end{align}
or shortened as:
\begin{align}
    \frac{\mathrm{d} u_j}{\mathrm{d} t} + \frac{1}{\Delta x} \left( h_{j+\frac{1}{2}} - h_{j-\frac{1}{2}} \right) = 0,
\end{align}
with the definition $h \left(x_{j+\frac{1}{2}}, t; \Delta x\right) = h_{j+\frac{1}{2}}$. We may also define the primitive function of $h(x, t; \Delta x)$:
\begin{equation}
    H(x, t; \Delta x) = \int^{x}_{x_a} h(\xi, t; \Delta x) d\xi.
\end{equation}
Therefore,
\begin{equation}
\begin{aligned}
	H(x_{j+\frac{1}{2}}, t; \Delta x) &= \int^{x_{j+\frac{1}{2}}}_{x_a} h(\xi, t; \Delta x) d\xi \\
         &= \Delta x \sum^{j}_{i=0} F(x_i, t) \\
         &= \Delta x \sum^{j}_{i=0} F_i. \label{eq:H_definition}
\end{aligned}
\end{equation}
Also,
\begin{equation}
	H^\prime\left(x_{j+\frac{1}{2}}, t; \Delta x\right) = h\left(x_{j+\frac{1}{2}}, t; \Delta x\right).
\end{equation}
Or for simplification, if we define $H^\prime_{j + \frac{1}{2}} = H^\prime\left(x_{j+\frac{1}{2}}, t; \Delta x\right)$, we get:
\begin{equation}
	H^\prime_{j + \frac{1}{2}} = h_{j+\frac{1}{2}}.
\end{equation}

Now, let us denote a $p^\text{th}$ order numerical representation of $H^\prime _{j+\frac{1}{2}}$ by $\widehat{F}_{j+\frac{1}{2}} \approx h_{j+\frac{1}{2}} + \mathcal{O}\left( \Delta x^p \right) = H^\prime _{j+\frac{1}{2}} + \mathcal{O}\left( \Delta x^p \right)$, which is a reconstructed form of the flux. We can get such an approximation using a $p^\text{th}$ order compact finite difference scheme for $H^\prime_{j+\frac{1}{2}}$ in general form:
\begin{equation}
\begin{aligned}
	\alpha \widehat{F}_{j - \frac{1}{2}} &+ \beta \widehat{F}_{j + \frac{1}{2}} + \gamma \widehat{F}_{j + \frac{3}{2}} = \\ 
    & \frac{1}{\Delta x} \left( a_{-\frac{3}{2}} H_{j-\frac{3}{2}} + a_{-1} H_{j-1} + a_{-\frac{1}{2}} H_{j-\frac{1}{2}} + a_{0} H_{j} \right. \\
    &\left. + a_{\frac{1}{2}} H_{j+\frac{1}{2}} +  a_{1} H_{j+1} +  a_{\frac{3}{2}} H_{j+\frac{3}{2}} +  a_{2} H_{j+2} +  a_{\frac{5}{2}} H_{j+\frac{5}{2}}\right). \label{eq:H_compact_finite_difference}
\end{aligned}
\end{equation}
Using equation~\eqref{eq:H_definition}, we get:
\begin{align}
	\alpha \widehat{F}_{j - \frac{1}{2}} &+ \beta \widehat{F}_{j + \frac{1}{2}} + \gamma \widehat{F}_{j + \frac{3}{2}} = \nonumber \\ 
    & \frac{1}{\Delta x} \left[ \left(a_{-\frac{3}{2}} + a_{-\frac{1}{2}} + a_{\frac{1}{2}} +  a_{\frac{3}{2}} +  a_{\frac{5}{2}} \right) H_{j-\frac{3}{2}} + \left( a_{-1} + a_{0} + a_{1} +  a_{2} \right) H_{j-1} \right. \nonumber \\
    & + a_{-\frac{1}{2}} F_{j-1} + a_{0} F_{j-\frac{1}{2}} + a_{\frac{1}{2}} \left( F_{j-1} + F_{j} \right) + a_{1} \left( F_{j-\frac{1}{2}} + F_{j+\frac{1}{2}}\right) \nonumber \\
    & \left. +  a_{\frac{3}{2}} \left( F_{j-1} + F_{j} + F_{j+1} \right) + a_{2} \left( F_{j-\frac{1}{2}} + F_{j+\frac{1}{2}} + F_{j+\frac{3}{2}} \right) +  a_{\frac{5}{2}} \left( F_{j-1} + F_{j} + F_{j+1} + F_{j+2} \right) \right].
\end{align}
After re-arranging,
\begin{align}
	\alpha \widehat{F}_{j - \frac{1}{2}} &+ \beta \widehat{F}_{j + \frac{1}{2}} + \gamma \widehat{F}_{j + \frac{3}{2}} \nonumber \\ 
    &= \left( a_{-1} + a_{0} + a_{1} + a_{2} \right) H_{j-1} + \left(a_{0} + a_{1} + a_{2}\right) F_{j-\frac{1}{2}} + \left(a_{1} + a_{2}\right) F_{j+\frac{1}{2}} + a_{2} F_{j+\frac{3}{2}} \nonumber \\
    &\quad + \left( a_{-\frac{3}{2}} + a_{-\frac{1}{2}} + a_{\frac{1}{2}} + a_{\frac{3}{2}} + a_{\frac{5}{2}} \right) H_{j-\frac{3}{2}} + \left( a_{-\frac{1}{2}} + a_{\frac{1}{2}} + a_{\frac{3}{2}} + a_{\frac{5}{2}} \right) F_{j-1} \nonumber \\
    &\quad + \left( a_{\frac{1}{2}} + a_{\frac{3}{2}} + a_{\frac{5}{2}} \right) F_j + \left(a_{\frac{3}{2}} + a_{\frac{5}{2}}\right) F_{j+1} + a_{\frac{5}{2}} F_{j+2}.
\end{align}
If $\left( a_{-1} + a_{0} + a_{1} + a_{2} \right) = 0$ and $\left( a_{-\frac{3}{2}} + a_{-\frac{1}{2}} + a_{\frac{1}{2}} + a_{\frac{3}{2}} + a_{\frac{5}{2}} \right) = 0$ which is always true for a central scheme, we get a compact stencil representation of the reconstructed flux as:
\begin{align}
	\alpha \widehat{F}_{j - \frac{1}{2}} + \beta \widehat{F}_{j + \frac{1}{2}} + \gamma \widehat{F}_{j + \frac{3}{2}} &= \left(a_{0} + a_{1} + a_{2}\right) F_{j-\frac{1}{2}} + \left(a_{1} + a_{2}\right) F_{j+\frac{1}{2}} + a_{2} F_{j+\frac{3}{2}} \nonumber \\
    &\quad + \left( a_{-\frac{1}{2}} + a_{\frac{1}{2}} + a_{\frac{3}{2}} + a_{\frac{5}{2}} \right) F_{j-1} \nonumber \\
    &\quad + \left( a_{\frac{1}{2}} + a_{\frac{3}{2}} + a_{\frac{5}{2}} \right) F_j + \left(a_{\frac{3}{2}} + a_{\frac{5}{2}}\right) F_{j+1} + a_{\frac{5}{2}} F_{j+2} \nonumber \\
    &= \widehat{a}_{-1} F_{j-1} + \widehat{a}_{-\frac{1}{2}} F_{j-\frac{1}{2}} + \widehat{a}_{0} F_j + \widehat{a}_{\frac{1}{2}} F_{j+\frac{1}{2}} + \widehat{a}_{1} F_{j+1} + \widehat{a}_{\frac{3}{2}} F_{j+\frac{3}{2}} \nonumber \\
    &\quad + \widehat{a}_{2} F_{j+2}, \label{eq:flux_reconstruction_form}
\end{align}
with $\widehat{a}_{-1} = \left( a_{-\frac{1}{2}} + a_{\frac{1}{2}} + a_{\frac{3}{2}} + a_{\frac{5}{2}} \right)$, $\widehat{a}_{0} = \left( a_{\frac{1}{2}} + a_{\frac{3}{2}} + a_{\frac{5}{2}} \right)$, $\widehat{a}_{1} = \left(a_{\frac{3}{2}} + a_{\frac{5}{2}}\right)$, $\widehat{a}_{2} = a_{\frac{5}{2}}$, $\widehat{a}_{-\frac{1}{2}} = \left( a_{0} + a_{1} + a_{2} \right)$, $\widehat{a}_{\frac{1}{2}} = \left(a_{1} + a_{2}\right)$, and $\widehat{a}_{\frac{3}{2}} = a_{2}$.

With this $p^\text{th}$ order approximation of $h_{j+\frac{1}{2}}$, we can solve the original conservation law in the conservation form as:
\begin{equation} \label{eq:reconstruction_form}
\frac{\partial u_j}{\partial t} + \frac{1}{\Delta x} \left( \widehat{F}_{j+\frac{1}{2}} - \widehat{F}_{j-\frac{1}{2}} \right) = 0.
\end{equation}
If we define the flux difference form for the numerical approximation of derivative:
\begin{equation}
	\left. \widehat{ \frac{\partial F}{\partial x} } \right|_{x=x_j} = \widehat{F}_{j}^\prime =  \frac{1}{\Delta x} \left( \widehat{F}_{j+\frac{1}{2}} - \widehat{F}_{j-\frac{1}{2}} \right). \label{eq:flux_derivative_definition}
\end{equation}

\noindent For a central scheme with $\alpha = \gamma$, $a_{0} = -a_{1}$, $a_{-1} = -a_{2}$, $a_{\frac{1}{2}} = 0$, $a_{-\frac{1}{2}} = -a_{\frac{3}{2}}$, and $a_{-\frac{3}{2}} = -a_{\frac{5}{2}}$, it can be easily proven that:
\begin{align}
	\alpha \widehat{F}_{j - 1}^\prime &+ \beta \widehat{F}_{j}^\prime + \alpha \widehat{F}_{j + 1}^\prime = \nonumber \\ 
    & \frac{1}{\Delta x} \left( - a_{\frac{5}{2}} F_{j-2} -a_{2} F_{j-\frac{3}{2}} - a_{\frac{3}{2}} F_{j-1} - a_{1} F_{j-\frac{1}{2}} + a_{1} F_{j+\frac{1}{2}} +  a_{\frac{3}{2}} F_{j+1} + a_{2} F_{j+\frac{3}{2}} + a_{\frac{5}{2}} F_{j+2}\right). \label{eq:flux_derivative_recovered}
\end{align}
Therefore, $\widehat{F}_{j}^\prime$ is $p^\text{th}$ order approximation of $F_{j}^\prime$ with the same compact finite difference scheme used in equation~\eqref{eq:H_compact_finite_difference} with the constraint that the scheme is central. 

Flux reconstruction equation~\eqref{eq:flux_reconstruction_form} relates any central finite difference scheme (compact or explicit) in form given by equation~\eqref{eq:flux_derivative_recovered} to the flux difference form (equation~\eqref{eq:flux_derivative_definition}). For instance, the flux reconstruction equation of the sixth order CMD scheme (equation~\eqref{eq:CMD}) is given by equation~\eqref{eq:flux_reconstruction_form_CMD} and that of the sixth order CND scheme (equation~\eqref{eq:CND}) is given by:
\begin{equation}
    \frac{1}{5} \widehat{F}_{j-\frac{1}{2}} + \frac{3}{5} \widehat{F}_{j+\frac{1}{2}} + \frac{1}{5} \widehat{F}_{j+\frac{3}{2}} = \frac{1}{60} {F}_{j-1} + \frac{29}{60} {F}_{j} + \frac{29}{60} {F}_{j+1} + \frac{1}{60} {F}_{j+2}.
\end{equation}

To derive boundary closures for the flux reconstruction equation given by equation~\eqref{eq:flux_reconstruction_form} such as the closure at the right boundary with $j = N-1$, we can define a boundary flux reconstruction equation:
\begin{equation}
	\alpha \widehat{F}_{j - \frac{1}{2}} + \left(\beta + \gamma\right) \widehat{F}_{j + \frac{1}{2}} = \dots \label{eq:flux_reconstruction_right_boundary}
\end{equation}
where the right hand side is constructed based on a choice of cell node and midpoint flux values (either ghost cells or only interior). Then subtract equation~\eqref{eq:flux_reconstruction_form} from the above and divide by $\Delta x$ to get:
\begin{equation}
	\alpha \widehat{F}^\prime_{j - 1} + \beta \widehat{F}^\prime_{j} = \dots
\end{equation}
Given a desired truncation error, we can use the above equation and standard Taylor series expansion to get the coefficients of the right hand side terms in equation~\eqref{eq:flux_reconstruction_right_boundary}. For example, the flux reconstruction equation of $\widehat{F}_{j+\frac{1}{2}}$ for the right boundary (equation~\eqref{eq:CMD_RB}), where $j=N-1$, is given by:
\begin{align}
\frac{9}{80} \widehat{F}_{j-\frac{1}{2}} + \frac{71}{80} \widehat{F}_{j+\frac{1}{2}} &= \frac{27233}{768000} {F}_{j-2} - \frac{80779}{336000} {F}_{j-\frac{3}{2}} + \frac{26353}{38400} {F}_{j-1} - \frac{7811}{8000} {F}_{j-\frac{1}{2}} + \frac{65699}{76800} {F}_{j} \nonumber \\
 &\quad + \frac{10989}{16000} {F}_{j+\frac{1}{2}} - \frac{9007}{192000} {F}_{j+1} - \frac{1633}{5376000} {F}_{j+2}.
\end{align}

\subsection{Conservation}

For a continuous problem in a non-periodic 1D domain, we have conservation of $u(x, t)$ given by:
\begin{equation}
\frac{\partial}{\partial t} \int_{x_0 - \frac{\Delta x}{2}}^{x_{N-1} + \frac{\Delta x}{2}} u(x, t) dx = F\left(x_0- \frac{\Delta x}{2}, t\right) - F\left(x_{N-1} + \frac{\Delta x}{2}, t\right).
\end{equation}
Note that $x_0 - \Delta x/2$ and $x_{N-1} + \Delta x/2$ are the boundaries of the domain.
If we choose a test function $\psi(x)$ given by:
\begin{equation}
\psi(x) = \sum_{j=0}^{N-1} \Delta x \cdot \delta\left(x - x_j\right),
\end{equation}
where $\delta(x)$ is the Dirac delta function, we have:
\begin{equation}
\frac{\partial}{\partial t} \int_{x_0 - \frac{\Delta x}{2}}^{x_{N-1} + \frac{\Delta x}{2}} \psi(x) u(x, t) dx = \Delta x \sum_{j=0}^{N-1} \frac{\partial u_j}{\partial t}.
\end{equation}
With the conservation form given by equation~\eqref{eq:reconstruction_form} after semi-discrete discretization, we have:
\begin{align}
\frac{\partial}{\partial t} \int_{x_0 - \frac{\Delta x}{2}}^{x_{N-1} + \frac{\Delta x}{2}} \psi(x) u(x, t) dx &= \widehat{F}\left(x_{-\frac{1}{2} }, t\right) - \widehat{F}\left(x_{N-\frac{1}{2}}, t\right) \nonumber \\
&= \widehat{F}\left(x_0 - \frac{\Delta x}{2}, t\right) - \widehat{F}\left(x_{N-1} + \frac{\Delta x}{2}, t\right).
\end{align}
Hence, the conservation form given by equation~\eqref{eq:reconstruction_form} guarantees discrete conservation under the test function $\psi(x)$. This form also proves that central compact or explicit finite difference schemes are discretely conservative for a periodic domain.

The main benefit of the conservation form and the corresponding flux reconstruction form of compact finite difference schemes, however, is the ability to derive boundary closures for compact finite difference schemes so that discrete conservation is guaranteed. The reconstruction form also has potential to allow the use of compact finite difference schemes with adaptive mesh refinement in order to get conservation across mesh levels with appropriately derived boundary schemes.

\section{HLLC, HLL, and HLLC-HLL Riemann solvers \label{appendix:HLLC_HLL}}

The flux in the $x$ direction from the HLLC Riemann solver, $\mathbf{F}_{\textnormal{HLLC}}$, for a 3D problem is given by:
\begin{equation}
	\mathbf{F}_{\textnormal{HLLC}} = \frac{1+\sign(s_{*})}{2} \left[ \bm{F}(\bm{Q}_L) + s_{-} \left( \bm{Q}_{*L} - \bm{Q}_L \right) \right] + \frac{1-\sign(s_{*})}{2} \left[ \bm{F}(\bm{Q}_R) + s_{+} \left( \bm{Q}_{*R} -\bm{Q}_R \right) \right],
\end{equation}
where $L$ and $R$ are the left and right states respectively, and $\bm{Q}_L$ and $\bm{Q}_R$ are the corresponding conservative variable vectors. With $K = L$ or $R$, the star state is defined as:
\begin{equation}
		\bm{Q}_{*K} = \chi_{*K}
    \begin{bmatrix}
    	\rho_K \\
        \rho_K s_* \\
        \rho_K v_K \\
        \rho_K w_K \\
        E_k + \left( s_* - u_K \right) \left( \rho_K s_* + \frac{p_K}{s_K - u_K} \right)
    \end{bmatrix},
\end{equation}

\noindent where
\begin{equation}
    \chi_{*K} = \frac{s_K - u_K}{s_K - s_*}.
\end{equation}

\noindent We use the waves speeds suggested by \citet{einfeldt1991godunov}:
\begin{equation}
	s_{-} = \min{\left( 0, s_L \right)}, \quad s_{+} = \max{\left( 0, s_R \right)},
\end{equation}
and
\begin{equation}
	s_{L} = \min{\left( \bar{u} - \bar{c}, u_L - c_L \right)}, \quad s_{R} = \max{\left( \bar{u} + \bar{c}, u_R + c_R \right)},
\end{equation}
where $\bar{u}$ and $\bar{c}$ are the averages from the left and right states. Roe averages are used in this paper. Following \citet{batten1997choice}, the wave speed for the star state is given by:
\begin{equation}
	s_{*} = \frac{p_R - p_L + \rho_L u_L \left( s_L - u_L \right) - \rho_R u_R \left( s_R - u_R \right)}{\rho_L \left( s_L - u_L \right) - \rho_R \left( s_R - u_R \right) }.
\end{equation}

The flux from the HLL Riemann solver proposed by \citet{harten1983upstream}, $\mathbf{F}_{\textnormal{HLL}}$, is given by:
\begin{equation}
	\mathbf{F}_{\textnormal{HLL}} =
	\begin{cases}
    	\bm{F}(\bm{Q}_L) , &\mbox{if } s_L \geq 0, \\
        \frac{s_R \bm{F}(\bm{Q}_L) - s_L \bm{F}(\bm{Q}_R) + s_R s_L \left( \bm{Q}_R - \bm{Q}_L \right) }{s_R - s_L} &\mbox{if } s_L \leq 0 \leq s_R, \\
        \bm{F}(\bm{Q}_R) , &\mbox{if } s_R \leq 0.
    \end{cases}
\end{equation}

The hybrid flux in the $x$ direction from the HLLC-HLL Riemann solver proposed by \citet{huang2011cures}, $\mathbf{F}_{\textnormal{HLLC-HLL}}$, for a 3D problem is given by:
\begin{equation}
    \mathbf{F}_{\textnormal{HLLC-HLL}} = \bm{\Theta} \mathbf{F}_{\textnormal{HLLC}} + \left( \bm{I} - \bm{\Theta} \right) \mathbf{F}_{\textnormal{HLL}},
\end{equation}

\noindent where 
\begin{equation}
    \bm{\Theta} =
        \begin{pmatrix}
         \tilde{\alpha}_1 & 0 & 0 & 0 & 0 \\
         0 & 1 & 0 & 0 & 0 \\
         0 & 0 & \tilde{\alpha}_1 & 0 & 0 \\
         0 & 0 & 0 & \tilde{\alpha}_1 & 0 \\
         0 & 0 & 0 & 0 & 1 \\
         \end{pmatrix},
\end{equation}
and $\bm{I}$ is the identity matrix. The weight, $\tilde{\alpha}_1 \in \left[0, 1\right]$, suggested by \citet{huang2011cures} is used:
\begin{align} \label{eq:HLLC_HLL_betas}
\alpha_1 &=
\begin{cases}
    1, &\mbox{if } \left| \bm{u}_R - \bm{u}_L \right| < \epsilon, \\
    \frac{\left| u_R - u_L \right|}{\left| \bm{u}_R - \bm{u}_L \right|}, &\mbox{otherwise},
\end{cases} \\
\alpha_2 &= \sqrt{1 - \alpha_1^{2}}, \\
\tilde{\alpha}_1 &= \frac{1}{2} + \frac{1}{2} \frac{\alpha_1}{\alpha_1 + \alpha_2}.
\end{align}

\noindent $\epsilon=1.0\mathrm{e}{-15}$ is the usual small constant close to machine epsilon. $\tilde{\alpha}_1$ is designed in the way such that when the shock normal direction is aligned with the grid surface normal direction, the hybrid flux is purely the HLLC flux. When the shock normal direction is perpendicular to the surface normal direction, HLL flux adds dissipation by sharing the same weight as the HLLC flux. In 1D problems, the HLLC-HLL Riemann solver is reduced to the regular HLLC Riemann solver since the shock normal direction is always perpendicular to the grid surface normal.

\section{Effect of postprocessing pipeline for velocity gradient statistics \label{sec:CHIT_postprocessing}}

Statistics of velocity gradient quantities like vorticity or dilatation are important in the analysis of turbulent flows. Any field with a power law energy spectrum exponent of $< 2$ has a gradient power spectrum that grows with the wavenumber, which is the case for velocity fields in turbulent flows. This amplifies the sensitivity of gradient statistics to the derivative scheme used to compute velocity gradients from the primitive velocity fields. In this paper, we use Fourier spectral derivatives which are exact up to the Nyquist wavenumber assuming that the solution represented on the grid is not aliased. The results here, as a result, are different from some previously published results. For the compressible homogeneous isotropic turbulence case presented in section~\ref{sec:CHIT}, we present our results for the DNS reference solution using different postprocessing derivative schemes. We also compare them to previously published results of \citet{johnsen2010assessment}.

\begin{figure}[!ht]
\begin{center}
\subfigure[Velocity variance]{\includegraphics[width=0.48\textwidth]{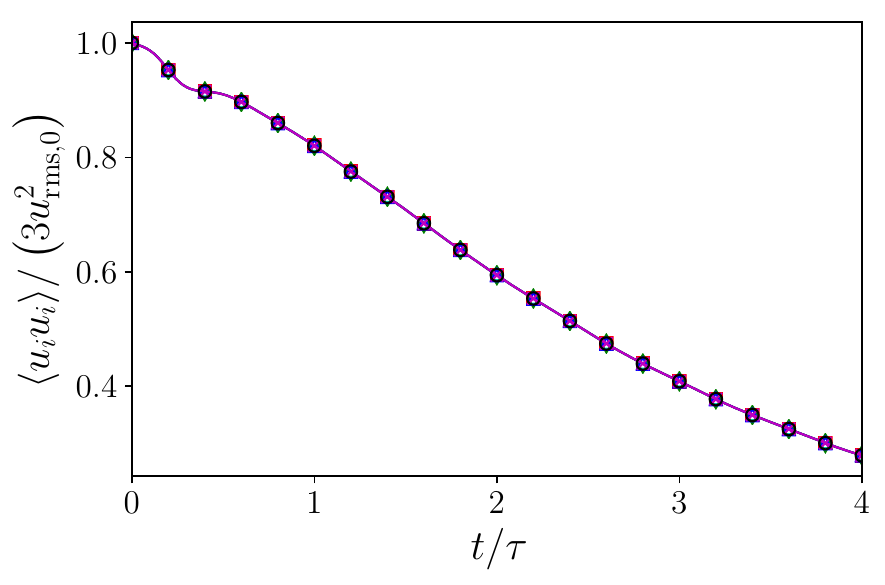} \label{fig:CHIT_postprocessing_VelocityVariance}} \\
\subfigure[Enstrophy]{\includegraphics[width=0.48\textwidth]{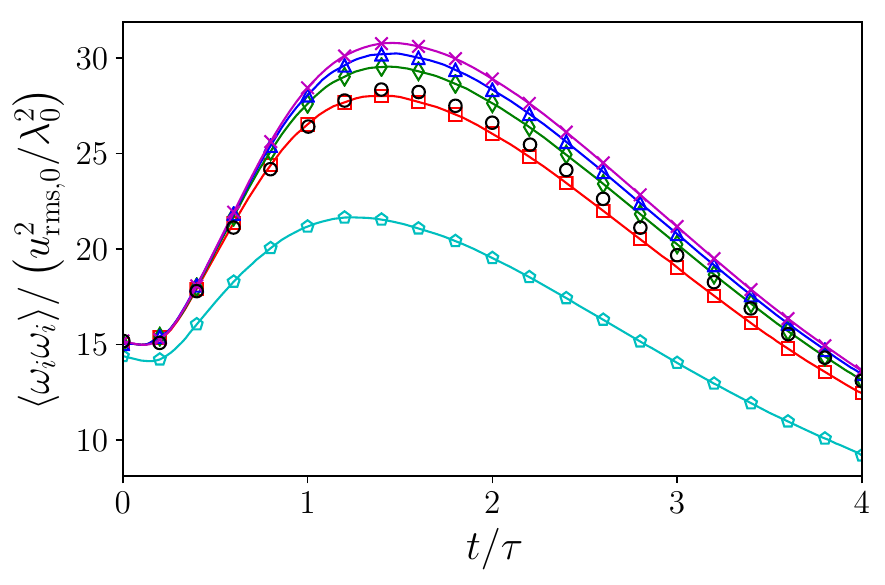} \label{fig:CHIT_postprocessing_VorticityVariance}}
\subfigure[Dilatation variance]{\includegraphics[width=0.48\textwidth]{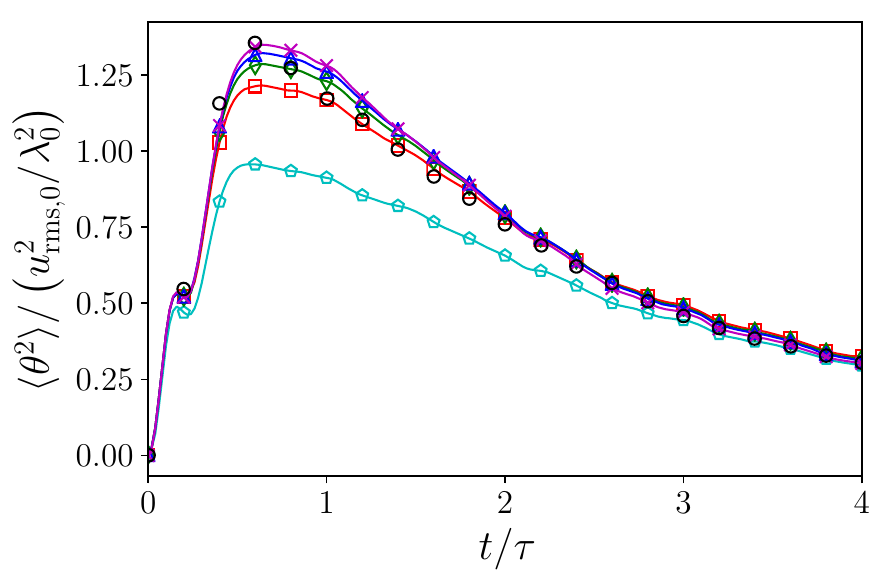} \label{fig:CHIT_postprocessing_DilatationVariance}}
\caption{Time evolution of statistical quantities for the compressible homogeneous isotropic turbulence problem presented in section~\ref{sec:CHIT} using different postprocessing derivative schemes on the DNS data. Black circles: \citet{johnsen2010assessment}; cyan pentagons: 2\textsuperscript{nd} order central explicit finite difference; red squares: 6\textsuperscript{th} order central explicit finite difference; green diamonds: 6\textsuperscript{th} order central compact finite difference; blue triangles: 10\textsuperscript{th} order central compact finite difference; magenta crosses: spectral derivatives.}
\label{fig:CHIT_postprocessing}
\end{center}
\end{figure}

Figure~\ref{fig:CHIT_postprocessing} shows the velocity statistics postprocessed using different derivative schemes. All results are obtained by spectrally filtering the velocity fields and then downsampling from the DNS resolution of $512^3$ down to $64^3$. The derivative operators are constructed on the downsampled $64^3$ grid and applied using the periodic boundary conditions of the problem. From figure~\ref{fig:CHIT_postprocessing_VorticityVariance}, we see that the velocity variance is the same for all the cases and match the results of \citet{johnsen2010assessment}. Figure~\ref{fig:CHIT_postprocessing_VorticityVariance} shows the enstrophy computed with different derivative schemes. From this, the effect of the postprocessing pipeline is evident. Using spectral derivatives which is the most accurate in the high wavenumber region has the highest enstrophy. For the other derivative schemes, the lower order derivatives capture much lower enstrophy. Also, compact derivatives are better than their explicit counterparts for the same order or accuracy. All of these results are in line with the modified wavenumbers of each derivative scheme. The same is true for the dilatation variance plotted in figure~\ref{fig:CHIT_postprocessing_DilatationVariance}. The plots also show the results of \citet{johnsen2010assessment} which are closest to the results using sixth order explicit finite difference. It was also confirmed by Larsson\footnote{Through private communication, 2018} (one of the authors of \citet{johnsen2010assessment}) that the sixth order explicit finite difference scheme was used for postprocessing. We see some difference between the sixth order explicit derivatives and the results of \citet{johnsen2010assessment} in the dilatation variance for $t/\tau < 1$. Since the initial conditions are solenoidal and generated randomly following a prescribed spectrum, the dilatation variance is sensitive to the initial conditions during the early acoustic transients and the disagreement between different initial conditions is to be expected.

% \section*{Appendix}
% 
% An appendix, if needed, should appear before the acknowledgments.
% Use the 'starred' version of the \verb|\section| commands to avoid
% section numbering.
% 

\section*{References}

% produces the bibliography section when processed by BibTeX
\bibliographystyle{abbrvnat}
\bibliography{bibtex_database.bib}

\begin{thebibliography}{49}
\providecommand{\natexlab}[1]{#1}
\providecommand{\url}[1]{\texttt{#1}}
\expandafter\ifx\csname urlstyle\endcsname\relax
  \providecommand{\doi}[1]{doi: #1}\else
  \providecommand{\doi}{doi: \begingroup \urlstyle{rm}\Url}\fi

\bibitem[Batten et~al.(1997)Batten, Clarke, Lambert, and
  Causon]{batten1997choice}
P.~Batten, N.~Clarke, C.~Lambert, and D.~Causon.
\newblock On the choice of wavespeeds for the {HLLC} {Riemann} solver.
\newblock \emph{SIAM Journal on Scientific Computing}, 18\penalty0
  (6):\penalty0 1553--1570, 1997.

\bibitem[Bhagatwala and Lele(2009)]{bhagatwala2009modified}
A.~Bhagatwala and S.~K. Lele.
\newblock A modified artificial viscosity approach for compressible turbulence
  simulations.
\newblock \emph{Journal of Computational Physics}, 228\penalty0 (14):\penalty0
  4965--4969, 2009.

\bibitem[Borges et~al.(2008)Borges, Carmona, Costa, and
  Don]{borges2008improved}
R.~Borges, M.~Carmona, B.~Costa, and W.~S. Don.
\newblock An improved weighted essentially non-oscillatory scheme for
  hyperbolic conservation laws.
\newblock \emph{Journal of Computational Physics}, 227\penalty0 (6):\penalty0
  3191--3211, 2008.

\bibitem[Brachet et~al.(1983)Brachet, Meiron, Orszag, Nickel, Morf, and
  Frisch]{brachet1983small}
M.~E. Brachet, D.~I. Meiron, S.~A. Orszag, B.~Nickel, R.~H. Morf, and
  U.~Frisch.
\newblock Small-scale structure of the {Taylor}--{Green} vortex.
\newblock \emph{Journal of Fluid Mechanics}, 130:\penalty0 411--452, 1983.

\bibitem[Chatterjee and Vijayaraj(2008)]{chatterjee2008multiple}
A.~Chatterjee and S.~Vijayaraj.
\newblock Multiple sound generation in interaction of shock wave with strong
  vortex.
\newblock \emph{AIAA journal}, 46\penalty0 (10):\penalty0 2558--2567, 2008.

\bibitem[Cook(2007)]{cook2007artificial}
A.~W. Cook.
\newblock Artificial fluid properties for large-eddy simulation of compressible
  turbulent mixing.
\newblock \emph{Physics of Fluids (1994-present)}, 19\penalty0 (5):\penalty0
  055103, 2007.

\bibitem[Cook and Cabot(2004)]{cook2004high}
A.~W. Cook and W.~H. Cabot.
\newblock A high-wavenumber viscosity for high-resolution numerical methods.
\newblock \emph{Journal of Computational Physics}, 195\penalty0 (2):\penalty0
  594--601, 2004.

\bibitem[Cook and Cabot(2005)]{cook2005hyperviscosity}
A.~W. Cook and W.~H. Cabot.
\newblock Hyperviscosity for shock-turbulence interactions.
\newblock \emph{Journal of Computational Physics}, 203\penalty0 (2):\penalty0
  379--385, 2005.

\bibitem[Deng and Zhang(2000)]{deng2000developing}
X.~Deng and H.~Zhang.
\newblock Developing high-order weighted compact nonlinear schemes.
\newblock \emph{Journal of Computational Physics}, 165\penalty0 (1):\penalty0
  22--44, 2000.

\bibitem[Einfeldt et~al.(1991)Einfeldt, Munz, Roe, and
  Sj{\"o}green]{einfeldt1991godunov}
B.~Einfeldt, C.-D. Munz, P.~L. Roe, and B.~Sj{\"o}green.
\newblock On {Godunov}-type methods near low densities.
\newblock \emph{Journal of Computational Physics}, 92\penalty0 (2):\penalty0
  273--295, 1991.

\bibitem[Fu et~al.(2016)Fu, Hu, and Adams]{fu2016family}
L.~Fu, X.~Hu, and N.~A. Adams.
\newblock A family of high-order targeted {ENO} schemes for compressible-fluid
  simulations.
\newblock \emph{Journal of Computational Physics}, 305:\penalty0 333--359,
  2016.

\bibitem[Fu et~al.(2017)Fu, Hu, and Adams]{fu2017targeted}
L.~Fu, X.~Hu, and N.~A. Adams.
\newblock Targeted {ENO} schemes with tailored resolution property for
  hyperbolic conservation laws.
\newblock \emph{Journal of Computational Physics}, 349:\penalty0 97--121, 2017.

\bibitem[Ghaisas et~al.(2018)Ghaisas, Subramaniam, and
  Lele]{ghaisas2018unified}
N.~S. Ghaisas, A.~Subramaniam, and S.~K. Lele.
\newblock A unified high-order {Eulerian} method for continuum simulations of
  fluid flow and of elastic-plastic deformations in solids.
\newblock \emph{Journal of Computational Physics}, 371:\penalty0 452--482,
  2018.

\bibitem[Ghosh and Baeder(2012)]{ghosh2012compact}
D.~Ghosh and J.~D. Baeder.
\newblock Compact reconstruction schemes with weighted {ENO} limiting for
  hyperbolic conservation laws.
\newblock \emph{SIAM Journal on Scientific Computing}, 34\penalty0
  (3):\penalty0 A1678--A1706, 2012.

\bibitem[Granet et~al.(2010)Granet, Vermorel, L{\'e}onard, Gicquel, and
  Poinsot]{granet2010comparison}
V.~Granet, O.~Vermorel, T.~L{\'e}onard, L.~Gicquel, and T.~Poinsot.
\newblock Comparison of nonreflecting outlet boundary conditions for
  compressible solvers on unstructured grids.
\newblock \emph{AIAA journal}, 48\penalty0 (10):\penalty0 2348--2364, 2010.

\bibitem[Harten et~al.(1983)Harten, Lax, and Leer]{harten1983upstream}
A.~Harten, P.~D. Lax, and B.~v. Leer.
\newblock On upstream differencing and {Godunov}-type schemes for hyperbolic
  conservation laws.
\newblock \emph{SIAM Review}, 25\penalty0 (1):\penalty0 35--61, 1983.

\bibitem[Henrick et~al.(2005)Henrick, Aslam, and Powers]{henrick2005mapped}
A.~K. Henrick, T.~D. Aslam, and J.~M. Powers.
\newblock Mapped weighted essentially non-oscillatory schemes: achieving
  optimal order near critical points.
\newblock \emph{Journal of Computational Physics}, 207\penalty0 (2):\penalty0
  542--567, 2005.

\bibitem[Hu and Adams(2011)]{hu2011scale}
X.~Hu and N.~Adams.
\newblock Scale separation for implicit large eddy simulation.
\newblock \emph{Journal of Computational Physics}, 230\penalty0 (19):\penalty0
  7240--7249, 2011.

\bibitem[Hu et~al.(2010)Hu, Wang, and Adams]{hu2010adaptive}
X.~Hu, Q.~Wang, and N.~Adams.
\newblock An adaptive central-upwind weighted essentially non-oscillatory
  scheme.
\newblock \emph{Journal of Computational Physics}, 229\penalty0 (23):\penalty0
  8952--8965, 2010.

\bibitem[Hu et~al.(2013)Hu, Adams, and Shu]{hu2013positivity}
X.~Hu, N.~A. Adams, and C.-W. Shu.
\newblock Positivity-preserving method for high-order conservative schemes
  solving compressible {Euler} equations.
\newblock \emph{Journal of Computational Physics}, 242:\penalty0 169--180,
  2013.

\bibitem[Huang et~al.(2011)Huang, Wu, Yu, and Yan]{huang2011cures}
K.~Huang, H.~Wu, H.~Yu, and D.~Yan.
\newblock Cures for numerical shock instability in {HLLC} solver.
\newblock \emph{International Journal for Numerical Methods in Fluids},
  65\penalty0 (9):\penalty0 1026--1038, 2011.

\bibitem[Inoue and Hattori(1999)]{inoue1999sound}
O.~Inoue and Y.~Hattori.
\newblock Sound generation by shock-vortex interactions.
\newblock \emph{Journal of Fluid Mechanics}, 380:\penalty0 81--116, 1999.

\bibitem[Jiang and Shu(1995)]{jiang1995efficient}
G.-S. Jiang and C.-W. Shu.
\newblock Efficient implementation of weighted {ENO} schemes.
\newblock Technical report, DTIC Document, 1995.

\bibitem[Johnsen et~al.(2010)Johnsen, Larsson, Bhagatwala, Cabot, Moin, Olson,
  Rawat, Shankar, Sj{\"o}green, Yee, Zhong, and Lele]{johnsen2010assessment}
E.~Johnsen, J.~Larsson, A.~V. Bhagatwala, W.~H. Cabot, P.~Moin, B.~J. Olson,
  P.~S. Rawat, S.~K. Shankar, B.~Sj{\"o}green, H.~C. Yee, X.~Zhong, and S.~K.
  Lele.
\newblock Assessment of high-resolution methods for numerical simulations of
  compressible turbulence with shock waves.
\newblock \emph{Journal of Computational Physics}, 229\penalty0 (4):\penalty0
  1213--1237, 2010.

\bibitem[Kawai et~al.(2010)Kawai, Shankar, and Lele]{kawai2010assessment}
S.~Kawai, S.~K. Shankar, and S.~K. Lele.
\newblock Assessment of localized artificial diffusivity scheme for large-eddy
  simulation of compressible turbulent flows.
\newblock \emph{Journal of Computational Physics}, 229\penalty0 (5):\penalty0
  1739--1762, 2010.

\bibitem[Larsson et~al.(2007)Larsson, Lele, and Moin]{larsson2007effect}
J.~Larsson, S.~Lele, and P.~Moin.
\newblock Effect of numerical dissipation on the predicted spectra for
  compressible turbulence.
\newblock In \emph{Annual Research Briefs}, pages 45--57. Center for Turbulence
  Research, Stanford University, 2007.

\bibitem[Lee et~al.(1991)Lee, Lele, and Moin]{lee1991eddy}
S.~Lee, S.~K. Lele, and P.~Moin.
\newblock Eddy shocklets in decaying compressible turbulence.
\newblock \emph{Physics of Fluids A: Fluid Dynamics}, 3\penalty0 (4):\penalty0
  657--664, 1991.

\bibitem[Lele(1992)]{lele1992compact}
S.~K. Lele.
\newblock Compact finite difference schemes with spectral-like resolution.
\newblock \emph{Journal of Computational Physics}, 103\penalty0 (1):\penalty0
  16--42, 1992.

\bibitem[Liu et~al.(2015)Liu, Zhang, Zhang, and Shu]{liu2015new}
X.~Liu, S.~Zhang, H.~Zhang, and C.-W. Shu.
\newblock A new class of central compact schemes with spectral-like resolution
  {II}: Hybrid weighted nonlinear schemes.
\newblock \emph{Journal of Computational Physics}, 284:\penalty0 133--154,
  2015.

\bibitem[Mart{\'\i}n et~al.(2006)Mart{\'\i}n, Taylor, Wu, and
  Weirs]{martin2006bandwidth}
M.~P. Mart{\'\i}n, E.~M. Taylor, M.~Wu, and V.~G. Weirs.
\newblock A bandwidth-optimized {WENO} scheme for the effective direct
  numerical simulation of compressible turbulence.
\newblock \emph{Journal of Computational Physics}, 220\penalty0 (1):\penalty0
  270--289, 2006.

\bibitem[Motheau et~al.(2017)Motheau, Almgren, and Bell]{motheau2017navier}
E.~Motheau, A.~Almgren, and J.~B. Bell.
\newblock {Navier}--{Stokes} characteristic boundary conditions using ghost
  cells.
\newblock \emph{AIAA Journal}, pages 1--10, 2017.

\bibitem[Nagarajan et~al.(2003)Nagarajan, Lele, and
  Ferziger]{nagarajan2003robust}
S.~Nagarajan, S.~K. Lele, and J.~H. Ferziger.
\newblock A robust high-order compact method for large eddy simulation.
\newblock \emph{Journal of Computational Physics}, 191\penalty0 (2):\penalty0
  392--419, 2003.

\bibitem[Nguetchue and Abelman(2008)]{nguetchue2008computational}
S.~N. Nguetchue and S.~Abelman.
\newblock A computational algorithm for solving nearly penta-diagonal linear
  systems.
\newblock \emph{Applied Mathematics and Computation}, 203\penalty0
  (2):\penalty0 629--634, 2008.

\bibitem[Nonomura and Fujii(2009)]{nonomura2009effects}
T.~Nonomura and K.~Fujii.
\newblock Effects of difference scheme type in high-order weighted compact
  nonlinear schemes.
\newblock \emph{Journal of Computational Physics}, 228\penalty0 (10):\penalty0
  3533--3539, 2009.

\bibitem[Nonomura and Fujii(2013)]{nonomura2013robust}
T.~Nonomura and K.~Fujii.
\newblock Robust explicit formulation of weighted compact nonlinear scheme.
\newblock \emph{Computers \& Fluids}, 2013.

\bibitem[Nonomura et~al.(2007)Nonomura, Iizuka, and
  Fujii]{nonomura2007increasing}
T.~Nonomura, N.~Iizuka, and K.~Fujii.
\newblock Increasing order of accuracy of weighted compact nonlinear scheme.
\newblock \emph{AIAA Paper}, 893, 2007.

\bibitem[Pirozzoli(2006)]{pirozzoli2006spectral}
S.~Pirozzoli.
\newblock On the spectral properties of shock-capturing schemes.
\newblock \emph{Journal of Computational Physics}, 219\penalty0 (2):\penalty0
  489--497, 2006.

\bibitem[Pirozzoli(2010)]{pirozzoli2010generalized}
S.~Pirozzoli.
\newblock Generalized conservative approximations of split convective
  derivative operators.
\newblock \emph{Journal of Computational Physics}, 229\penalty0 (19):\penalty0
  7180--7190, 2010.

\bibitem[Sedov(1993)]{sedov1993similarity}
L.~I. Sedov.
\newblock \emph{Similarity and dimensional methods in mechanics}.
\newblock CRC press, 1993.

\bibitem[Shankar et~al.(2010)Shankar, Kawai, and Lele]{shankar2010numerical}
S.~Shankar, S.~Kawai, and S.~Lele.
\newblock Numerical simulation of multicomponent shock accelerated flows and
  mixing using localized artificial diffusivity method.
\newblock In \emph{48th AIAA Aerospace Sciences Meeting Including the New
  Horizons Forum and Aerospace Exposition}, page 352, 2010.

\bibitem[Shu and Osher(1988)]{shu1988efficient}
C.-W. Shu and S.~Osher.
\newblock Efficient implementation of essentially non-oscillatory
  shock-capturing schemes.
\newblock \emph{Journal of Computational Physics}, 77\penalty0 (2):\penalty0
  439--471, 1988.

\bibitem[Sod(1978)]{sod1978survey}
G.~A. Sod.
\newblock A survey of several finite difference methods for systems of
  nonlinear hyperbolic conservation laws.
\newblock \emph{Journal of Computational Physics}, 27\penalty0 (1):\penalty0
  1--31, 1978.

\bibitem[Spiteri and Ruuth(2002)]{spiteri2002new}
R.~J. Spiteri and S.~J. Ruuth.
\newblock A new class of optimal high-order strong-stability-preserving time
  discretization methods.
\newblock \emph{SIAM Journal on Numerical Analysis}, 40\penalty0 (2):\penalty0
  469--491, 2002.

\bibitem[Subramaniam et~al.(2018)Subramaniam, Ghaisas, and
  Lele]{subramaniam2018high}
A.~Subramaniam, N.~S. Ghaisas, and S.~K. Lele.
\newblock High-order {Eulerian} simulations of multimaterial elastic-plastic
  flow.
\newblock \emph{Journal of Fluids Engineering}, 140\penalty0 (5):\penalty0
  050904, 2018.

\bibitem[Wong and Lele(2017)]{wong2017high}
M.~L. Wong and S.~K. Lele.
\newblock High-order localized dissipation weighted compact nonlinear scheme
  for shock-and interface-capturing in compressible flows.
\newblock \emph{Journal of Computational Physics}, 339:\penalty0 179--209,
  2017.

\bibitem[Woodward and Colella(1984)]{woodward1984numerical}
P.~Woodward and P.~Colella.
\newblock The numerical simulation of two-dimensional fluid flow with strong
  shocks.
\newblock \emph{Journal of Computational Physics}, 54\penalty0 (1):\penalty0
  115--173, 1984.

\bibitem[Zhang et~al.(2005)Zhang, Zhang, and Shu]{zhang2005multistage}
S.~Zhang, Y.-T. Zhang, and C.-W. Shu.
\newblock Multistage interaction of a shock wave and a strong vortex.
\newblock \emph{Physics of Fluids (1994-present)}, 17\penalty0 (11):\penalty0
  116101, 2005.

\bibitem[Zhang et~al.(2008)Zhang, Jiang, and Shu]{zhang2008development}
S.~Zhang, S.~Jiang, and C.-W. Shu.
\newblock Development of nonlinear weighted compact schemes with increasingly
  higher order accuracy.
\newblock \emph{Journal of Computational Physics}, 227\penalty0 (15):\penalty0
  7294--7321, 2008.

\bibitem[Zhang and Shu(2012)]{zhang2012positivity}
X.~Zhang and C.-W. Shu.
\newblock Positivity-preserving high order finite difference {WENO} schemes for
  compressible euler equations.
\newblock \emph{Journal of Computational Physics}, 231\penalty0 (5):\penalty0
  2245--2258, 2012.

\end{thebibliography}

\end{document}